\documentclass[useAMS,usenatbib]{mn2e}
\usepackage[normalem]{ulem}
\usepackage{graphicx}
\usepackage{amsmath}
\usepackage{color}
\usepackage{amsfonts}
\usepackage{hyperref}
\usepackage{supertabular}
\begin{document}

%\title{How well do stellar population models reproduce u,g,r,i,z-band photometry of the local galaxy population?}
\title[Observed versus modelled $u$,$g$,$r$,$i$,$z$-band photometry of local galaxies]{Observed versus modelled $u$,$g$,$r$,$i$,$z$-band photometry of local galaxies - Evaluation of model performance}
%\author{K. S. Alexander Hansson$^{ 1}$, Thorsten Lisker$^{ 1}$, Eva K. Grebel$^{ 1}$ and John S. Gallagher, III$^{ 2}$\\
%   $^{ 2}$University of Wisconsin-Madison}
\author[Hansson et al.]
{K. S. Alexander Hansson\thanks{E-mail: alexander@x-astro.net}, Thorsten Lisker and Eva K. Grebel\\
Astronomisches Rechen-Institut, Zentrum f\"ur Astronomie der Universit\"at Heidelberg, M\"onchhofstr.\ 12-14, 69120 Heidelberg, Germany\\
}
%\author{hej}
\date{Accepted 4 July 2012}
\maketitle
%__________________________________________________________
\begin{abstract}

We test how well available stellar population models can reproduce observed $u$,$g$,$r$,$i$,$z$-band photometry of the local galaxy population ($0.02\leq z\leq 0.03$) as probed by the SDSS. Our study is conducted from the perspective of a user of the models, who
has observational data in hand and seeks to convert them into physical
quantities. Stellar population models for galaxies are created by synthesizing star formations histories and chemical enrichments using single stellar populations from several groups (Starburst99, GALAXEV, Maraston2005, GALEV).
The role of dust is addressed through a simplistic, but observationally motivated, dust model that couples the amplitude of the extinction to the star formation history, metallicity and the viewing angle. Moreover, the influence of emission lines is considered (for the subset of models for which this component is included).
The performance of the models is investigated by: 1) comparing their prediction with the observed galaxy population in the SDSS using the ($u$-$g$)-($r$-$i$) and ($g$-$r$)-($i$-$z$) color planes, 2) comparing predicted stellar mass and luminosity weighted ages and metallicities, specific star formation rates, mass to light ratios and total extinctions with literature values from studies based on spectroscopy.
Strong differences between the various models are seen with several models occupying regions in the color-color diagrams where no galaxies are observed. We would therefore like to emphasize the importance of the choice of model. Using our preferred model we find that the star formation history, metallicity and also dust content can be constrained over a large part of the parameter space through the use of $u$,$g$,$r$,$i$,$z$-band photometry. However, strong local degeneracies are present due to overlap of models with high and low extinction in certain parts of color space.

\end{abstract}


%_____________________INTRODUCTION
\section{INTRODUCTION}

Optical broad band colors have proven to be a powerful tool in studying galaxies. Their dependence on luminosity and environment have greatly increased our knowledge of these systems \citep{1977ApJ...216..214V,2007ApJ...658..898P,2008AJ....135..380L}. Colors are also used to derive quantities such as star formation histories and stellar masses \citep{1968ApJ...151..547T,1973ApJ...179..427S,1991ApJ...367..126C,2001ApJ...550..212B,2003MNRAS.344.1000B,2007AJ....133..734B}, both of which are key properties for understanding galaxy formation and evolution. For example, strong correlations between stellar mass and galaxy structure \citep{2003MNRAS.341...54K}, star formation history \citep{2007MNRAS.378.1550P}, chemical enrichment \citep{2008MNRAS.391.1117P} and gas content \citep{2009MNRAS.397.1243Z} have been presented suggesting that mass is the main driver behind galaxy evolution. These relations reflect the importance of gravity on galactic scales and, moreover, provide further evidence concerning the expected connection between stellar mass and dark matter \citep[cf.][]{2010ApJ...710..903M}. The derivation of star formation histories and stellar masses are made through a modelling of the light emission from the galaxy and, if necessary, modelling of obscuration by dust. The method enables a comparison between observational quantities and galaxy formation models \citep[e.g.][]{2006MNRAS.366..499D,2006MNRAS.370..645B,2011MNRAS.413..101G}. The quality of the derivation of star formation histories and stellar masses from colors naturally depends on the quality of the models. A straightforward test is to check how well the models can reproduce the ensemble of observables. In this paper we therefore take a closer look at how successful various models are in reproducing $u$,$g$,$r$,$i$,$z$-band photometry of the $local$\footnote{By focusing on local galaxies ($0.02<z<0.03$) we circumvent the shift in the spectral energy distribution caused by redshift.} galaxy population.

The base ingredient of stellar population models of galaxies are single stellar populations (SSPs), which are combined into star formation histories by linear combinations \citep{1968ApJ...151..547T,1973ApJ...179..427S}. SSPs can consequently be seen as a basic ingredient of the models having a great impact on the emergent spectral energy distribution. A large number of single stellar population models are available in the literature, including those of \cite{1999ApJS..123....3L,2003MNRAS.344.1000B,2005MNRAS.362..799M,2009MNRAS.396..462K}, which we use in this paper. We do not aim at being complete considering the numerous options available. However, we intend to cover some of the most widely used models predicting $u$,$g$,$r$,$i$,$z$-band photometry.

We treat dust obscuration in a simple way. The amplitude of the extinction is coupled to a galaxy's star formation history, metallicity and viewing angle, as motivated by works of \cite{2005MNRAS.358..363C,2008ApJ...678..804E,2010MNRAS.403.1894D,2010MNRAS.404..792M}, and a power law form of the the wavelength dependence of the extinction is assumed \citep[see, e.g.,][]{2000ApJ...539..718C}. This method can be seen as an easy, but simplistic, approach with respect to detailed modelling of dust \citep[e.g][]{2004AA...419..821T,2011AA...527A.109P}, though it has the nice feature that it only requires the axis ratio apart from colors.

As spectroscopic data carry a lot more information than $u$,$g$,$r$,$i$,$z$-band photometry a comparison with these kinds of data can serve as an important test. We therefore carry out a detailed comparison between literature data based on spectroscopy and our photometric model in order to validate the latter.

We note that a different approach to the testing of stellar population models has been performed by \cite{2009ApJ...699..486C,2010ApJ...708...58C} and \cite{2010ApJ...712..833C}. We refer the reader to this series of papers if they are interested in more details on the impact and uncertainties of various model ingredients rather than overall performance for optical broad band photometry.

This paper is organized as follows. In Section \ref{sec:obs} the observational sample used for the model evaluation is presented. The SSPs are introduced in Section \ref{sec:ssp} and how these are combined into star formation histories is described in Section \ref{sec:sfh}. How we treat dust is described in Section \ref{sec:dust}. The performance of the models is presented in Section \ref{sec:per}. The results are discussed in Section \ref{sec:dis}, and summarized in Section \ref{sec:sum}.

\section{OBSERVATIONAL SAMPLE}
\label{sec:obs}

Our source of observational imaging data is the Sloan Digital Sky Survey (SDSS) data release 7 \citep[DR7][]{2009ApJS..182..543A}. Galaxies with measured spectroscopic redshifts from the SDSS in the range $0.02<z<0.03$ are chosen to minimize the shifts in the spectral energy distribution caused by redshifts, while keeping a large sample with photometry of sufficient quality (note that, in particular, nearby galaxies have inaccurate photometric measurements from the SDSS pipeline \citep{2005AJ....129.2562B,2007ApJ...662..808L,2007AJ....133.1741B,2007ApJ...660.1186L,2011AJ....142...31B}.) Our sample consists of 24120 galaxies. We downloaded model magnitudes (modelMag) for these galaxies, which have been corrected for Galactic extinction using the maps of \cite{1998ApJ...500..525S}. The sample spans an absolute magnitude range of $-17\ge M_{r}\ge-23$, and thus only the brightest of the dwarf galaxies are included. The photometry is k-corrected using the code of \cite{2010MNRAS.405.1409C}, but the small amplitude of the corrections ($\leq0.1$) suggests that the errors introduced in this step are negligible. In this paper we only make use of magnitudes in the AB system, including the SDSS to AB conversion factors\footnote{See www.sdss.org/dr7/algorithms/fluxcal} of $u_{AB}=u_{SDSS}-0.04\mbox{mag}$ and $z_{AB}=z_{SDSS}+0.02\mbox{mag}$. A quality assessment of the galaxy colors, as described in the Appendix \ref{sec:a}, shows that the distribution of colors for the sample appears to be consistent with independent measurements.

\begin{figure}
\includegraphics[width=4.0cm]{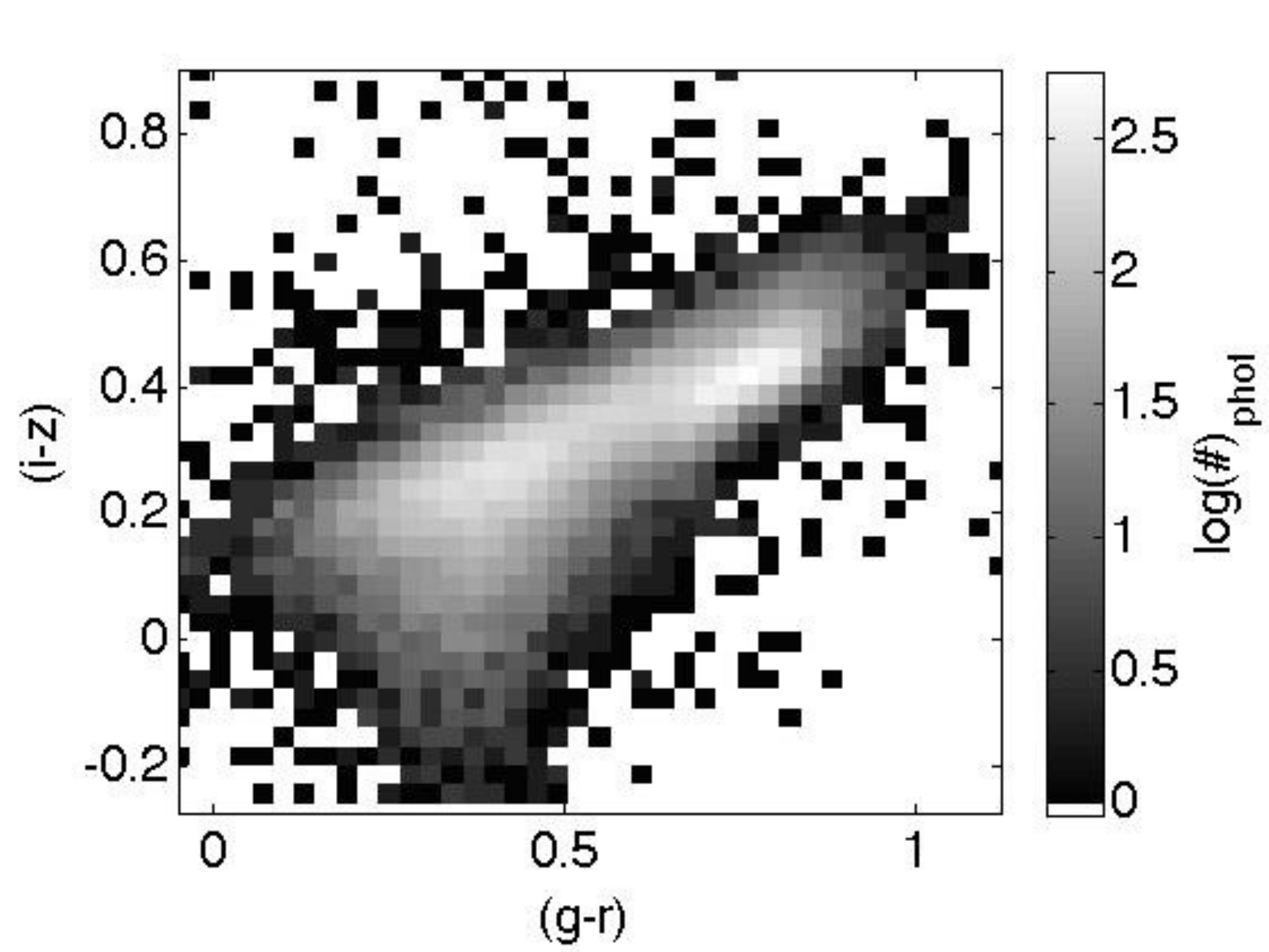}
\includegraphics[width=4.0cm]{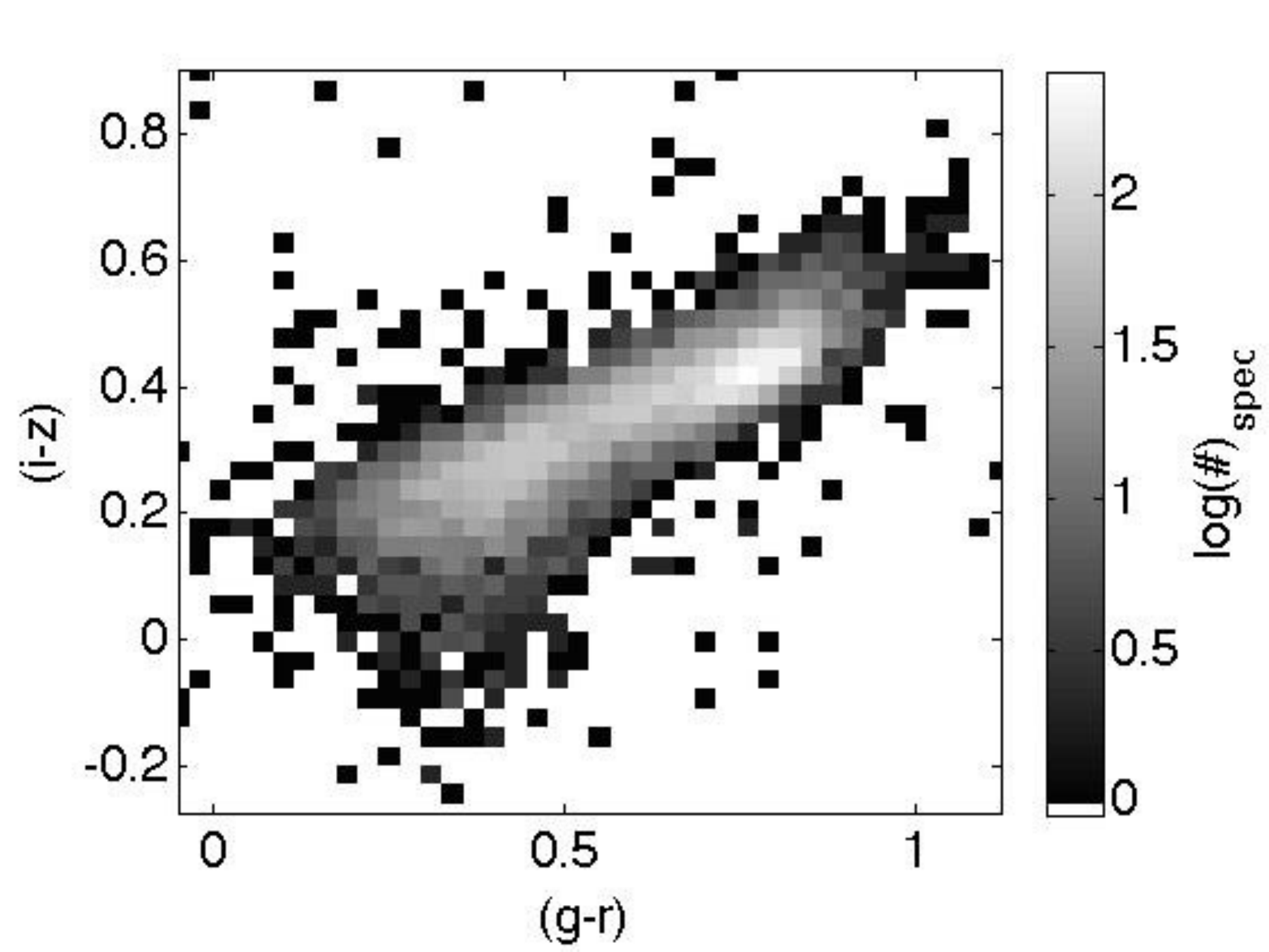}\\
\includegraphics[width=4.0cm]{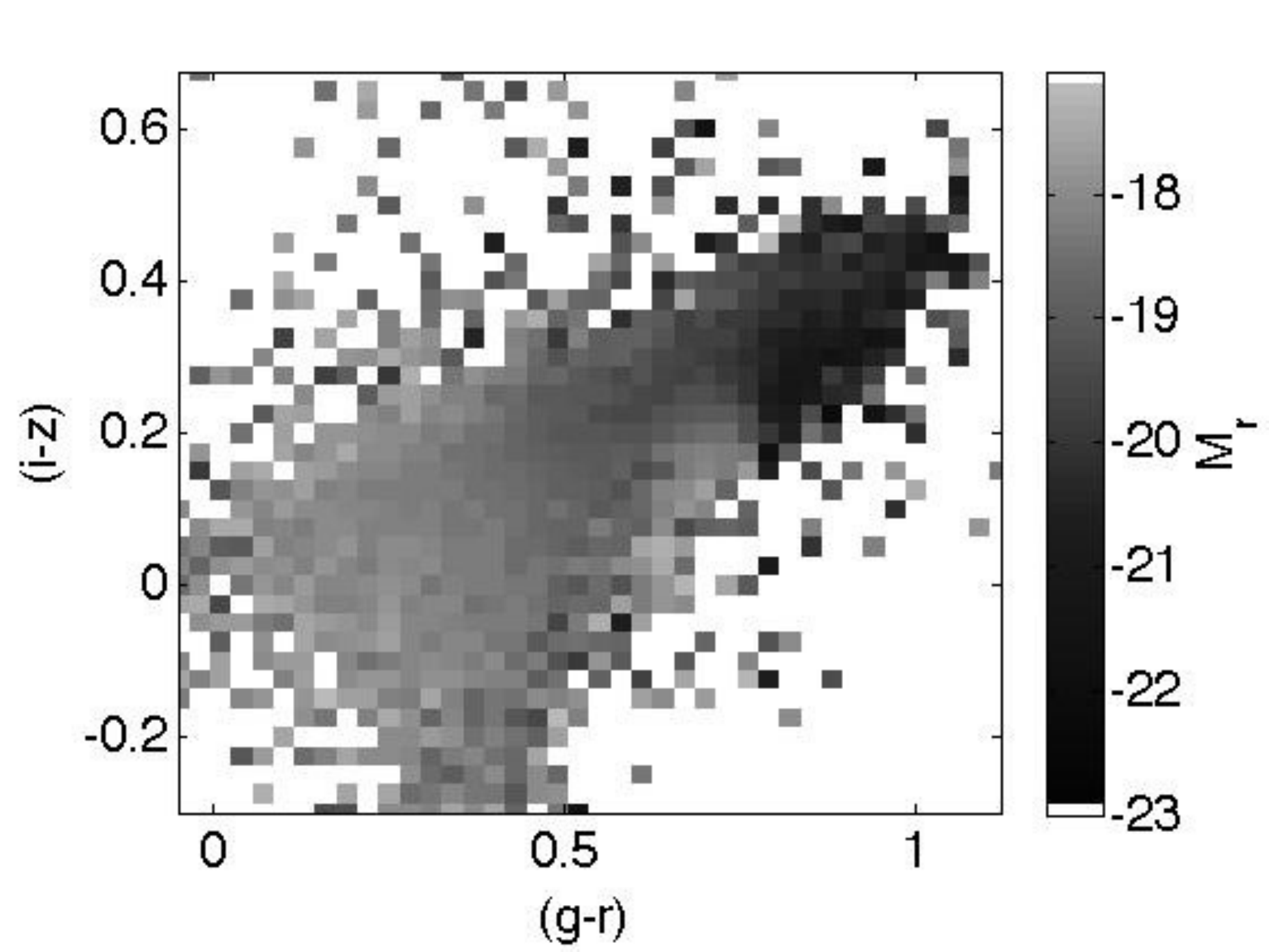}
\includegraphics[width=4.0cm]{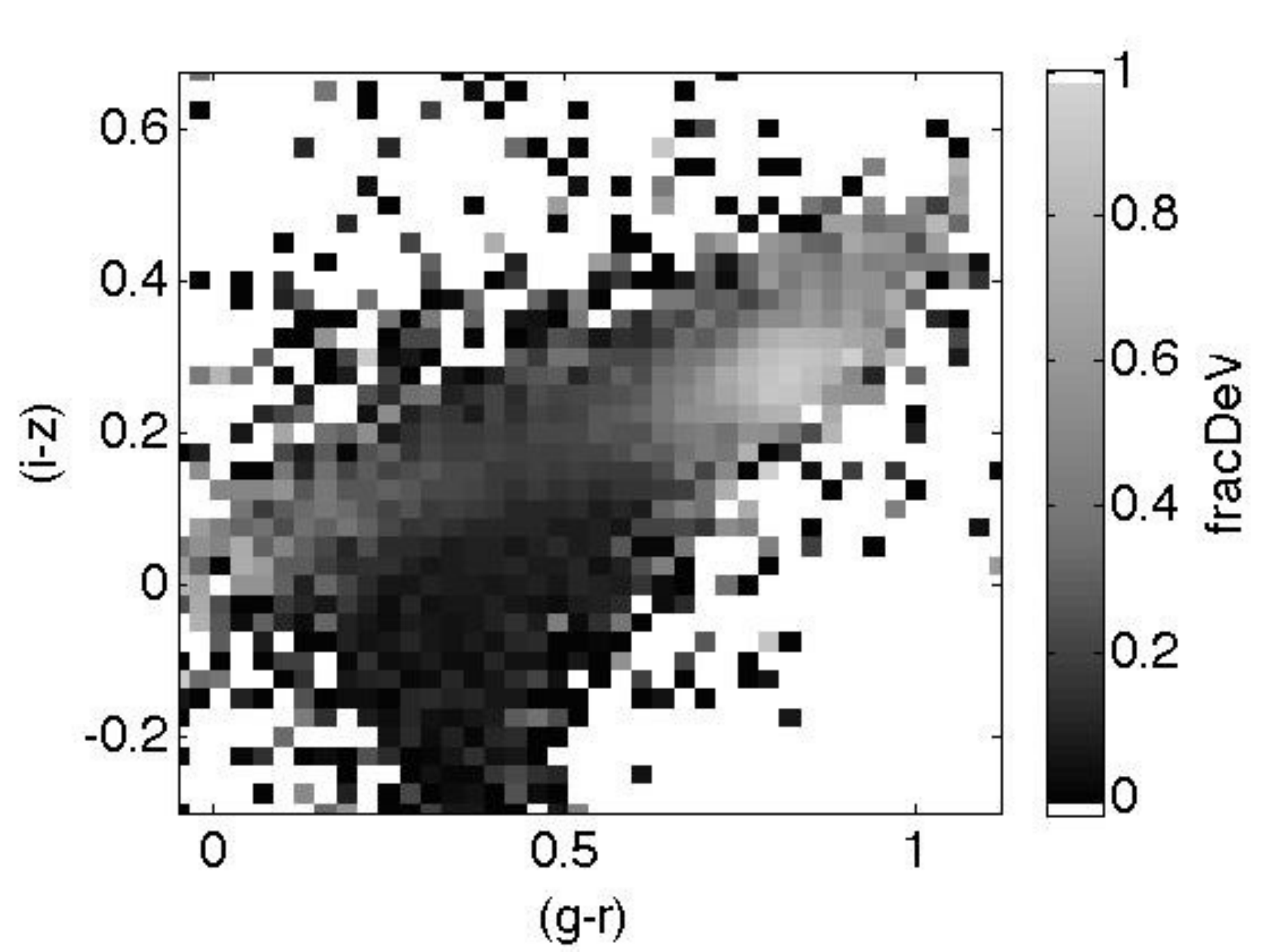}\\
\includegraphics[width=4.0cm]{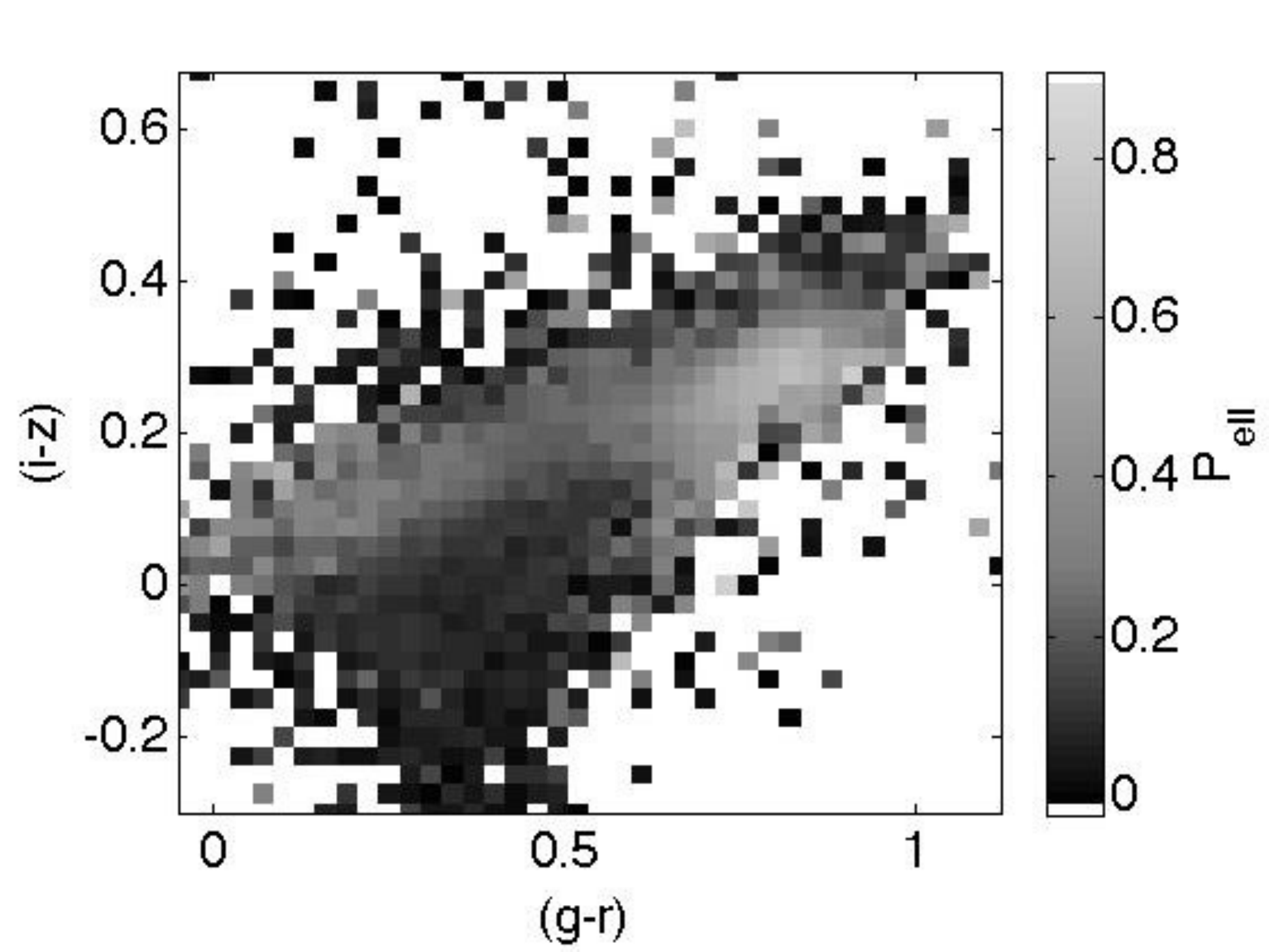}
\includegraphics[width=4.0cm]{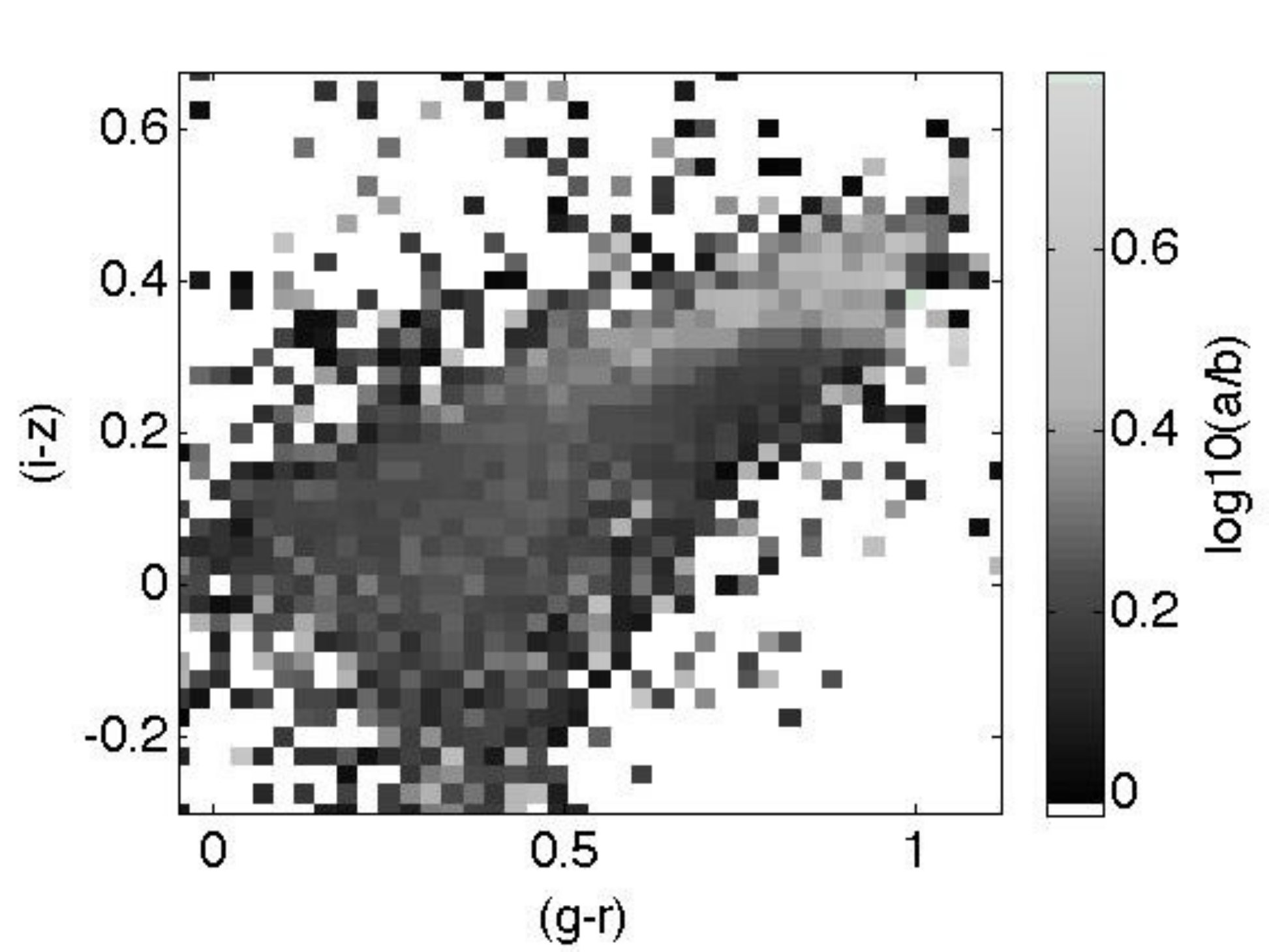}
\caption{Properties of the galaxy samples used in this work. {\bf Top panels:} 2D histograms in the ($i$-$z$) vs. ($g$-$r$) plane of the number of galaxies in the full sample selected from SDSS imaging data (left) and the spectroscopic subsample (right). {\bf Middle and bottom panels:} Average values of absolute $r$-band magnitudes, $M_{r}$, $r$-band de Vaucouleur fractions, $fracDeV$, debiased probabilities of having elliptical morphology, $P_{ell}$, and axis ratios, $a/b$, for the full sample in bins of ($i$-$z$) vs. ($g$-$r$). Note that $P_{ell}$ is only available for about 90\% of the full sample.}
\label{fig:osamp}
\end{figure}

Information about galaxy structure can aid in the interpretation of the modelling and is therefore of interest. Fig. \ref{fig:osamp} illustrates the properties of our galaxy sample, relying on structural parameters from the SDSS pipeline and the Galaxy Zoo project \citep{2008MNRAS.389.1179L}. The SDSS pipeline \citep{2002AJ....123..485S} models galaxy light profiles through a best-fitting linear combination of a de Vaucouleurs and an exponential profile. The de Vaucouleurs fractions, $fracDeV$, is a measure of the fraction of light in the de Vaucouleurs profile of the two component fit. We further make use of the ratio of the semi major, $a$, and semi minor axis, $b$, at the object's 25mag/arcsec$^{2}$ isophote which we denote as $a/b$. From the Galaxy Zoo project \citep{2011MNRAS.410..166L} we obtain $P_{ell}$ which gives the probability that the objects are assigned elliptical morphology through a visual classification. We chose the values that were debiased from resolution effects (see \cite{2009MNRAS.393.1324B}). Moreover, we make use of absolute $r$-band magnitudes, $M_{r}$, computed from the SDSS modelMags and redshifts using $H_{0}$=71km/s.

To test the performance of the photometric modelling we compare the outcome with results based on spectroscopic data from the SDSS. As the spectra are obtained using optical fibers with a diameter of 3'' only a fraction of the light from each galaxy is sampled. Stellar mass weighted ages and metallicities as well as total $V$-band extinctions, $A_{V}$, are taken from the SEAGal/STARLIGHT project \citep{2005MNRAS.358..363C}. These data have been derived from the SDSS fiber spectra using a spectral synthesis technique in which dust is treated as a foreground screen with a \cite{1989ApJ...345..245C} extinction law for $R_{V}$=3.1. Luminosity weighted ages and metallicities are compiled from \cite{2005MNRAS.362...41G}, who derived ages also by a spectral synthesis technique, while metallicities are instead derived through an analysis of absorption line indices, see \cite[][]{1994ApJS...95..107W,2003MNRAS.344.1000B,2003MNRAS.339..897T}. We moreover make use of specific star formation rates based on \cite{2004MNRAS.351.1151B} with an improved aperture correction \citep{2007ApJS..173..267S}. All data are based on SDSS DR7 except those compiled from \cite{2005MNRAS.362...41G} which are based on SDSS data release 4 \citep{2006ApJS..162...38A}. For convenience we only use spectroscopic data for galaxies in common between these studies. The spectroscopic sample used throughout is therefore smaller than the photometric sample and contains 7542 galaxies in the range $0.02<z<0.03$. To allow a fair comparison with properties derived from photometry we make use of magnitudes measured within the spectroscopic fibers when comparing with the data from \cite{2005MNRAS.358..363C} and \cite{2005MNRAS.362...41G}, while the modelMags are used for the specific star formation rates, since aperture corrections based on integrated SDSS photometry have been applied to these.

\section{SINGLE STELLAR POPULATION MODELS}
\label{sec:ssp}

The stellar population models we test in this paper are summarized in Table \ref{tab:t1}. We retrieve SSPs for each of these models which we combine into star formation histories as described in the following section. As most of the models have several options regarding specific model ingredients we decided to constrain the choice to avoid ending up with an impractical number of models. Details of the model sub-selection are given in Table \ref{tab:t1} and the model ingredients are briefly commented below.
The total amount of spectral information contained in $u$,$g$,$r$,$i$,$z$-band photometry can be fully represented in four dimensions by, e.g., the ($u$-$r$), ($g$-$r$), ($i$-$r$) and ($z$-$r$) colors. However, for practical reasons we make use of 2D projections in the form of color-color diagrams, and for simplicity we constrain ourselves to the use of the ($u$-$g$)-($r$-$i$) and ($g$-$r$)-($i$-$z$) planes. Note that these projections alone can not capture all available information, though they ought to include the majority since they contain information from all five bands.

\subsection{Stellar spectra}

The models fall in two different categories regarding how the stellar spectra are generated: theoretical models based on stellar atmosphere models \citep{1997AAS..125..229L} and models based on libraries of observed stellar spectra \citep{2003AA...402..433L}.

\subsection{Stellar evolution}
In terms of stellar evolution several different sets of models are used, including \cite{1992AAS...96..269S,1994AAS..106..275B,2000MNRAS.315..679C,2000AAS..141..371G,2007AA...469..239M,2008AA...482..883M}.

\subsection{Initial mass function}

\begin{figure}
\includegraphics[width=8.5cm]{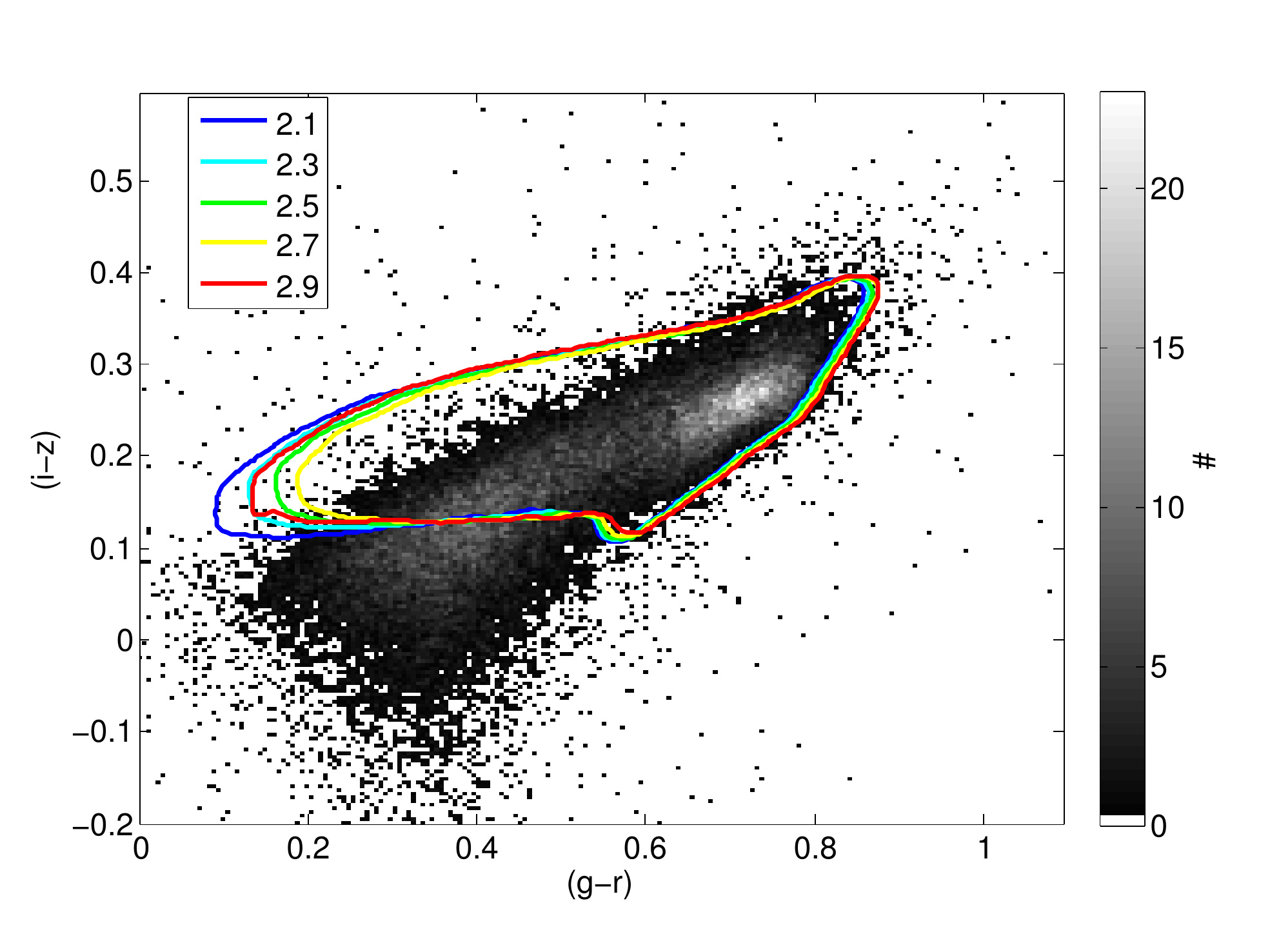}
\includegraphics[width=8.5cm]{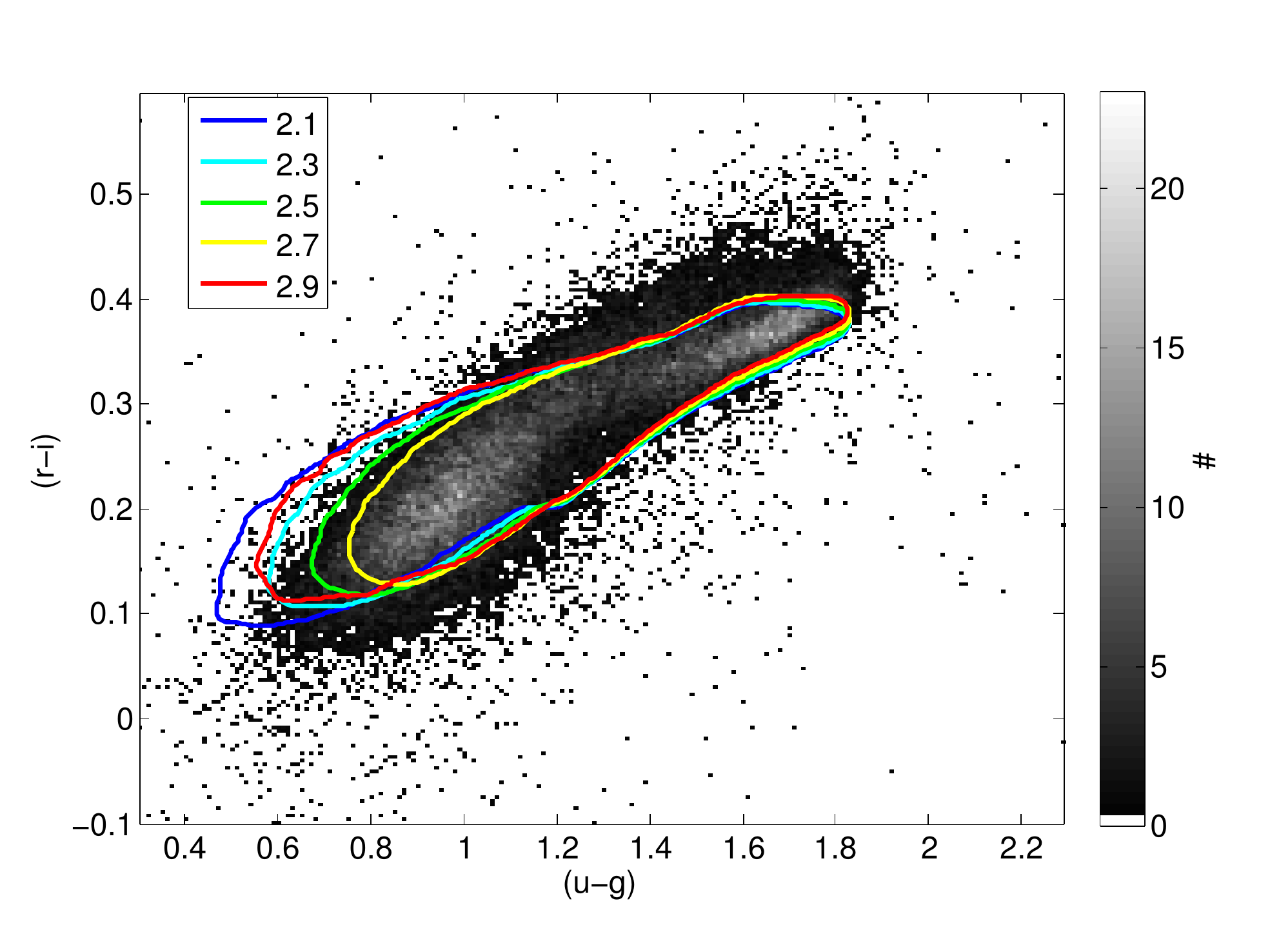}
\caption{Top panel: ($i$-$z$) versus ($g$-$r$) for our sample of SDSS galaxies (number density map) along with model boundary contours (colored lines) for the S99P models with various IMF slopes from 2.1-2.9 (Salpeter=2.35). Bottom panel: Same as above but for  ($r$-$i$) versus ($u$-$g$).}
\label{fig:imf}
\end{figure}

In a recent review \cite{2010ARAA..48..339B} conclude that there is no clear evidence that the initial mass function (IMF) varies strongly and systematically, and that the majority of systems on galactic scales are consistent with having a \cite{2001MNRAS.322..231K} or \cite{2003PASP..115..763C} initial mass function. However, an IMF that evolves with time cannot be excluded and recent studies do suggest a steeper IMF in massive elliptical galaxies \citep{2010Natur.468..940V,2011ApJ...735L..13V,2011MNRAS.417.3000S}. Here we investigate the the behavior of our models with respect to single slope IMFs. The \cite{1999ApJS..123....3L} model (S99) has the option of freely choosing the IMF. Figure \ref{fig:imf} shows the behavior of the model boundaries as a function of IMF slope. Clearly, the S99 models do not reproduce the observed distribution of colors very well regardless of IMF slope. As expected, the colors only change mildly with the IMF and we therefore chose to apply either a \cite{2001MNRAS.322..231K} or a \cite{2003PASP..115..763C} IMF. For each of the models one can choose between one of these two IMF presciptions.

\subsection{Emission lines}
\label{sec:line}

\begin{figure}
\centering
\includegraphics[width=8.5cm]{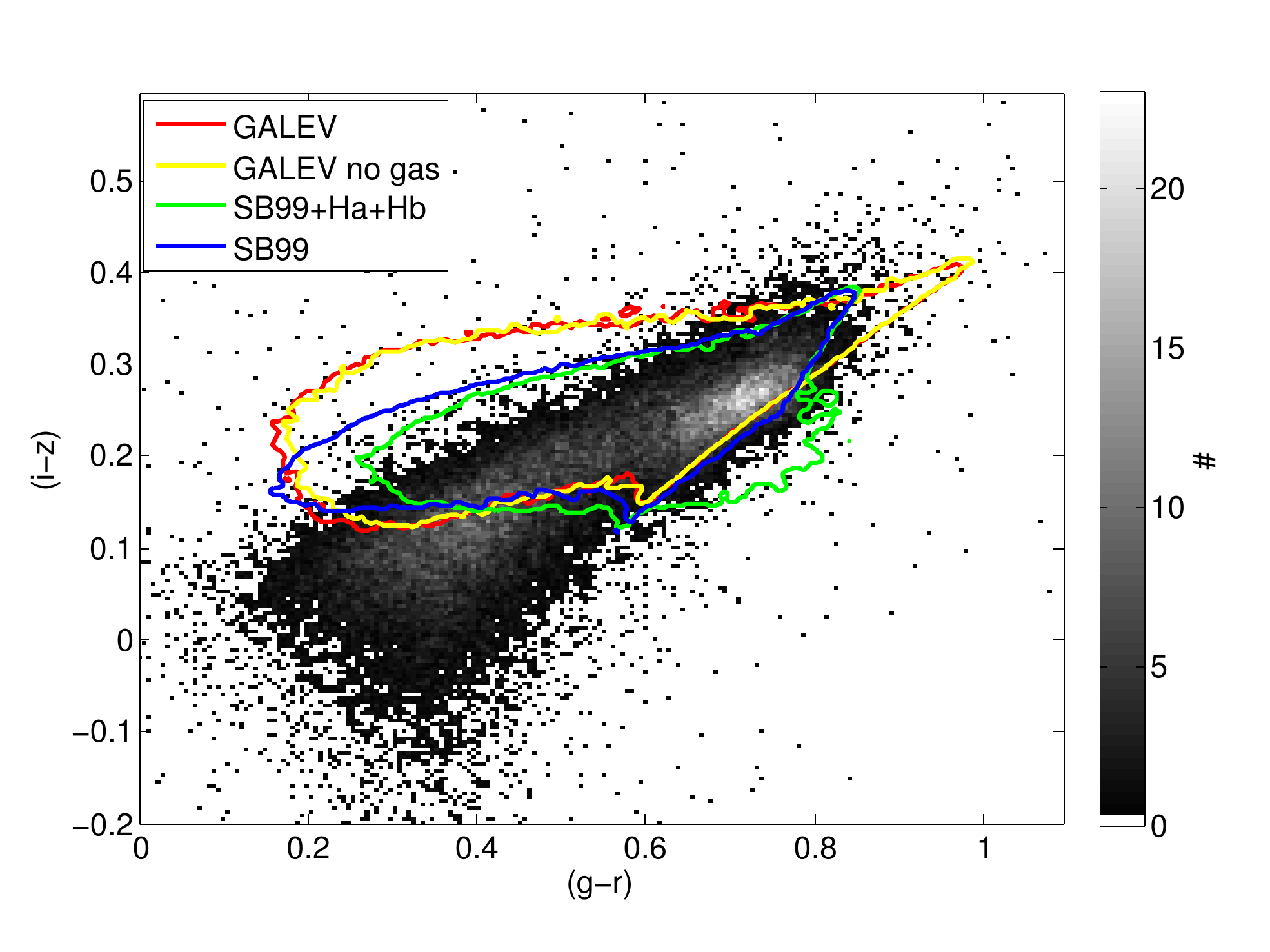}
\includegraphics[width=8.5cm]{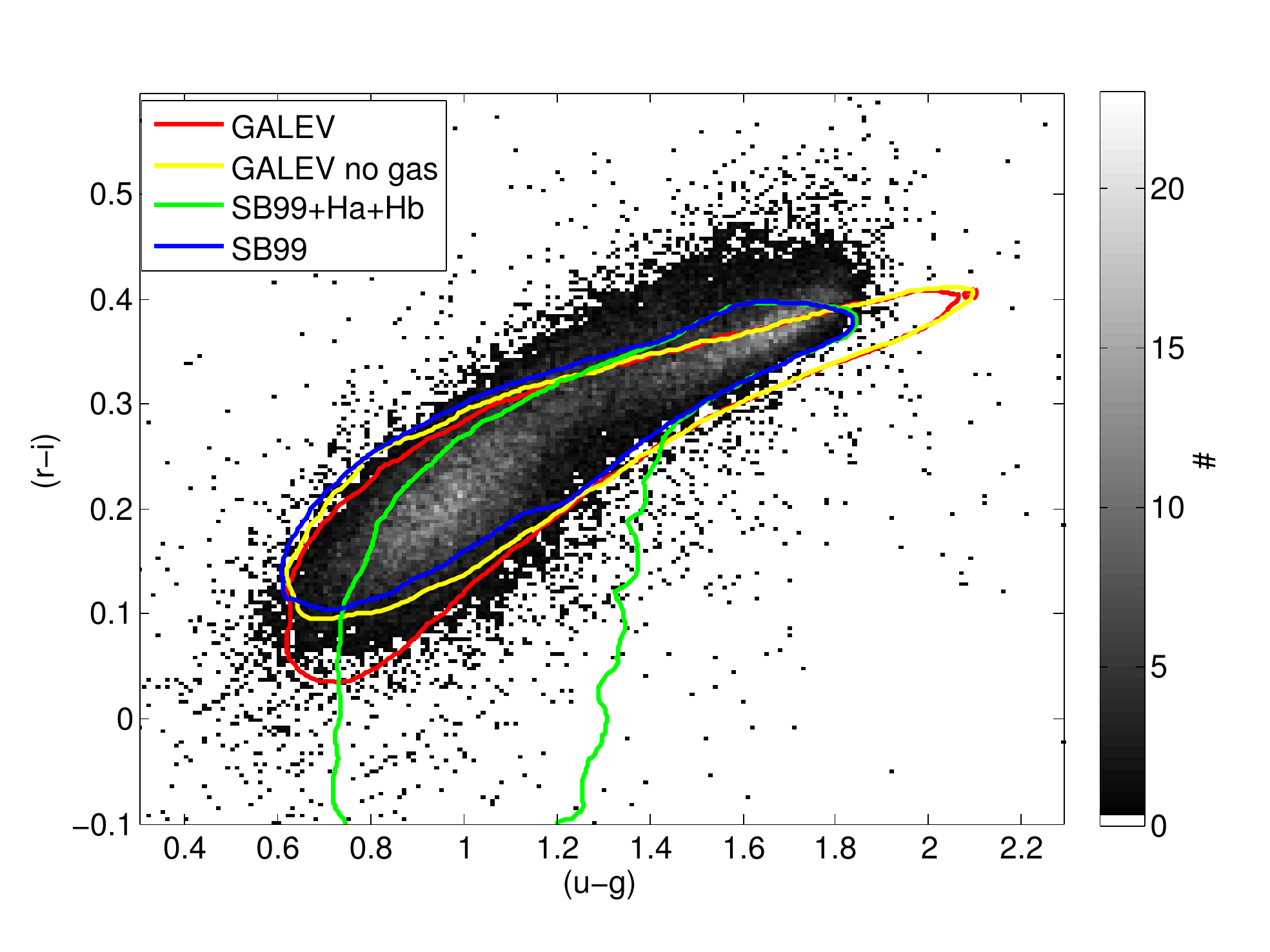}
\caption{Top panel: ($i$-$z$) versus ($g$-$r$) for our sample of SDSS galaxies (number density map) along with model boundary contours for the GALEV (red and yellow lines) and the SB99 (blue and green lines) models, with and without the inclusion of emission line fluxes. Bottom panel: Same as above but for ($r$-$i$) versus ($u$-$g$).}
\label{fig:em1}
\end{figure}

Emission lines are prominent features in optical spectra of star forming galaxies. It is therefore desirable to include emission lines in the modelling of photometric properties. Two of the models we use incorporate gas emission, namely S99 and GALEV \citep{2009MNRAS.396..462K}. The former predicts the strength of $\mbox{H}\alpha$ and $\mbox{H}\beta$ while the latter predicts several other lines in the optical, which are expected to be important in star forming galaxies \citep{2003AA...401.1063A}. Modelling of line emission from SSPs inevitably suffers from some uncertainty, but can be sufficiently accurate for predicting broad-band colors \citep{2011AJ....141..133G}. From the color evolution of the SSPs it is clear that gas emission in the models only has a significant effect on the colors at ages of about $10^{7}\mbox{yr}$ or younger. Fig. \ref{fig:em1} shows how the outlines of the model grids shift with the inclusion of emission line modelling. H$\alpha$ and H$\beta$, the only emission lines included in S99, only affect the $r$ and $g$ bands while additional prominent emission lines included in GALEV also affect $u$, $i$ and $z$ causing large discrepancies between the two models in ($r$-$i$). Modelling of line emission as in S99 or GALEV can be included in any of the models we consider by adding the corresponding change in the colors of S99 or GALEV. This can only be done for SSPs of the same age and metallicity under the assumption that the stellar emission in $u$,$g$,$r$,$i$ and $z$ is similar in the two models.

\begin{table*}
\begin{minipage}{\textwidth}
\centering
\begin{tabular}{*{1}{lllll}l}
\hline
Model ID & reference & stellar spectra & stellar evolution& IMF & line emission\\
\hline
BC03hr    & \cite{2003MNRAS.344.1000B}  & \cite{2003AA...402..433L} & \cite{1994AAS..106..275B} & \cite{2003PASP..115..763C} & no\\
BC03lr    & \cite{2003MNRAS.344.1000B}  & \cite{1997AAS..125..229L} & \cite{1994AAS..106..275B} & \cite{2003PASP..115..763C} &no\\
CB07hr    & in prep.$^{c}$              & \cite{2003AA...402..433L} & \cite{2008AA...482..883M}$^{a}$ & \cite{2003PASP..115..763C} &no\\
CB07lr    & in prep.$^{c}$              & \cite{1997AAS..125..229L} & \cite{2008AA...482..883M}$^{a}$ &  \cite{2003PASP..115..763C} &no\\
S99P$^{f}$ & \cite{2005ApJ...621..695V}  & \cite{1997AAS..125..229L} & \cite{2000AAS..141..371G} & \cite{2001MNRAS.322..231K}& yes\\
S99G$^{f}$ & \cite{1999ApJS..123....3L}  & \cite{1997AAS..125..229L} & \cite{1992AAS...96..269S} & \cite{2001MNRAS.322..231K}& yes\\
GALEV$^{d}$ & \cite{2009MNRAS.396..462K}  & \cite{1997AAS..125..229L} & \cite{1994AAS..106..275B}$^{e}$ & \cite{2001MNRAS.322..231K}&yes\\
M05        & \cite{2005MNRAS.362..799M}  & \cite{1997AAS..125..229L} & \cite{2000MNRAS.315..679C}$^{ab}$& \cite{2001MNRAS.322..231K}&no\\
\hline
\footnote{Additional references left out, see model reference.}
\footnote{These models have the option of choosing between populations with or without blue horizontal branch. We chose the latter.}
\footnote{S. Charlot private communication.}
\footnote{Updated SDSS filter responses from \cite{2010AJ....139.1628D} used.}
\footnote{GALEV uses the 1999 version of isocrones from the Padova group.}
\footnote{Magnitudes obtained by convolving the low resolution spectra with the updated SDSS filter response from \cite{2010AJ....139.1628D}.}
\end{tabular}
\end{minipage}
\caption{Sources of SSP models}
\label{tab:t1}
\end{table*}

\section{STAR FORMATION HISTORY}
\label{sec:sfh}

We synthesize star formation histories from the SSPs through the use of smooth models to which some stochastic sampling has been added. Such Monte Carlo libraries of star formation histories have previously been employed by several authors \citep{2003MNRAS.341...54K,2008MNRAS.388.1595D,2009MNRAS.400.1181Z} and have proven versatile in modelling stellar population properties. The stellar population synthesis method we employ is thus not new, but the details differ between our and these previous works. The addition of some random component in the modelling is motivated by the knowledge that star formation to some extent occurs stochastically. Processes such as galaxy merging \citep{2007AA...468...61D} and ram pressure stripping \citep{1972ApJ...176....1G} often occur on short timescales and can strongly influence a galaxy's star formation history.

\cite{2002ApJ...576..135G}, inspired by the work of \cite{1986AA...161...89S}, showed that a ``delayed exponential'' star formation history does a better job in reproducing colors of Virgo cluster galaxies than the classical exponential model. We therefore use this model as a starting point for our star formation histories
\begin{equation} 
SFR=\frac{T}{\psi^{2}}\exp\left(-\frac{T^{2}}{2\psi^{2}}\right)
\label{eq:gav}
\end{equation}
where SFR is the star formation rate, $T$ is the time from the initial onset of star formation and $\psi$ is a parameter governing the decline of star formation over time. We take the galaxy formation time to be 13.5Gyr ago. The exact starting point is not crucial due to the slow evolution of spectral properties at old ages, but it is to some extent motivated by current estimates of the onset of reionization in the Universe \citep{2006ARAA..44..415F}. 

Star formation histories are created through a sampling of Eq. 1 using SSPs for values of $\psi$ in the range 1Gyr to 20Gyr. The sampling is done logarithmically, since the color evolution is roughly proportional to the logarithm of the age. Thus, the age of each SSP, $t_{i}$, is randomly drawn from the following distribution
\begin{equation}
\frac{T^{2}}{\psi^{2}}\exp\left(-\frac{T^{2}}{2\psi^{2}}\right)
\end{equation}
To compensate for the $logarithmic$ sampling the strength, i.e. mass, of each SSP is multiplied by its age. In total $5\psi$ SSPs are used for each star formation history (more SSPs are needed to sample the larger range in log(age) spanned by models with higher values of $\psi$.). Stochasticity is introduced in the model through the sampling and by further modifying the strength of each SSP by multiplying with a random component. The random component is drawn from a distribution which is taken to be the absolute value of a normal distribution with ($\mu=0$,$\sigma=1$). To include many different metallicity configurations the SSPs constituting each SFH are randomly divided into five groups and each of these groups is assigned a random metallicity from the options available for that particular SSP.

The chosen input SSPs limit the lowest ages that can be included\footnote{Note than \cite{2005MNRAS.362..799M} do not offer young ages at the lowest metallicity.}. However, note that in the model by \cite{2000ApJ...539..718C} the light from populations younger than $10^{7}\mbox{yr}$ is heavily obscured by the dust in the birth clouds and does not contribute significantly to the integrated light at optical wavelengths. We do therefore not need to consider these missing models at young ages.

Using optical spectral energy distributions the resolution in terms of the age of a stellar population seems to be limited to at least $\sim10\%$ \citep{2010MNRAS.403..797G}. Any gaps introduced by a discrete sampling of a star formation history can thus safely be neglected if they are smaller than about $10$\%.

\begin{figure}
\includegraphics[width=8.5cm]{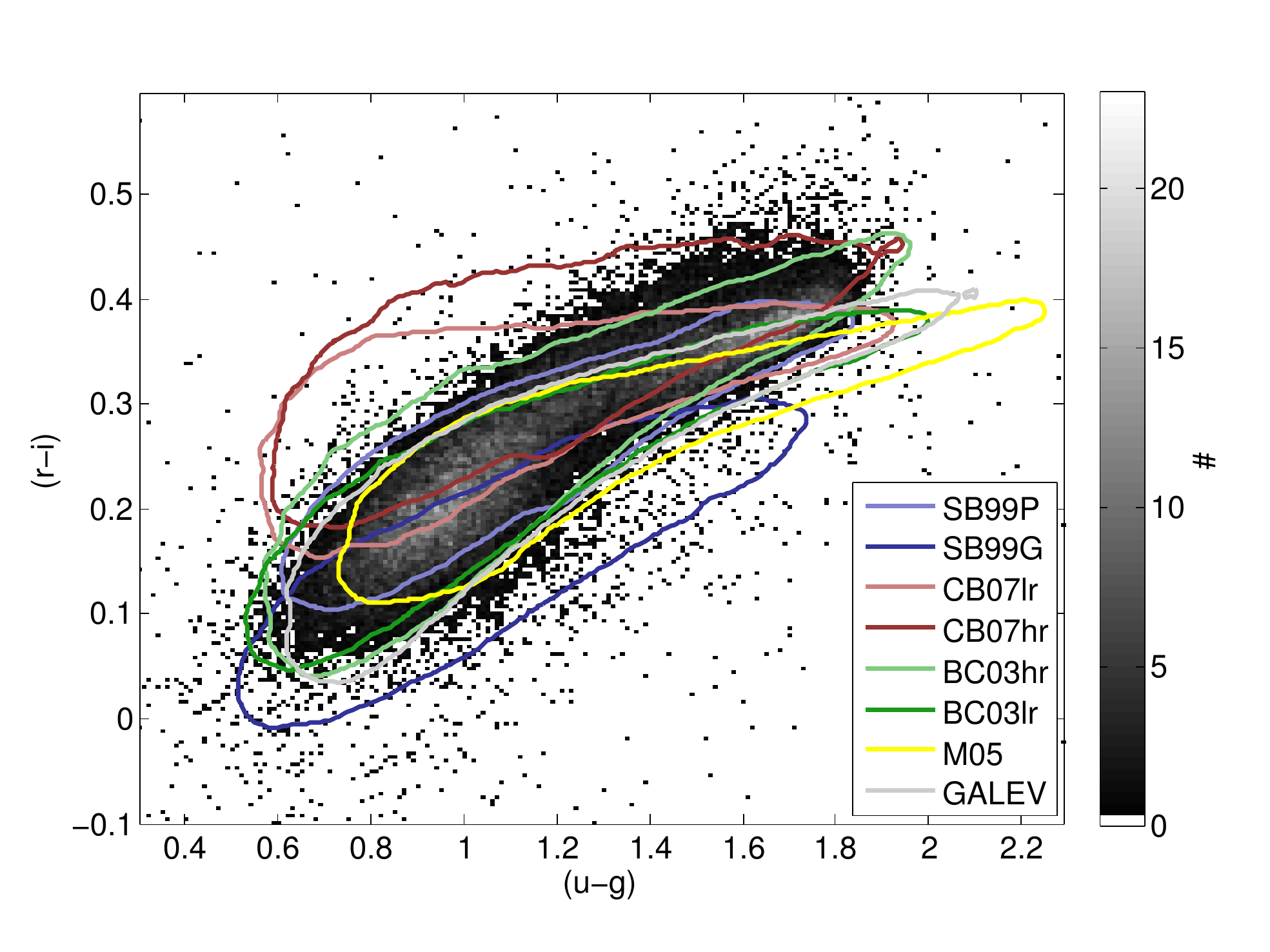}
\caption[ugri]{($r$-$i$) versus ($u$-$g$) for our sample of SDSS galaxies (number density map) along with model boundary contours for eight different sets of stellar population models (color coded contours). Details of the models are summarized in Table \ref{tab:t1} and described in Sect. \ref{sec:ssp} and \ref{sec:sfh}.}
\label{fig:ugri}
\end{figure}

\begin{figure}
\includegraphics[width=8.5cm]{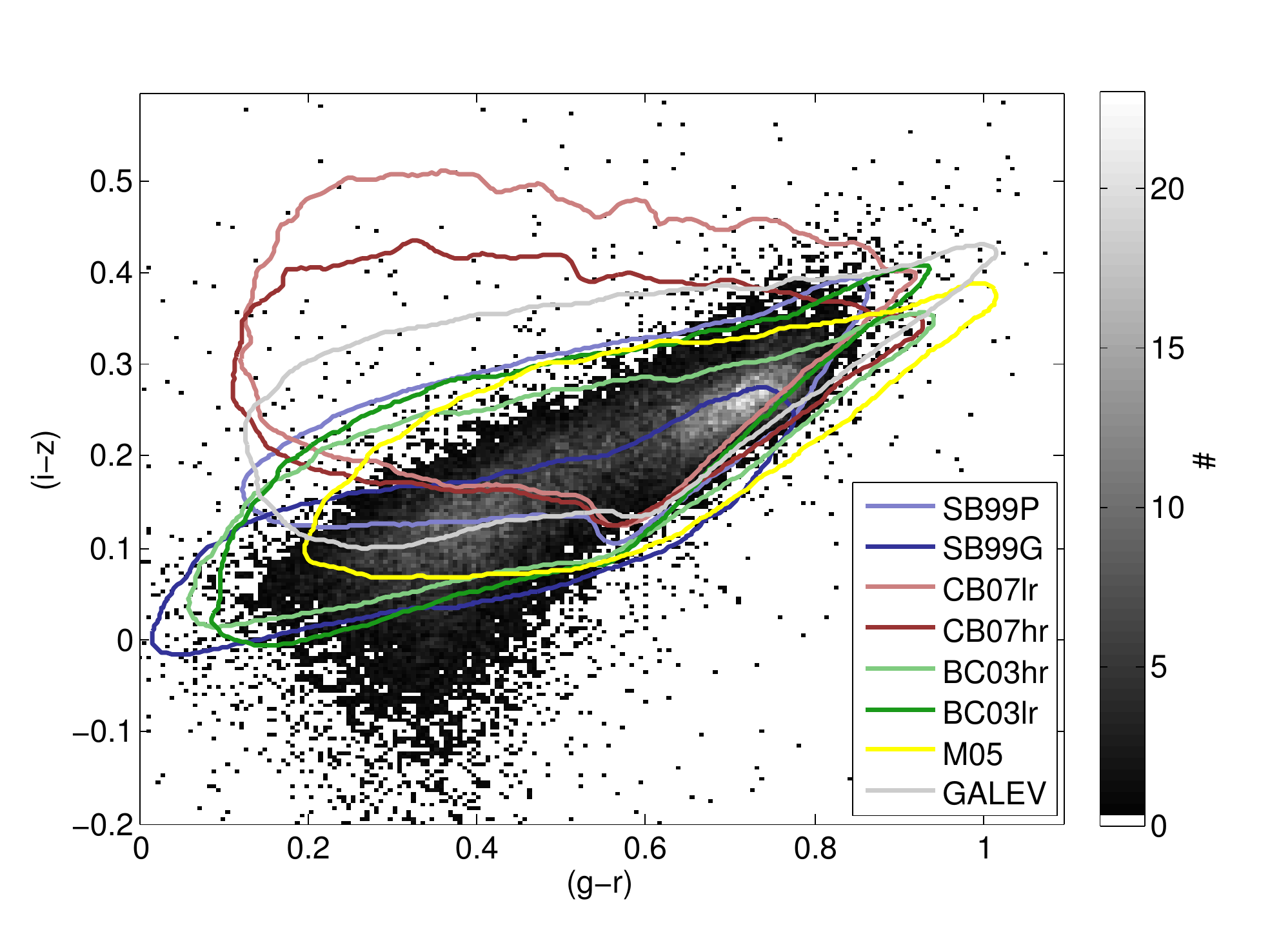}
\caption[griz]{($i$-$z$) versus ($g$-$r$) for our sample of SDSS galaxies (number density map) along with model boundary contours for eight different sets of stellar population models (color coded contours). Details of the models are summarized in Table \ref{tab:t1} and described in Sect. \ref{sec:ssp} and \ref{sec:sfh}.}
\label{fig:griz}
\end{figure}

For a range of values\footnote{$\psi=1.0,1.5,...3.0,3.25,...,9.0,10.0,...,20.0$ in $\psi$. A variation in step size turned out to be useful for better coverage in the ($u$-$g$)-($r$-$i$) and ($g$-$r$)-($i$-$z$) planes.} we create 1250 realizations of the different SFHs, which results in 50000 models for each set of SSPs.
Contours of the model boundaries in color-color space are created in the following way. 2D-histograms in  ($g$-$r$)-($i$-$z$) and ($u$-$g$)-($r$-$i$) are made with bin sizes of 0.0055, 0.0040, 0.0100 and 0.0035 for ($g$-$r$), ($i$-$z$), ($u$-$g$), and ($r$-$i$), respectively. These histograms are convolved with a Gaussian kernel ($\sigma=3\mbox{pixel}$) to remove the noise at the edges after which contours are drawn at a density of 0.2 models/pixel\footnote{The model boundaries in Fig. \ref{fig:imf}-\ref{fig:new} are created in the same way.}. Figure \ref{fig:ugri} and \ref{fig:griz} shows the outlines of these stellar population models (as summarized in Table \ref{tab:t1})  in the ($u$-$g$)-($r$-$i$) and ($g$-$r$)-($i$-$z$) planes.

\section{DUST}
\label{sec:dust}

In this paper we employ a simplistic model for dust extinction in galaxies that has, by construction, no free parameters. Although several assumptions are made we expect our model to work reasonably well since it is constructed to capture the most important features of dust extinction in galaxies. It is motivated by:

1) The connection between the presence of dust and star formation/young stellar populations. \citet{2010MNRAS.403.1894D} showed that the dust mass in galaxies can be estimated remarkably well using the average star formation rate during the last $10^{8}$yr (both quantities are determined from a model fit to photometry over a large wavelength range, $\sim0.1-100\mu \mbox{m}$). They use a relation in the form of a power law with index 1.1 and find a scatter of about 0.5dex over at least three orders of magnitude in star formation rate. If this relation is normalized by the stellar mass, then the mass fraction of the dust is proportional to a power of the recent specific star formation rate, a property that can be constrained by optical colors alone \citep{2004MNRAS.351.1151B}.

We take the effective $r$-band extinction, $A_{r}$, to be
\begin{equation}
\label{eq:1}
A_{r}=A_{0}\left(\frac{\sum\limits_{i=0}^{n} s_{i}\mbox{e}^{-(t_{i}/t_{0})}/\sum\limits_{i=0}^{n} s_{i}}{\int_0^{13.5Gyr}\mbox{e}^{-(t/t_{0})}\mbox{d}t}\right)^{1.1}
\end{equation}
where $A_{0}$ and $t_{0}$ are constants, $t_{i}$ is the age of the $i$th SSP, $s_{i}$ is the strength of the $i$th SSP, the summation is done over all $n$ SSPs in the star formation history and the power index of 1.1 is taken directly from \cite{2010MNRAS.403.1894D}. This equation makes $A_{r}$ dependent on the fraction of young stars by comparing the strengths of the SSPs weighted by the factor $\mbox{e}^{-(t_{i}/t_{0})}$ (numerator) with the corresponding value for a constant star formation history (denominator). This is essentially a smoother version of the $10^{8}$yr cut adopted by \cite{2010MNRAS.403.1894D} assuming that $A_{r}$ is proportional to the dust to stellar mass ratio. $A_{0}$ is constrained by the reddest galaxies observed. A value of $A_{0}=0.40$ keeps the models within or close to the observed cloud of galaxies. $t_{0}$ should be chosen in accordance with the work of \cite{2010MNRAS.403.1894D} and we adopt $t_{0}$=$3\cdot10^{8}$yr.

2) An observed correlation exists between metallicity and dust to gas mass ratio. We base the metallicity dependence of our dust models on the observational results of \cite{2008ApJ...678..804E}. Their relation between the nebular metallicity, $Z$, and HI to dust mass fraction ($M_{\text{HI}}/M_{\text{dust}}$) can be rather well fitted by a power law (the fit has a scatter of 0.5dex of which at least half can be explained by measurement errors) with low metallicity galaxies having higher $M_{\text{HI}}/M_{\text{dust}}$. Assuming that the nebular metallicity is the same as the average metallicity of the stellar populations contributing to the dust (see Eq. \ref{eq:1}), z, this leads to a modification of $A_{r}$ according to
\begin{equation}
\label{eq:2}
A_{r}=A_{r}(z/z_{\odot})^{m}
\end{equation}
where $m$ is a constant. Here we have further assumed that the metallicity is the only parameter governing $M_{\text{HI}}/M_{\text{dust}}$ for a fixed star formation history and that $M_{\text{HI}}/M_{\text{dust}}$ is proportional to $A_{r}$. We use $m$=1.7 as measured from the data presented in \cite{2008ApJ...678..804E}. We note that \cite{2007MNRAS.375....2D} also use a power law to model the metallicity dependence of the effective extinction with a power law index very similar to the one we adopt.

3) At optical wavelengths, the wavelength dependence of the extinction appears to be rather well modelled by a power law of index $k$ \citep{2000ApJ...539..718C}
\begin{equation}
A_{\lambda} \propto \lambda^{k}
\end{equation}
where $A_{\lambda}$ is the effective extinction at a wavelength $\lambda$. \cite{2000ApJ...539..718C} use a different proportionality constant for stellar populations older and younger than $10^{7}$yr. However, the contribution of optical light from young populations ($<10^{7}$yr) is small due to obscuration from the dust clouds in which the stars were born. We therefore apply a single power law to model the wavelength dependence of the extinction. This should be seen as a simple approach, as we do not know how the detailed effective extinction curve varies over a range of galactic environments. The colors of dusty models depend on the constant $k$. We use $k$=-1.1 which is an appropriate choice considering the width of the model cloud in ($r$-$i$) at ($u$-$g$)$>$1.3 as compared to the observations. The effective extinction in $u$,$g$,$r$,$i$ and $z$ is computed using the effective wavelength of these filters. The use of effective wavelengths rather than a flux weighted average only introduces errors in $A$ of a few percent (as estimated using the S99 models $without$ emission lines).\\

4) The extinction is dependent on the angle under which the galaxy is seen. Empirical relations between colors and inclination for spiral galaxies in the SDSS were studied by \cite{2010MNRAS.404..792M}. Here we adopt a similar functional form to express how the effective $r$-band extinction is modified depending on the isophotal axis ratio, $a/b$,
\begin{equation}
\label{eq:ab}
A_{r} = A_{r} + \gamma_{r} \text{log}_{10}(a/b)
\end{equation}
where $\gamma_{r}$ is a constant. To determine $\gamma_{r}$ we select spiral galaxies ($P_{ell}<0.2$) from our spectroscopic sample and plot $A_{V}$ from \cite{2005MNRAS.358..363C} versus $ \text{log}_{10}(a/b)$. For four bins in $fracDeV_{r}$ we determine $\gamma_{V}$ (in analogy to $\gamma_{r}$) through least square fits. The bins were chosen such that they contain equal numbers of galaxies and we find ($fracDeV_{r}$,$\gamma_{V}$) = (0.44-1.00,0.51),(0.17-0.44,0.58),(0.02-0.17,0.57) and (0.00-0.02,0.58). Given the small bin to bin variations we adopt $\gamma_{r}$=0.47 (i.e. $\gamma_{V}$=0.55) and apply Eq. \ref{eq:ab} for all galaxies having $A_{r}>0.05$mag according to Eq. \ref{eq:1} and \ref{eq:2}. The extinction for galaxies with little or no dust are thus assumed to be inclination independent while dusty galaxies are treated as spirals.

We have constructed our model so that it does not have any free parameters: The extinction is completely coupled to the galaxy's star formation history, chemical enrichment and the observed axis ratio. The constants included in the model have values that can be inferred from previous studies, but some were adjusted in order to produce as realistic an ensemble of models as possible as judged by comparing the distribution of observed and model colors. We adopted $A_{0}=0.40$, $t_{0}=0.30$Gyr, $m=1.7$ and $k=-1.1$.

Note that models which do not have any recent ($t \la t_{0}$) star formation are essentially unaffected by extinction and devoid of line emission. These models are simply linear combinations of single stellar populations.

\section{PERFORMANCE}
\label{sec:per}

\begin{figure}
\includegraphics[width=8.5cm]{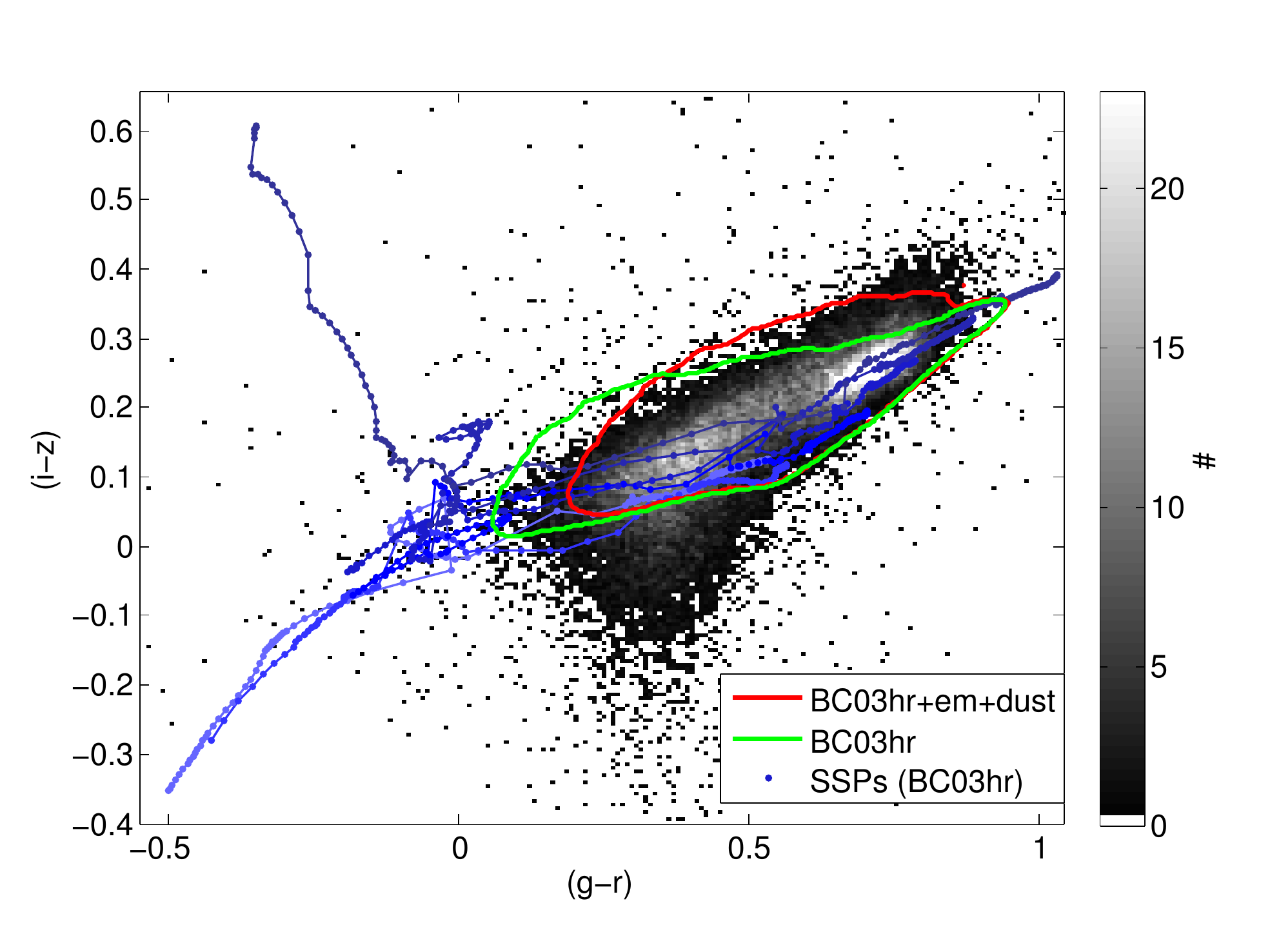}
\includegraphics[width=8.5cm]{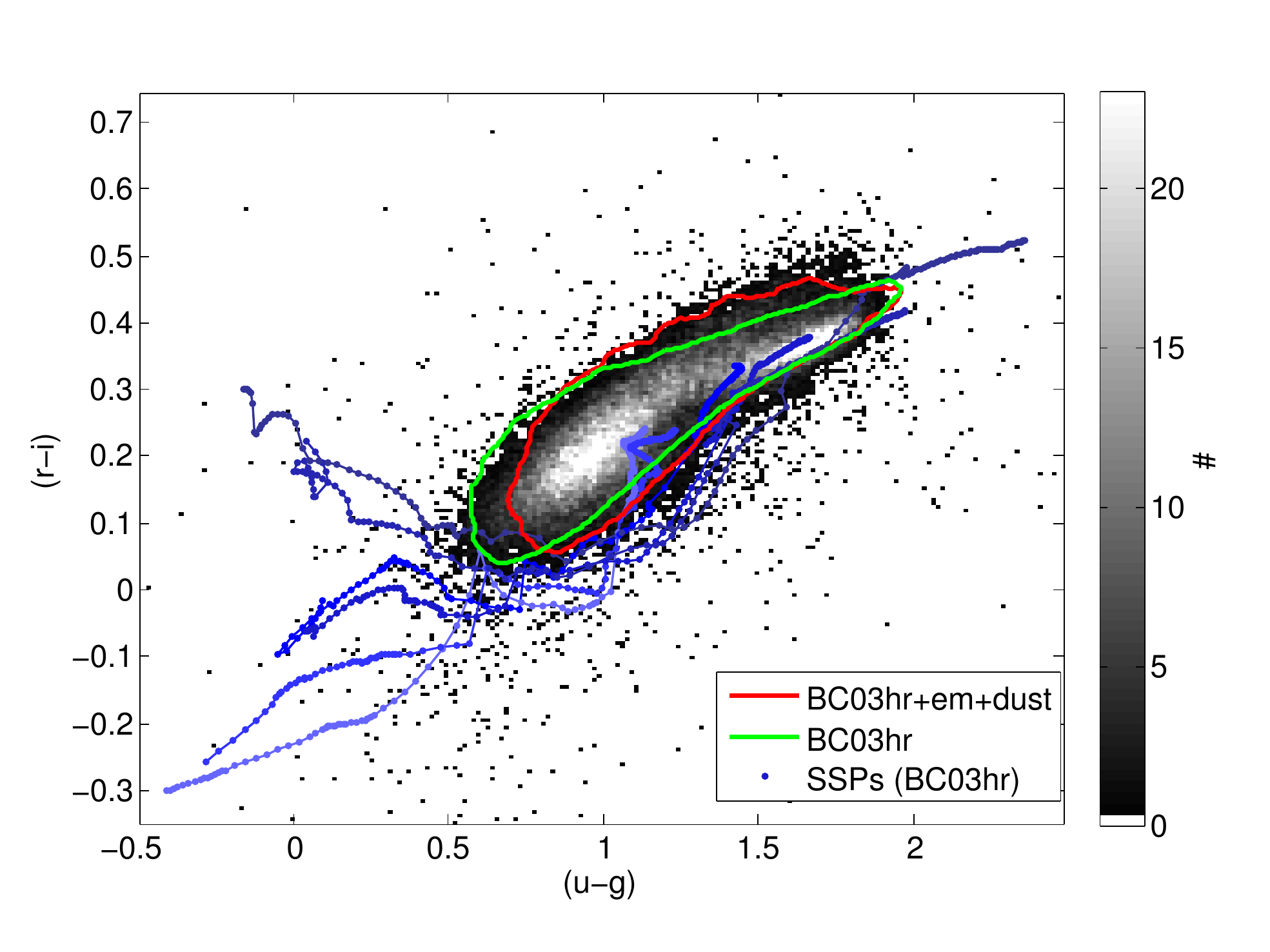}
\caption{($i$-$z$) versus ($g$-$r$), upper panel, and ($r$-$i$) versus ($u$-$g$), lower panel, for our sample of SDSS galaxies (number density map) along with model boundary contours for the BC03hr model (green) and the BC03hr model with dust and line emission (red). The SSPs from BC03hr with ages older than $10^{7}$yr are shown (blue dots, metallicity increase from light to darker blue) and these are connected along lines of constant metallicity.}
\label{fig:new}
\end{figure}

We test the models described above using an observational sample of galaxies from the Sloan Digital Sky Survey DR7 \citep{2009ApJS..182..543A} as described in Sect. \ref{sec:obs}. Given the boundaries of the model cloud for dust-free models in Fig. \ref{fig:ugri} and \ref{fig:griz} in comparison with the observations we see large differences between the various models. In particular the models of Charlot and Bruzual in preparation (CB07) and GALEV occupy a large region in ($g$-$r$)-($i$-$z$) where no galaxies are found to reside and the same is true for CB07 in ($u$-$g$)-($r$-$i$). The overall best agreement between observations and models is shown by the \cite{2003MNRAS.344.1000B} models based on observed stellar spectra (BC03hr) and spectra from stellar atmosphere models (BC03lr). Of these two BC03hr performs significantly better in the reddest region of both color-color diagrams. In the following we have therefore adopted BC03hr as our preferred model. A missing ingredient in this model is the line emission. As we showed in Fig. \ref{fig:em1} inclusion of line emission from GALEV only changes the model track slightly. We can therefore safely include line emission from GALEV into the BC03hr models as discussed in Sect. \ref{sec:line}. GALEV is preferred over SB99 since the former takes more emission lines into account. Moreover, we apply our dust models to BC03hr. The resulting change in the model boundaries is shown in Fig. \ref{fig:new}. Considering the photometric errors the model performs very well in reproducing the colors of the galaxy sample, except at about ($g$-$r$,$i$-$z$)=(0.4,0.0), where models appear to be missing. The colors of the SSPs in Fig. \ref{fig:new} show that galaxies dominated by recent star formation can have ($i$-$z$)$<$0.0. However, these objects would be very blue in ($g$-$r$) and thus cannot explain the ``missing models''. Considering Fig. \ref{fig:new}, our recipe for the star formation histories combines the blue points into models within the green contours, which thereafter are shifted to the red contours by dust and line emission. Fig. \ref{fig:new} clearly shows that a mixture of SSPs of different ages and metallicities is needed to explain the locus of observed galaxies. Additionally, dust is needed to explain some of the reddest galaxies.

\subsection{Derived parameters}
\label{sec:bay}

\begin{figure}
\includegraphics[width=4cm]{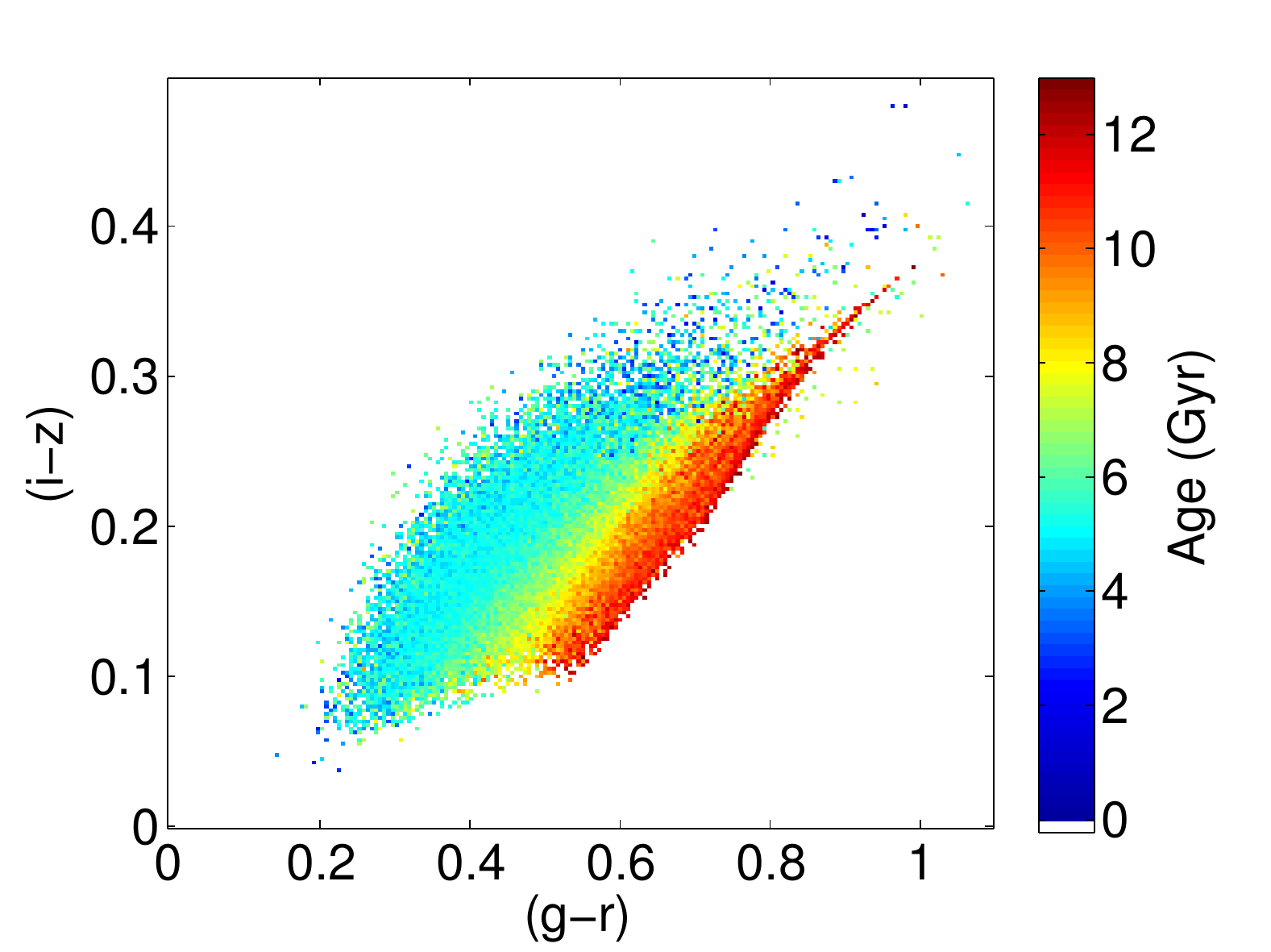}
\includegraphics[width=4cm]{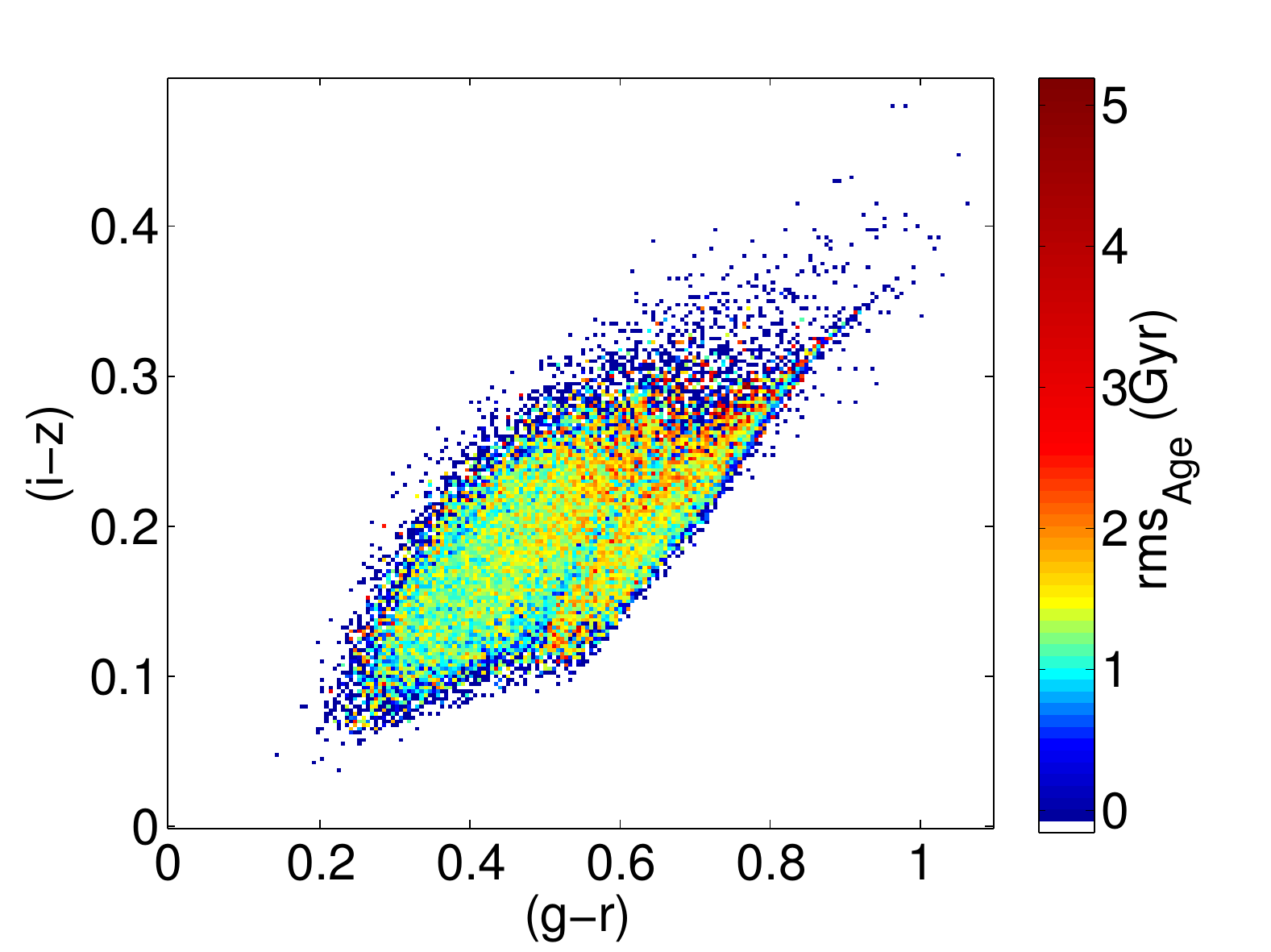}\\
\includegraphics[width=4cm]{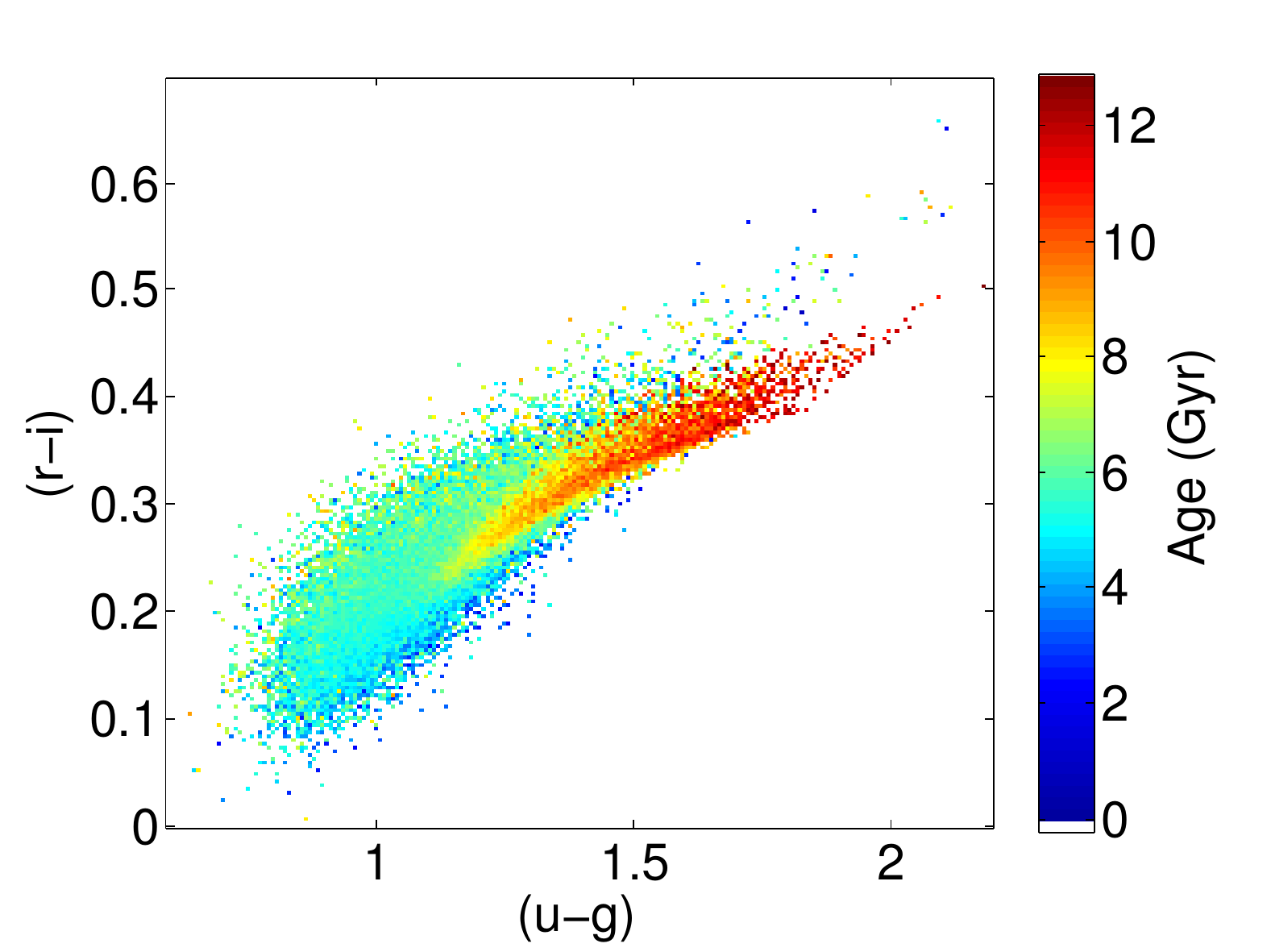}
\includegraphics[width=4cm]{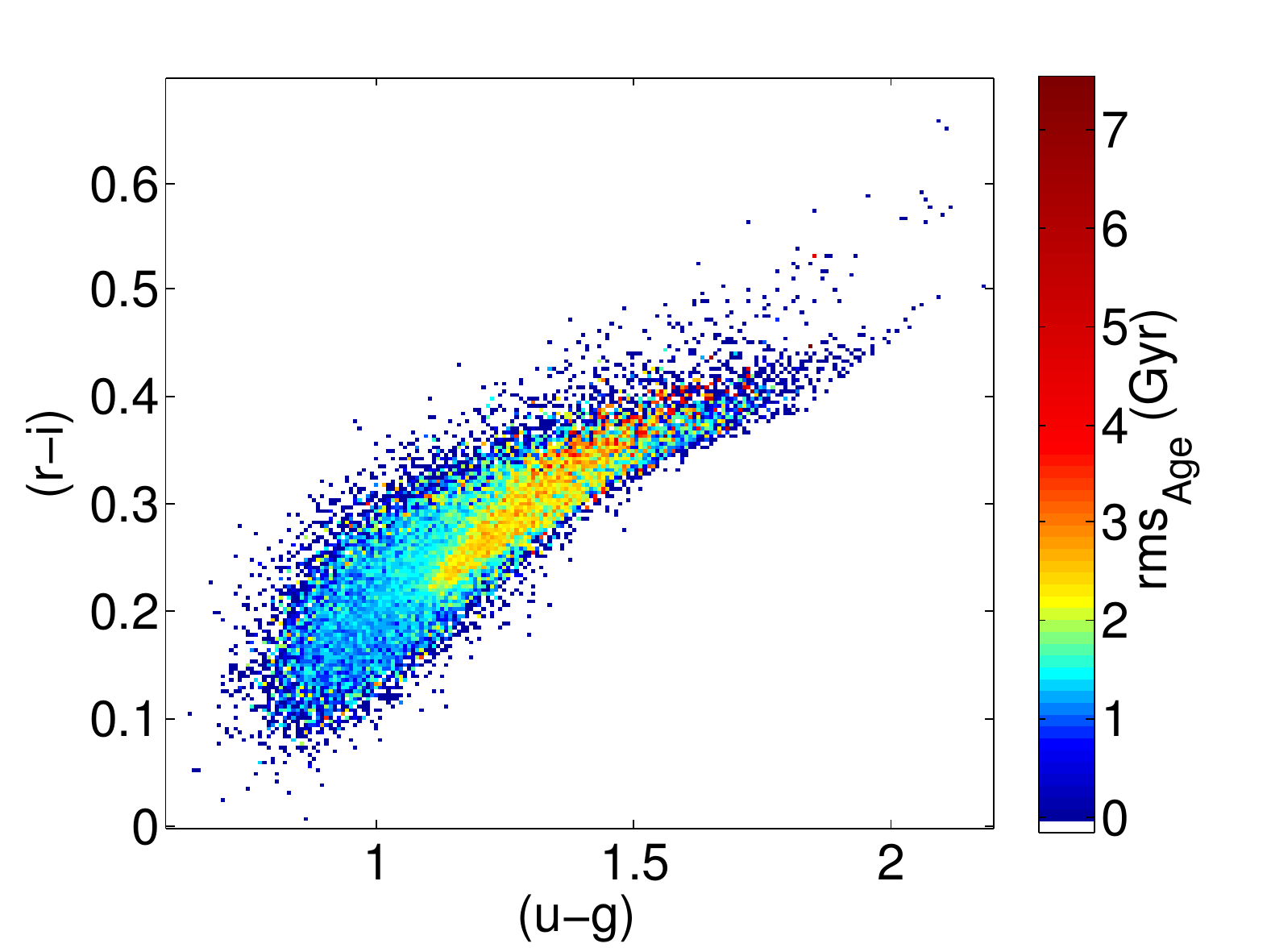}
\caption{Left panels: stellar mass weighted age (color coded pixels) as a function of ($i$-$z$) versus ($g$-$r$), as well as ($r$-$i$) versus ($u$-$g$) for our 50000 models based on BC03hr with dust. Right panels: standard deviation in stellar mass weighted age (color coded pixels) as a function of ($i$-$z$) versus ($g$-$r$), as well as ($r$-$i$) versus ($u$-$g$) for our 50000 models based on BC03hr with dust.}
\label{fig:age}
\end{figure}

\begin{figure}
\includegraphics[width=4cm]{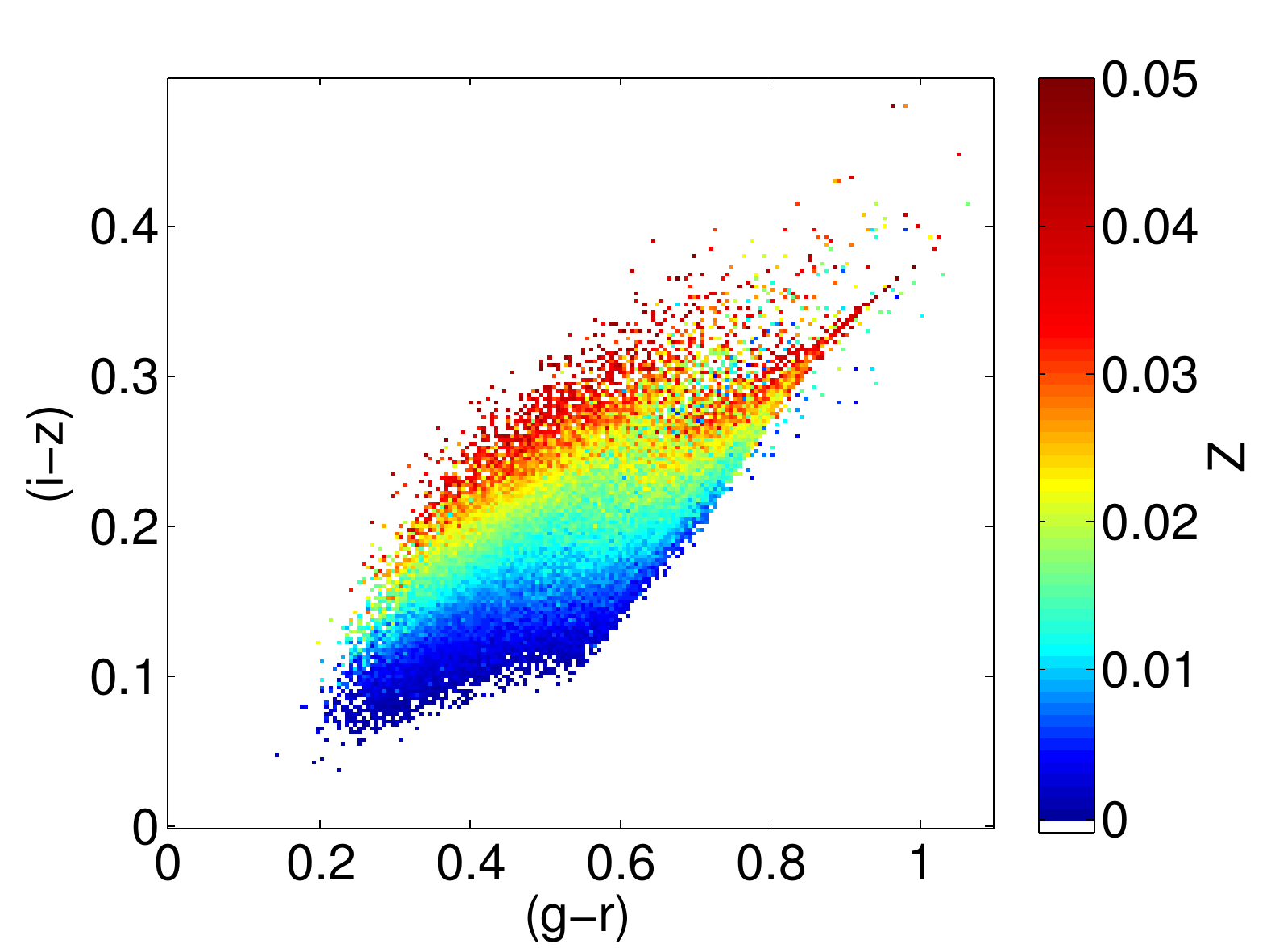}
\includegraphics[width=4cm]{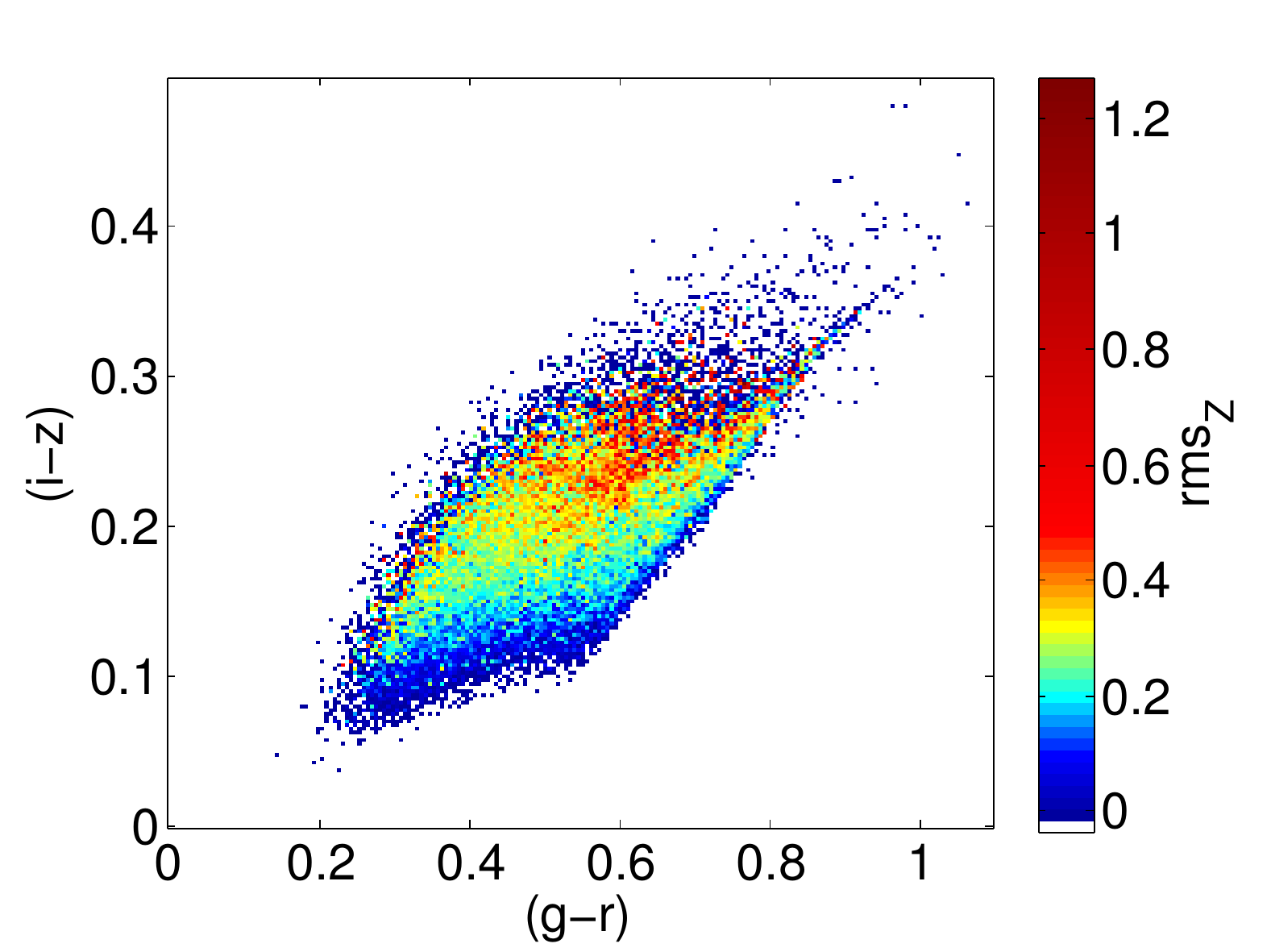}\\
\includegraphics[width=4cm]{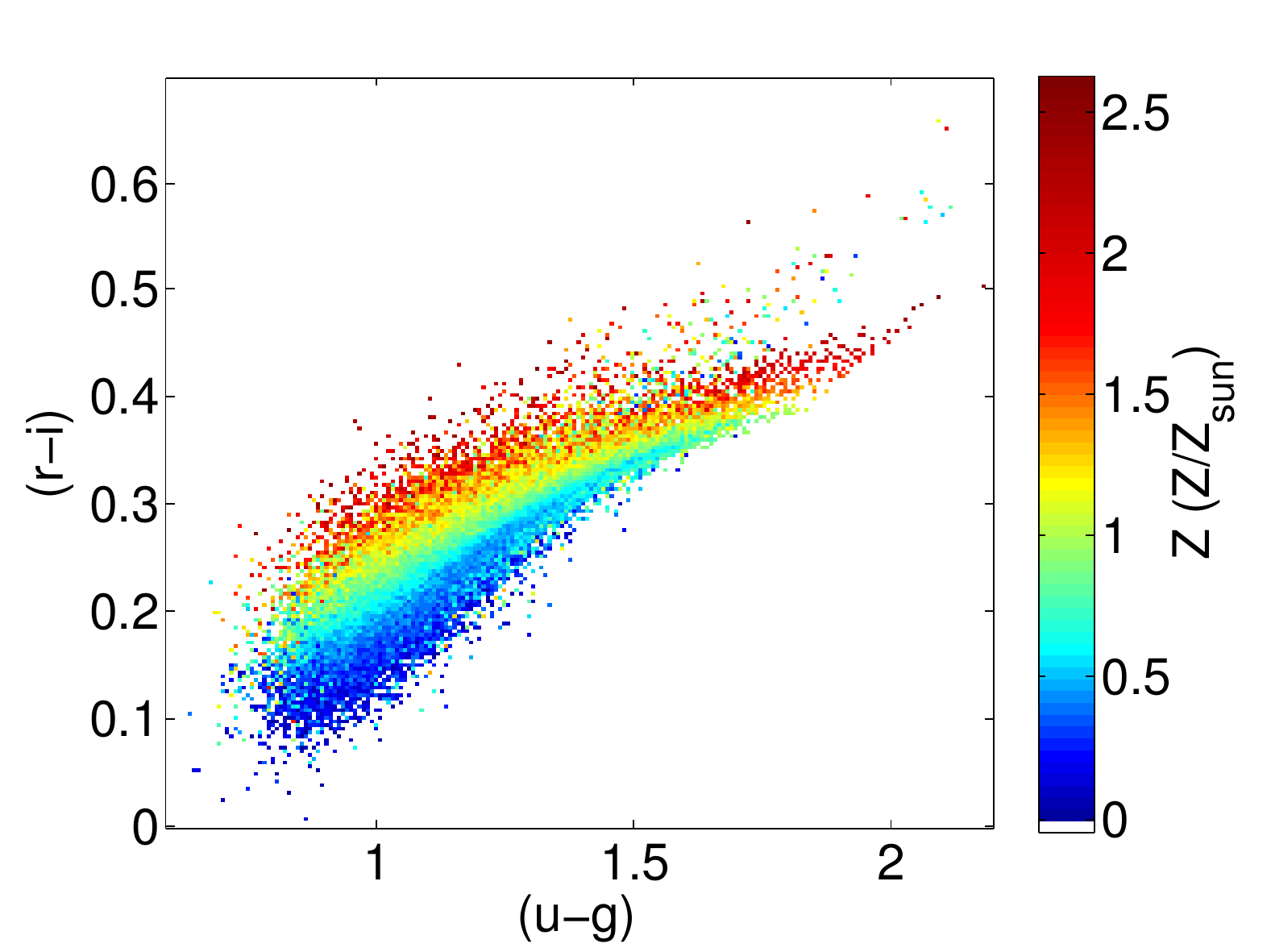}
\includegraphics[width=4cm]{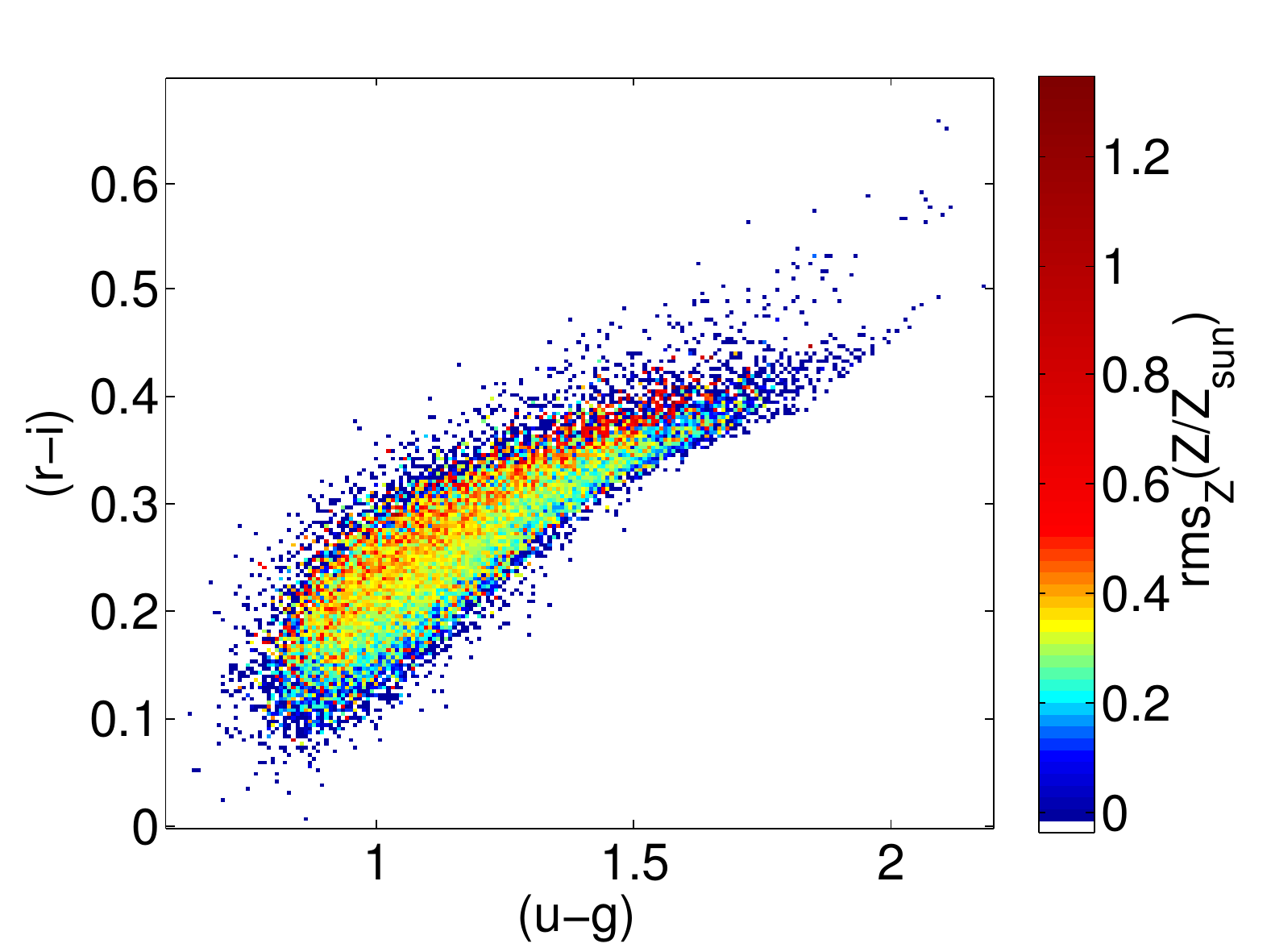}
\caption{Left panels: stellar mass weighted metallicity (color coded pixels) as a function of ($i$-$z$) versus ($g$-$r$), as well as ($r$-$i$) versus ($u$-$g$) for our 50000 models based on BC03hr with dust. Right panels: standard deviation in stellar mass weighted metallicity (color coded pixels) as a function of ($i$-$z$) versus ($g$-$r$), as well as ($r$-$i$) versus ($u$-$g$) for our 50000 models based on BC03hr with dust.}
\label{fig:z}
\end{figure}

\begin{figure}
\includegraphics[width=4cm]{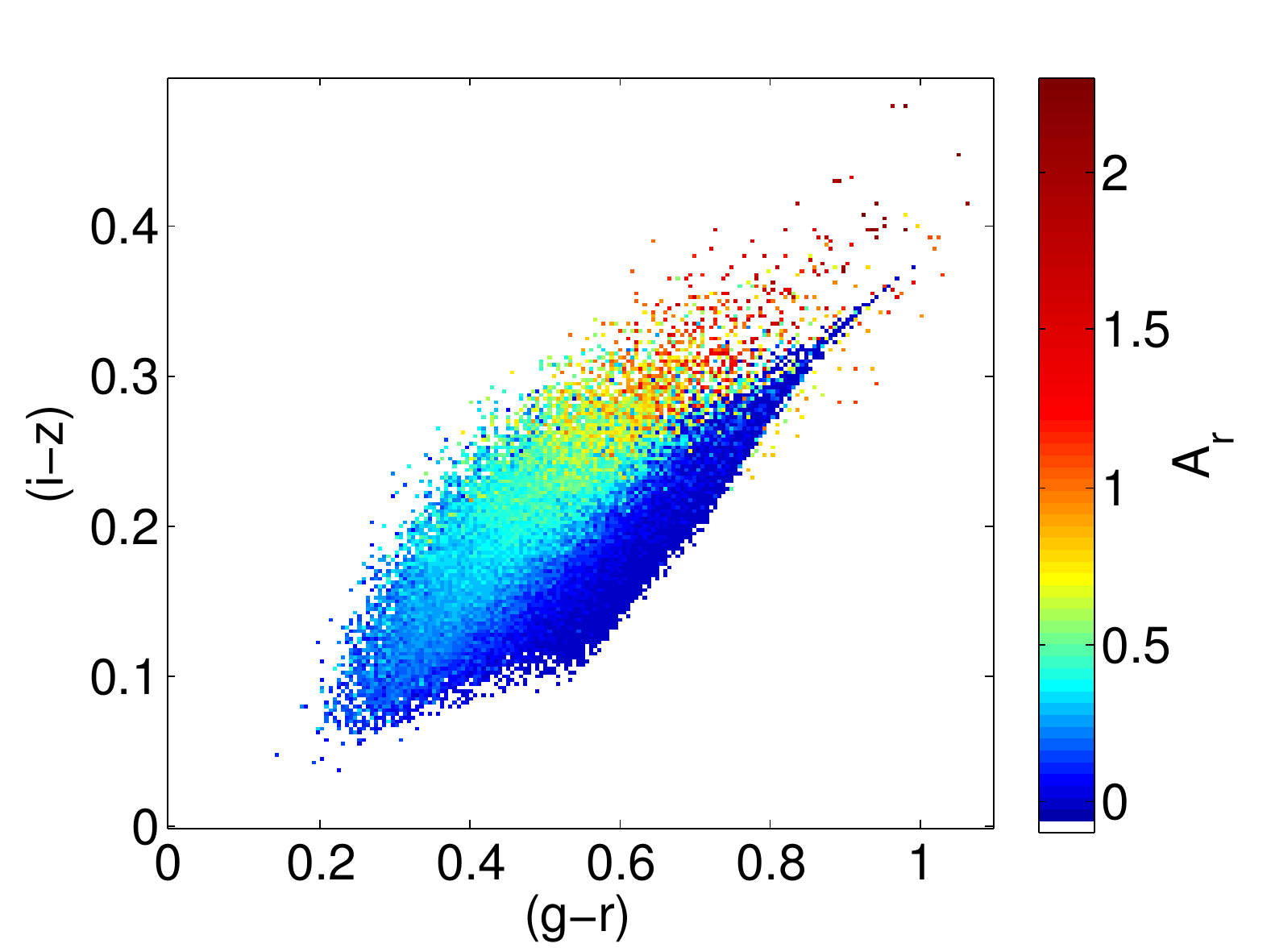}
\includegraphics[width=4cm]{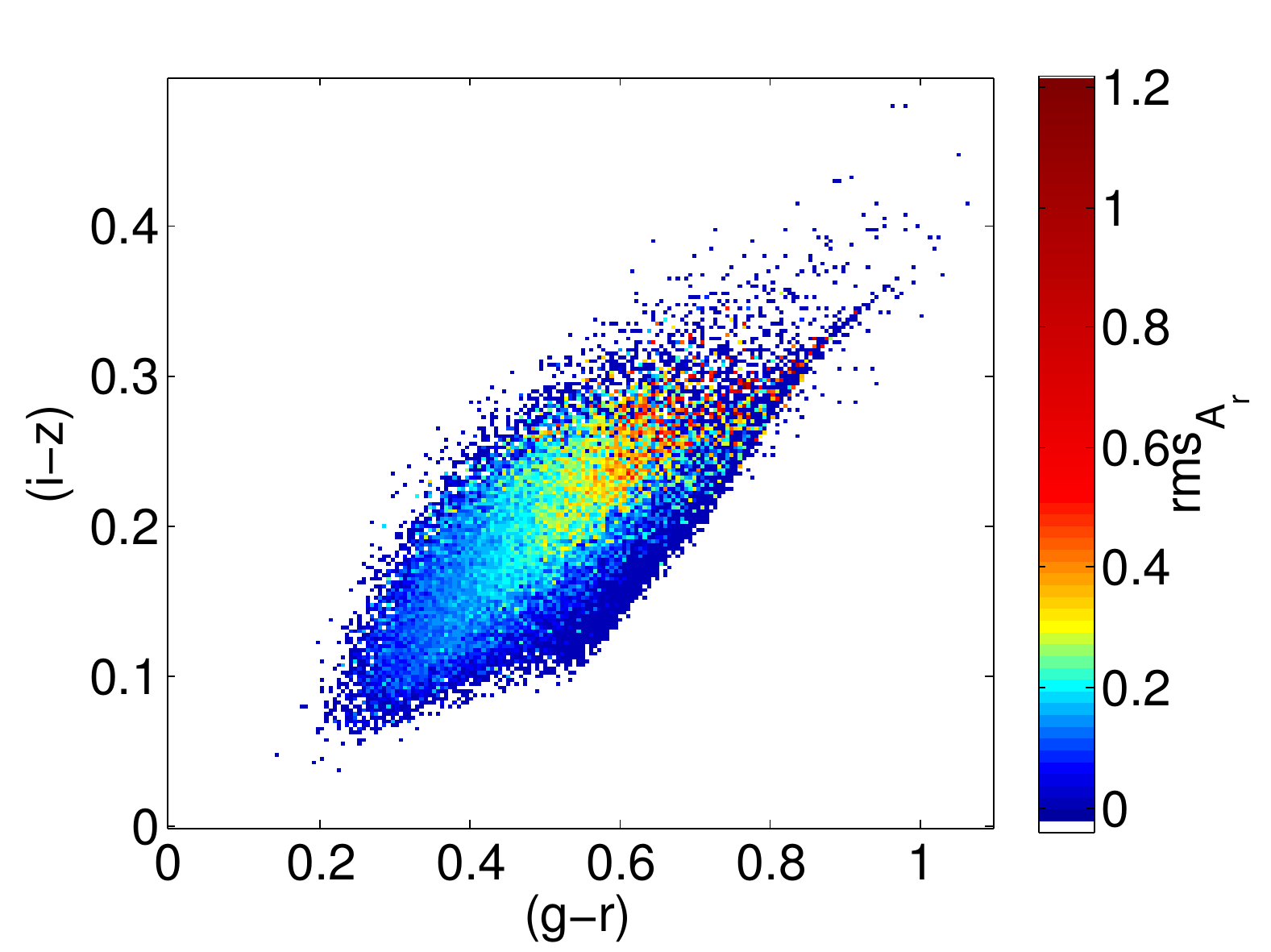}\\
\includegraphics[width=4cm]{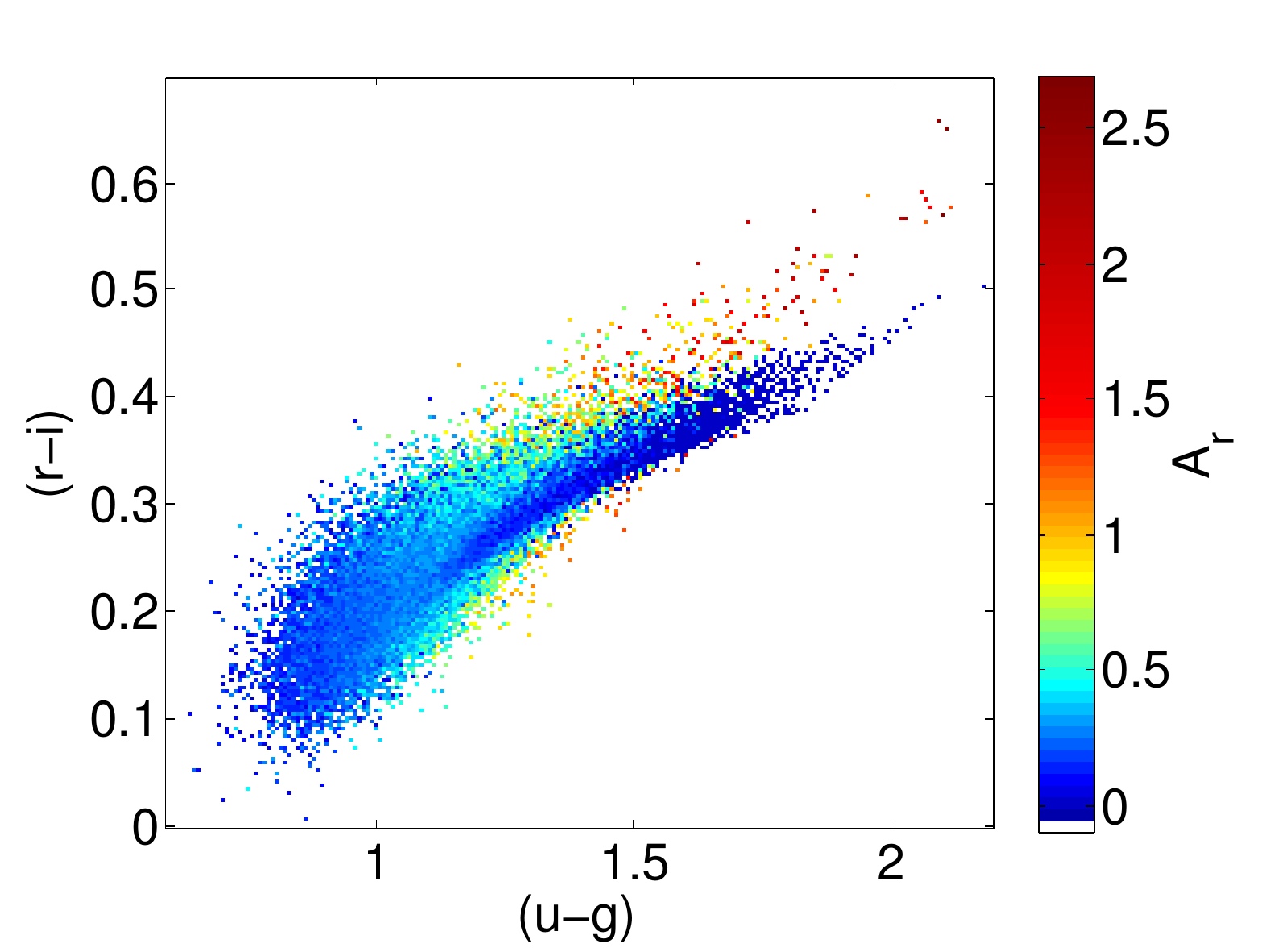}
\includegraphics[width=4cm]{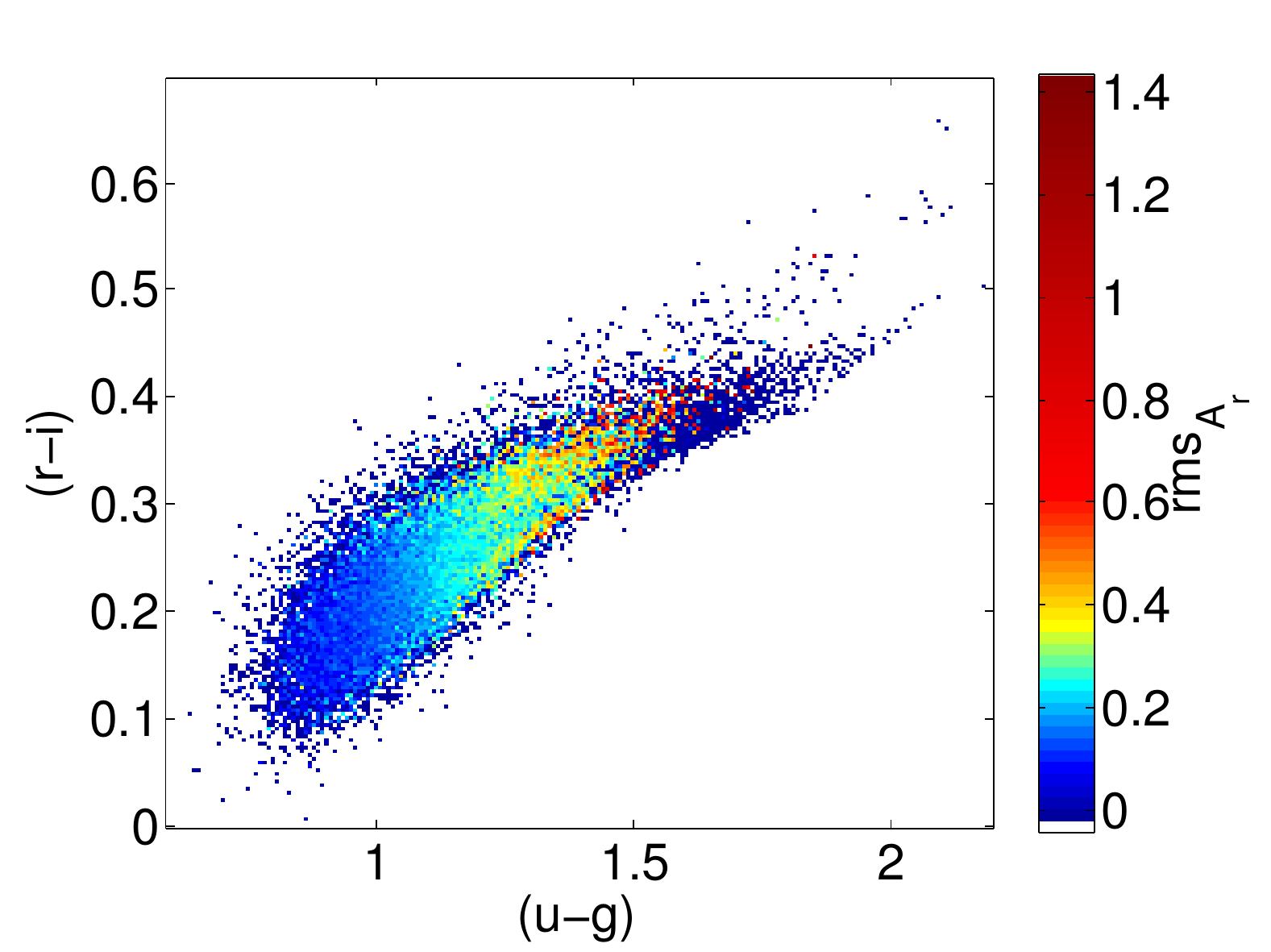}
\caption{Left panels: total $r$-band extinction (color coded pixels) as a function of ($i$-$z$) versus ($g$-$r$), as well as ($r$-$i$) versus ($u$-$g$) for our 50000 models based on BC03hr with dust. Right panels: standard deviation in total $r$-band extinction (color coded pixels) as a function of ($i$-$z$) versus ($g$-$r$), as well as ($r$-$i$) versus ($u$-$g$) for our 50000 models based on BC03hr with dust.}
\label{fig:d}
\end{figure}

\begin{figure*}
\caption{Images (left): 50'x50' SDSS $u$,$g$,$r$,$i$,$z$ color composite images \citep{2002cs........2013S} of 10 randomly selected galaxies from our sample. Diagrams (middle): ($i$-$z$) vs. ($g$-$r$) colors for the galaxies (red dots) and the model boundaries (black lines). Diagrams (right): Normalized star formation rates (red lines) with errors (shaded region) for the corresponding galaxies as a function of stellar age. These star formation histories have been derived through a Bayesian maximum likelihood using all 10 colors and our models based on BC03hr with emission lines and dust (see Appendix \ref{sec:e}). Stellar mass weighted ages and metallicities, z, as well as effective $r$-band optical depths are indicated above each image+diagram pair along with absolute r-band magnitudes, $M_{r}$, and $\mbox{H}\alpha$ equivalent widths from the SDSS (positive values indicate emission and negative values absorption).}
\includegraphics[width=17.0cm]{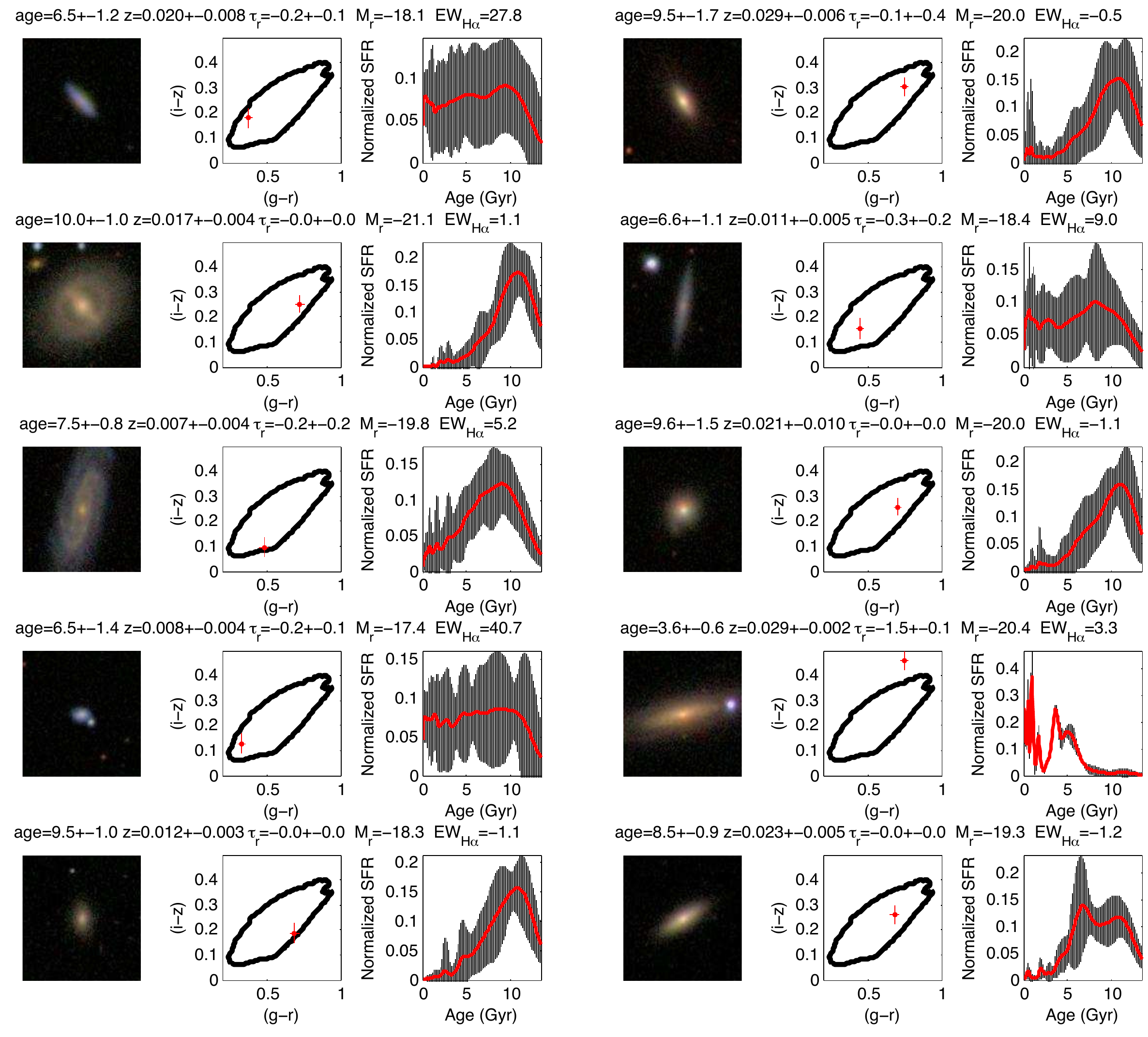}
\label{fig:sfh}
\end{figure*}

To illustrate the predictions of the models in terms of galaxy properties Fig. \ref{fig:age}-\ref{fig:d} show the stellar mass weighted\footnote{In this work stellar masses are derived using a \cite{2003PASP..115..763C} IMF and the stellar mass are taken to be the mass in present day stars and remnants.} ages, metallicities and total $r$-band extinction in the color-color planes along with their standard deviations. This is shown for the BC03hr models with dust and line emission which are in good agreement with the observations (Fig. \ref{fig:ugri} and \ref{fig:griz}). A strong, rather orthogonal dependence is seen between age and metallicity. However, for a region of color space roughly from ($u$-$g$,$r$-$i$)=(1.1,0.2) to ($u$-$g$,$r$-$i$)=(1.6,0.4) and ($g$-$r$,$i$-$z$)=(0.5,0.2) to ($g$-$r$,$i$-$z$)=(0.8,0.3) all properties are rather poorly constrained due to a mix of models with high and low dust extinction. The model library was constructed to cover the ($u$-$g$,$r$-$i$) and ($u$-$g$,$r$-$i$) diagrams and does not tell anything about the probability that a specific model is a good representation of the observations. Further improvement can be achieved using a Bayesian approach. With help from the semi-analytic galaxy formation models of \cite{2011MNRAS.413..101G} we compute model weights to match the distribution of stellar population models with semi-analytic galaxies in terms of density in stellar mass weighted age versus stellar mass weighted metallicity, see Appendix \ref{sec:e}. For 10 randomly chosen example galaxies we show the SDSS color composite image\footnote{Images are taken from http://cas.sdss.org/dr5/en/tools/chart/list.asp} along with with a star formation history derived through a Bayesian maximum likelihood for the BC03hr models with emission lines and dust, see Fig. \ref{fig:sfh}.

\subsection{Visual checks}

\begin{figure}
\center
\includegraphics[width=8.5cm]{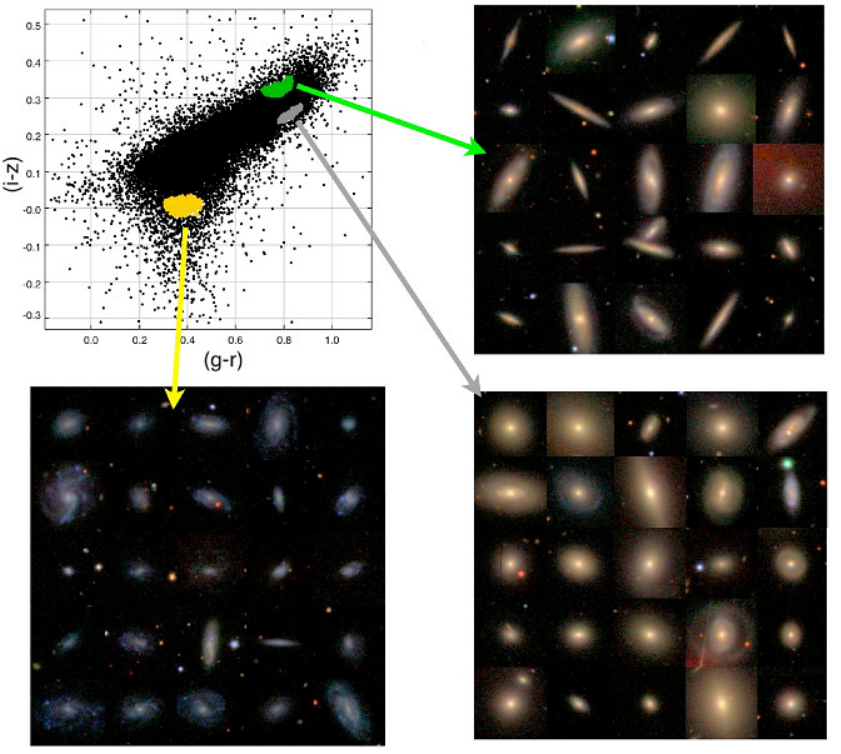}
\caption[]{Illustration of galaxies with special characteristics according to our BC03hr model with dust extinction. From all galaxies (black dots) objects with high expected extinction (green dots), low expected extinction (grey dots) and galaxies falling off the model grid (yellow dots) are selected. SDSS color composites are shown for randomly chosen examples from each group. Note the high frequency of highly inclined disks in the high extinction sample.}
\label{fig:off}
\end{figure}

As a test we selected random samples of galaxies for three different parts in ($g$-$r$)-($i$-$z$) space, which according to our models have certain special characteristics. The first group contains galaxies that belong to a well populated region around ($g$-$r$,$i$-$z$)=(0.4,0.0) that can not be reached by any models (Fig. \ref{fig:off}, yellow dots). This group seems to be populated by star forming galaxies that in many cases exhibit Sm type morphology. The second group at ($g$-$r$,$i$-$z$)$\sim$(0.8,0.3) contains galaxies that, according to our model, are expected to have strong extinction by dust (Fig. \ref{fig:off}, green dots). This group contains many edge on or highly inclined disk galaxies. The last group at ($g$-$r$,$i$-$z$)$\sim$(0.8,0.2) is expected to contain dust-free galaxies (Fig. \ref{fig:off}, grey dots). With a few exceptions this group contains early-type galaxies with little or no visible dust and star formation. SDSS multicolor composites of individual representative galaxies are shown in Fig. \ref{fig:off}.

\subsection{Comparison with spectroscopic studies}
\label{sec:ifs1}

\begin{figure*}
\includegraphics[width=5cm]{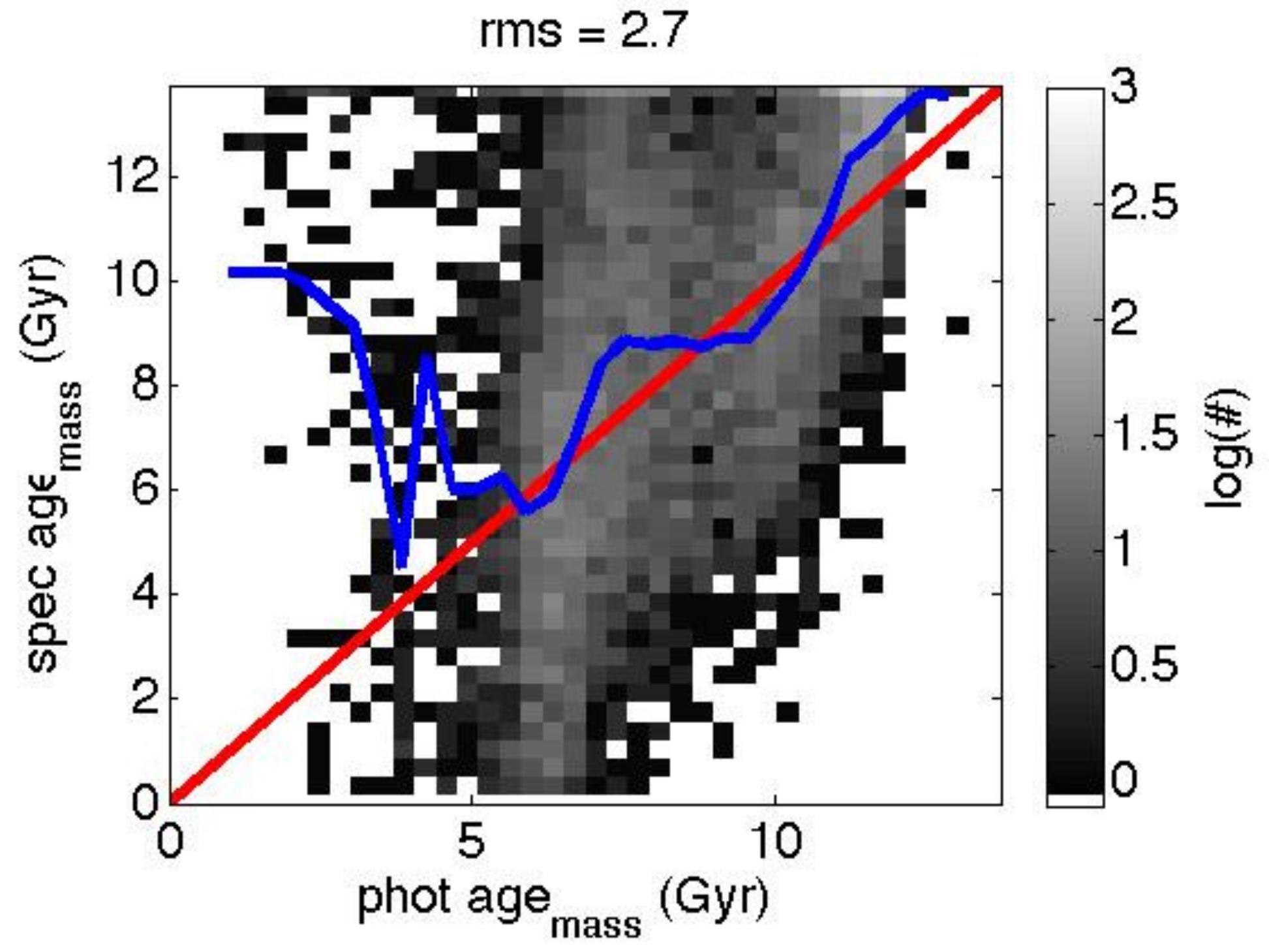}
\includegraphics[width=5cm]{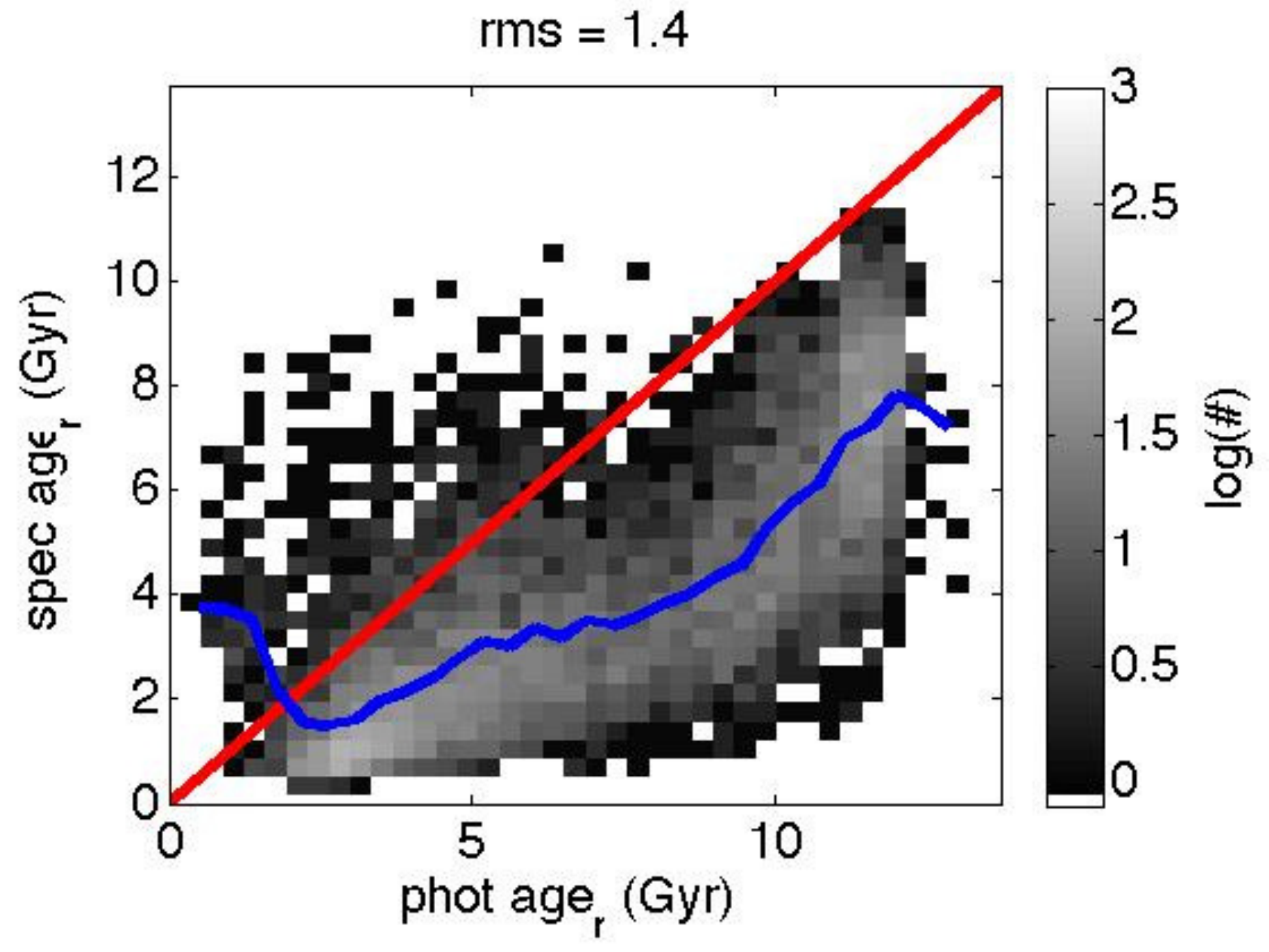}
\includegraphics[width=5cm]{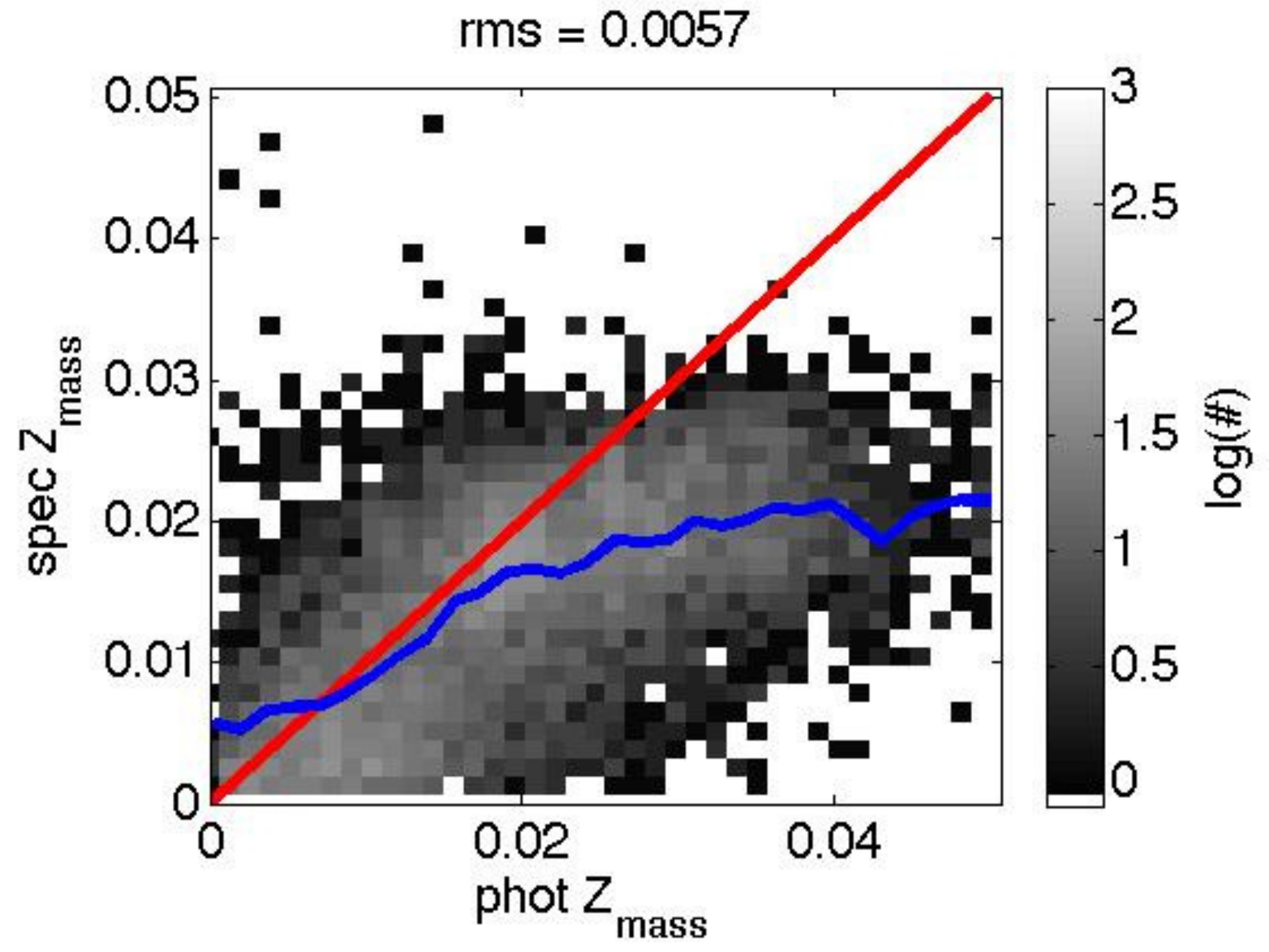}\\
\includegraphics[width=5cm]{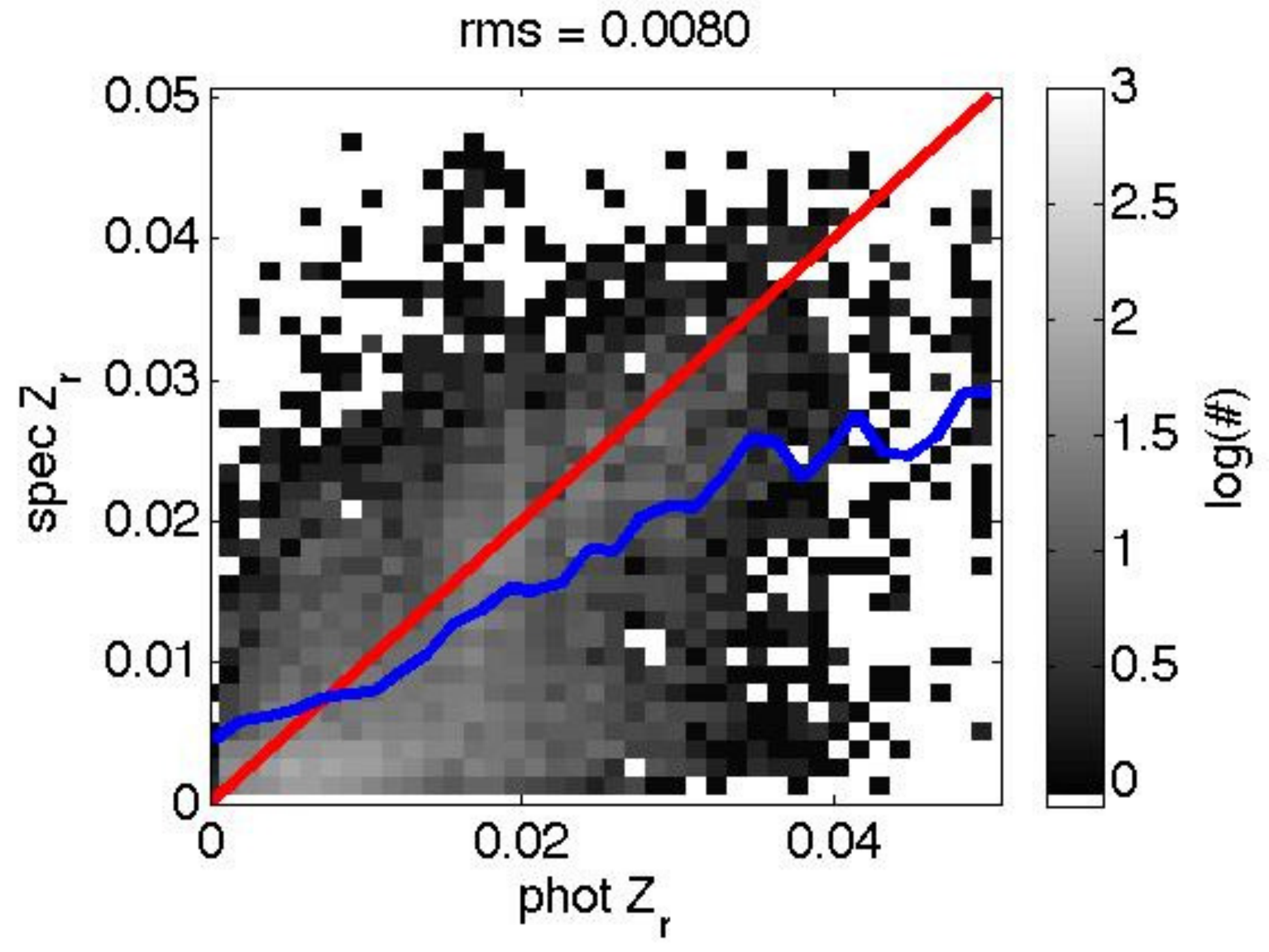}
\includegraphics[width=5cm]{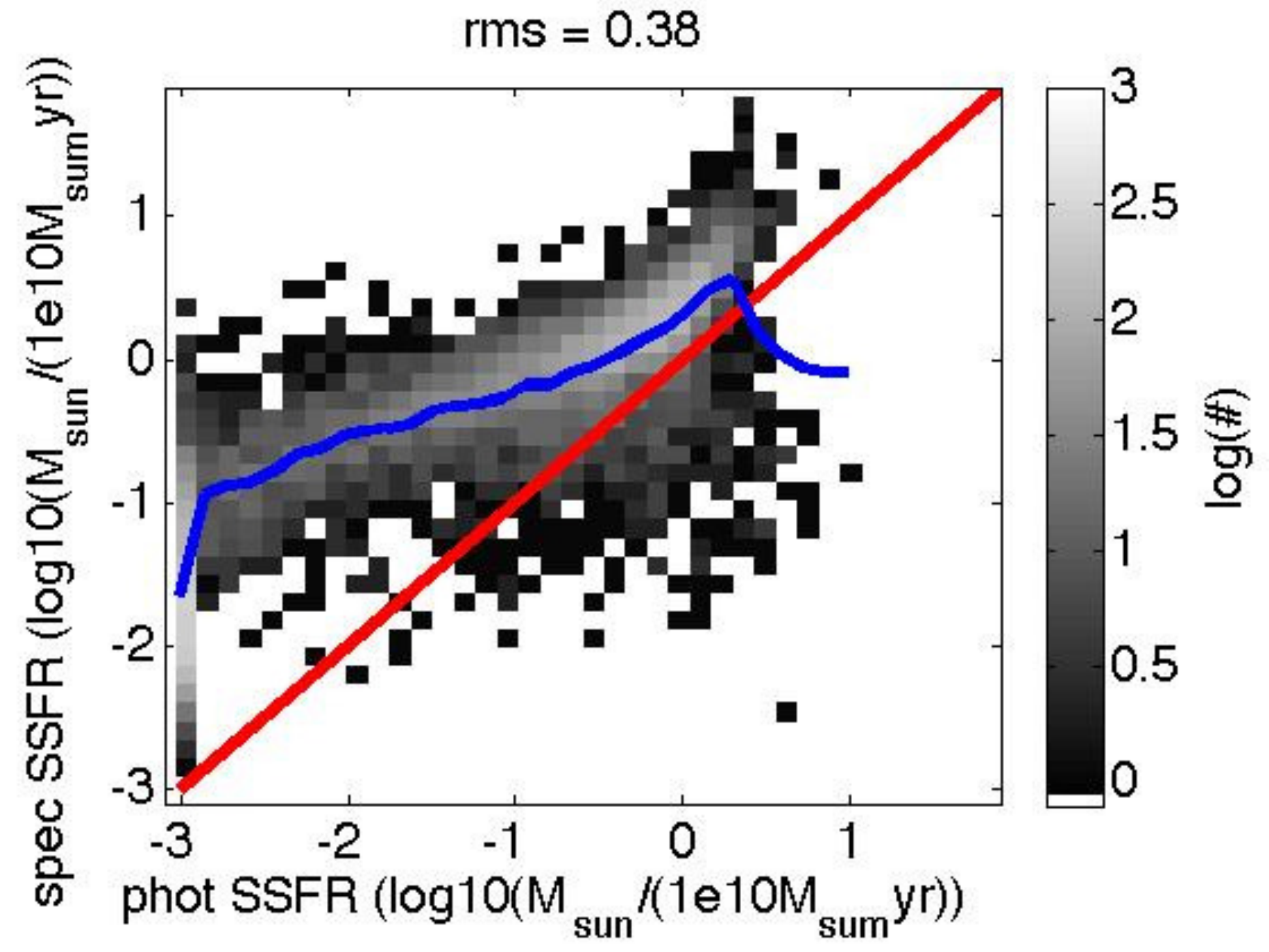}
\includegraphics[width=5cm]{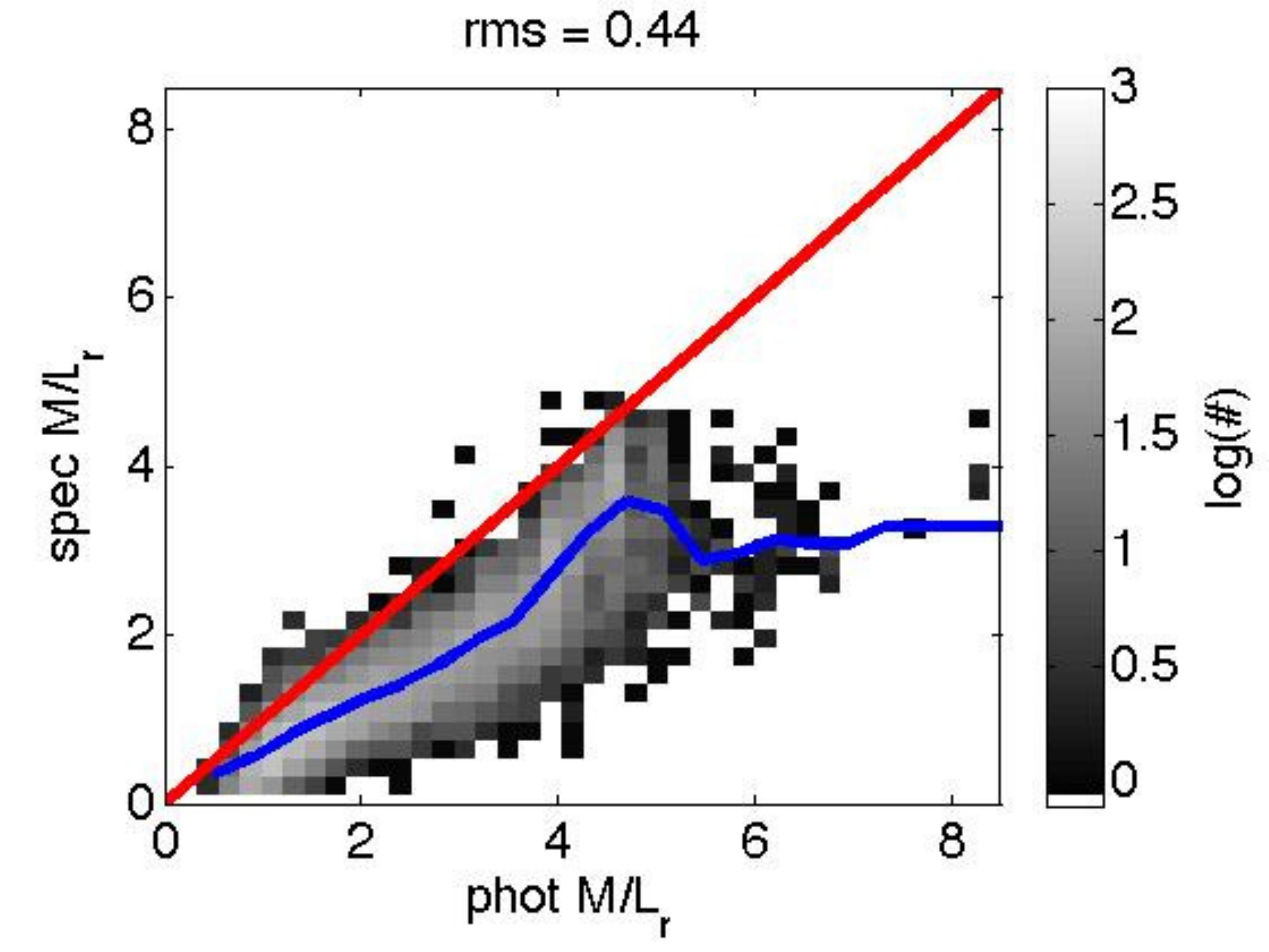}\\
\includegraphics[width=5cm]{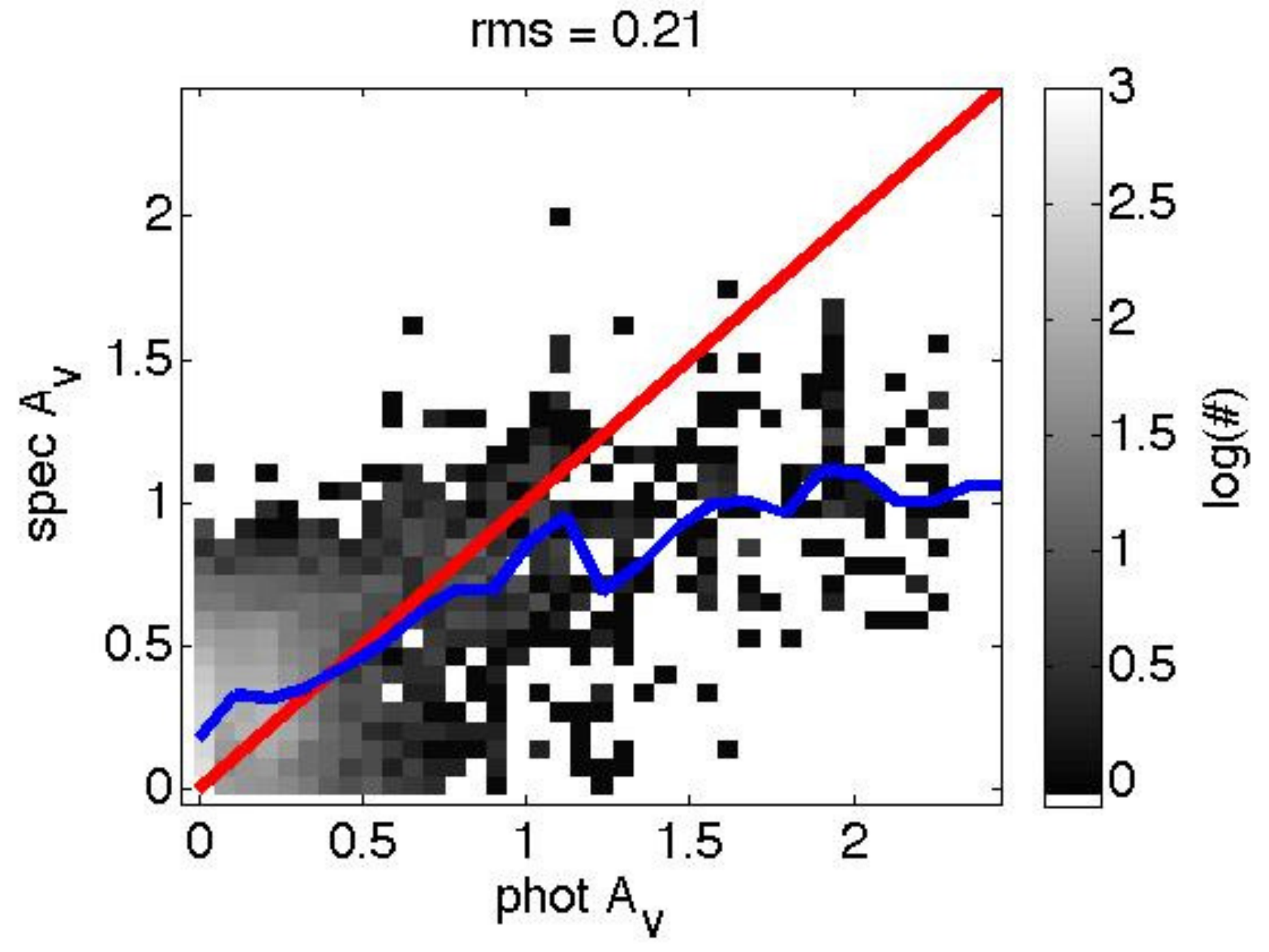}
\caption{Comparison between stellar mass weighted ages, $\text{age}_{mass}$, $r$-band luminosity weighted ages, $\mbox{age}_{r}$, stellar mass weighted metallicities, $\mbox{z}_{mass}$, luminosity weighted metallicities, $\mbox{z}_{lum}$, specific star formation rates, $SSFR$, $r$-band mass to light ratios, $M/L_{r}$, and total $V$-band extinctions, $A_{V}$, for the spectroscopic comparison sample (greyscale histogram) as derived from $u$,$g$,$r$,$i$,$z$-band photometry (phot) and spectra (spec) \citep{2005MNRAS.358..363C,2005MNRAS.362...41G}. The red lines show the 1:1 relations, blue lines indicate the mean in the spectroscopic values in bins of the photometric values. The bin widths are 1/30 of the entire range displayed, however, these are in some cases adjusted to enclose at least 50 galaxies in each bin. On top of each figure the standard deviation in the spectroscopic values around the mean, i.e. the blue line, is indicated. }
\label{fig:es}
\end{figure*}

\begin{figure*}
\includegraphics[width=5cm]{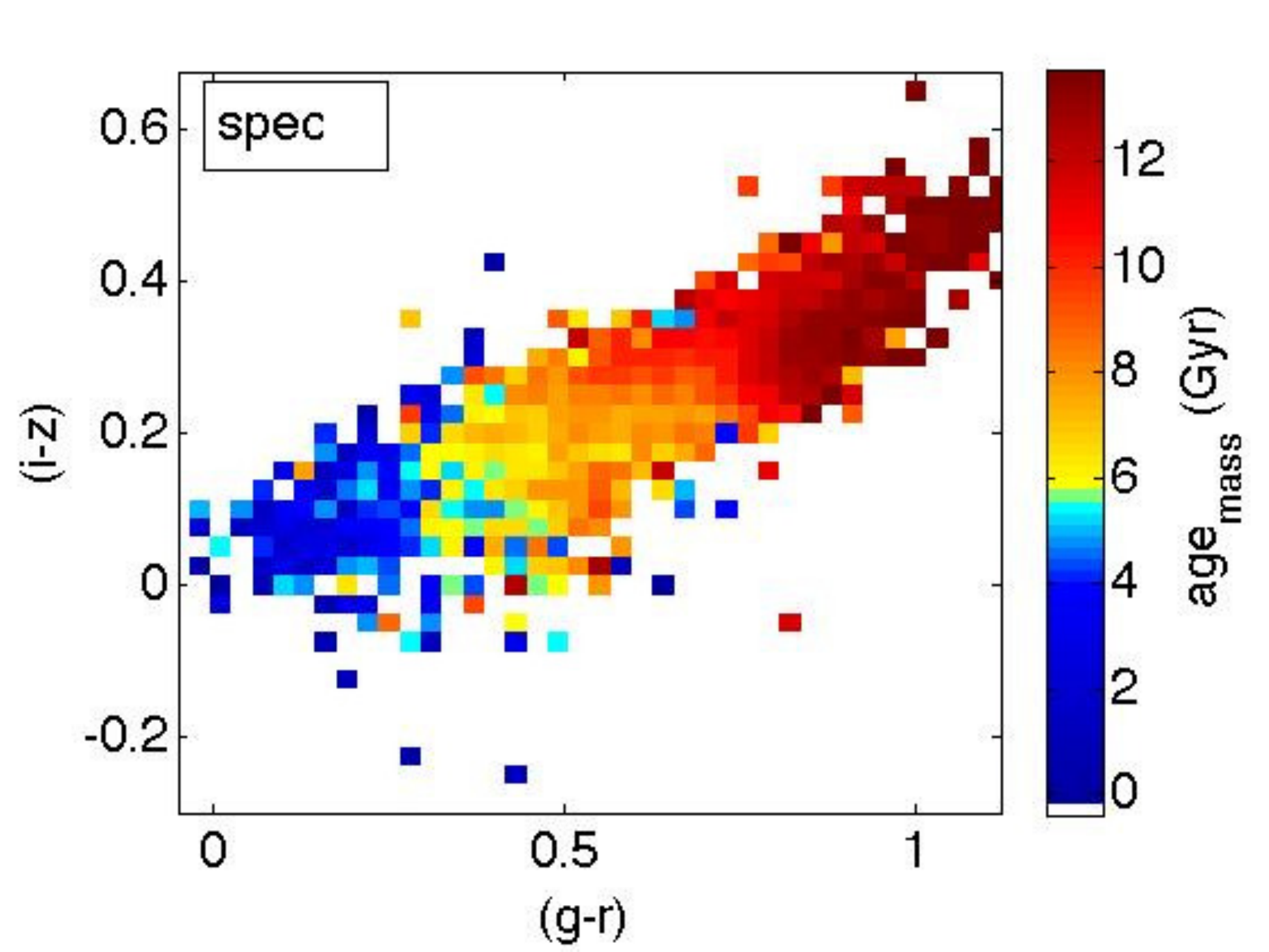}
\includegraphics[width=5cm]{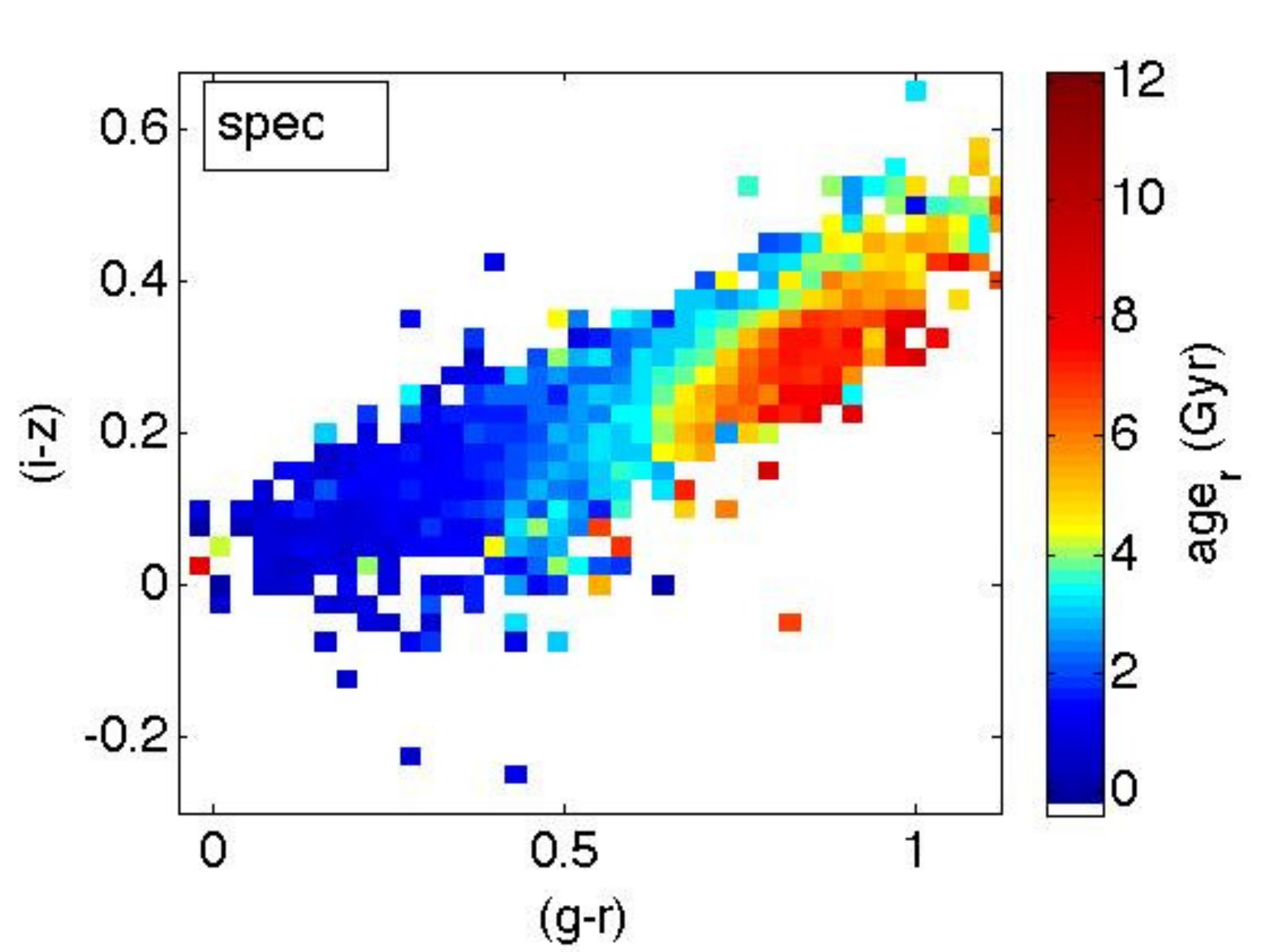}
\includegraphics[width=5cm]{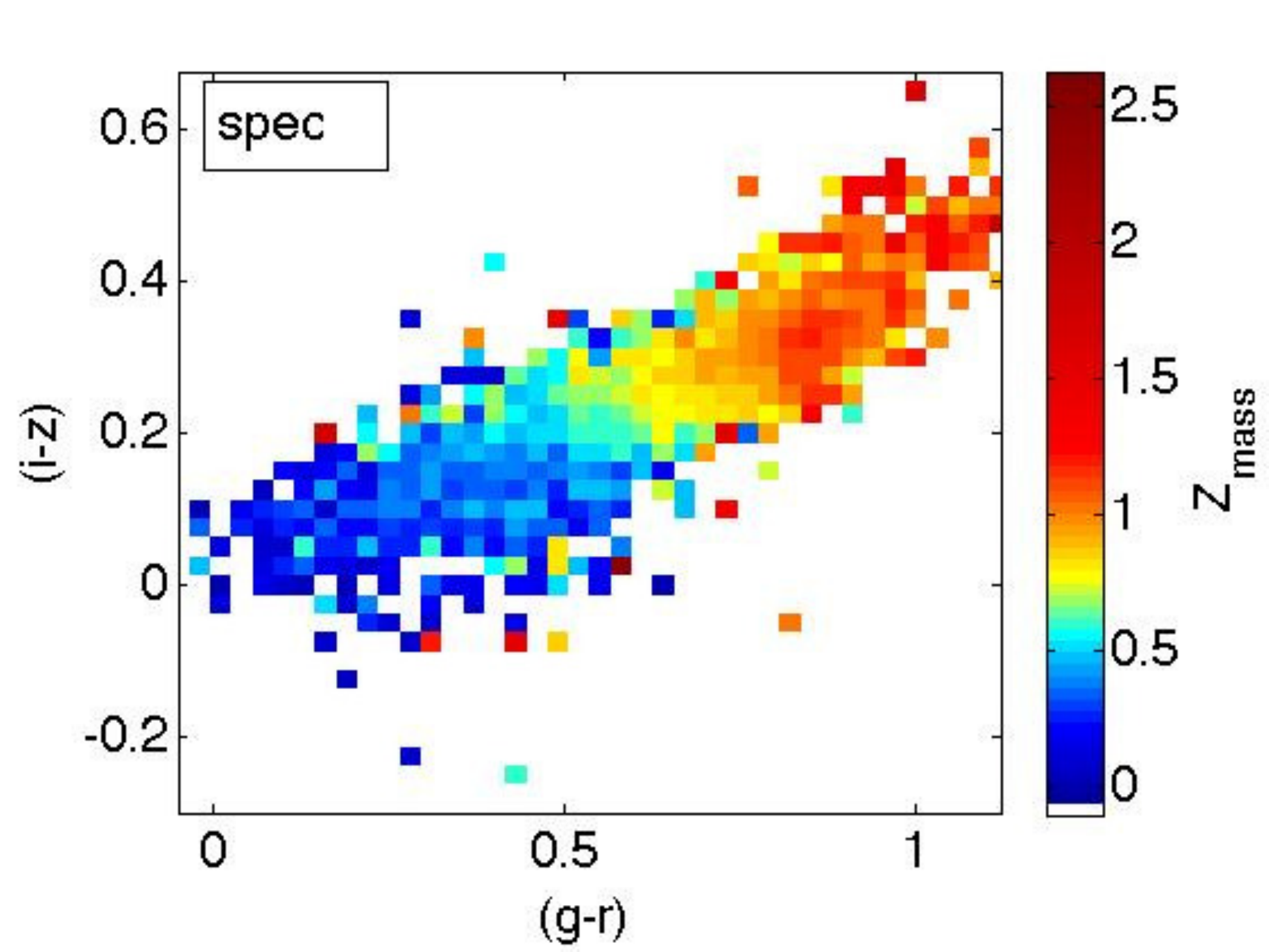}\\
\includegraphics[width=5cm]{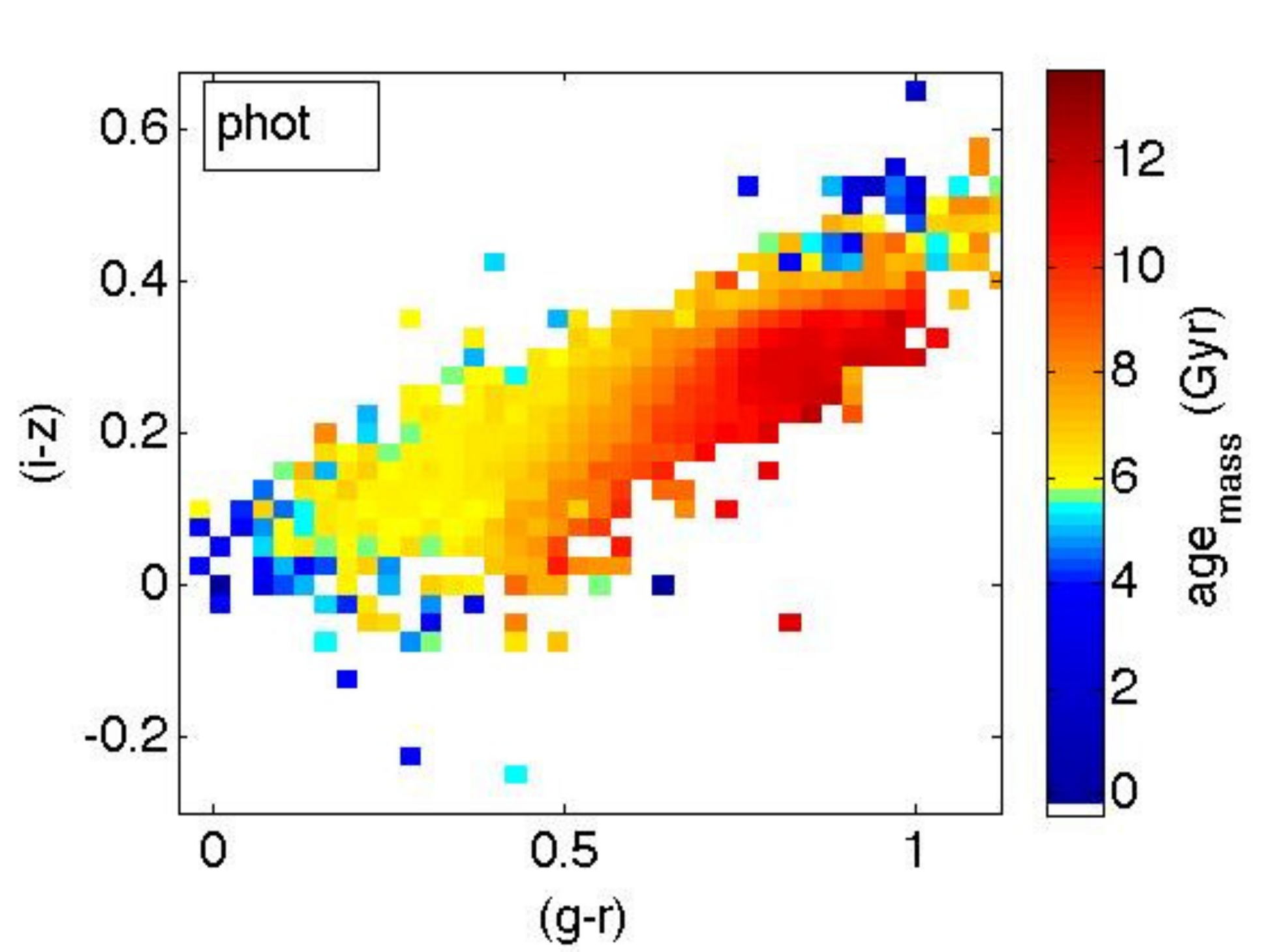}
\includegraphics[width=5cm]{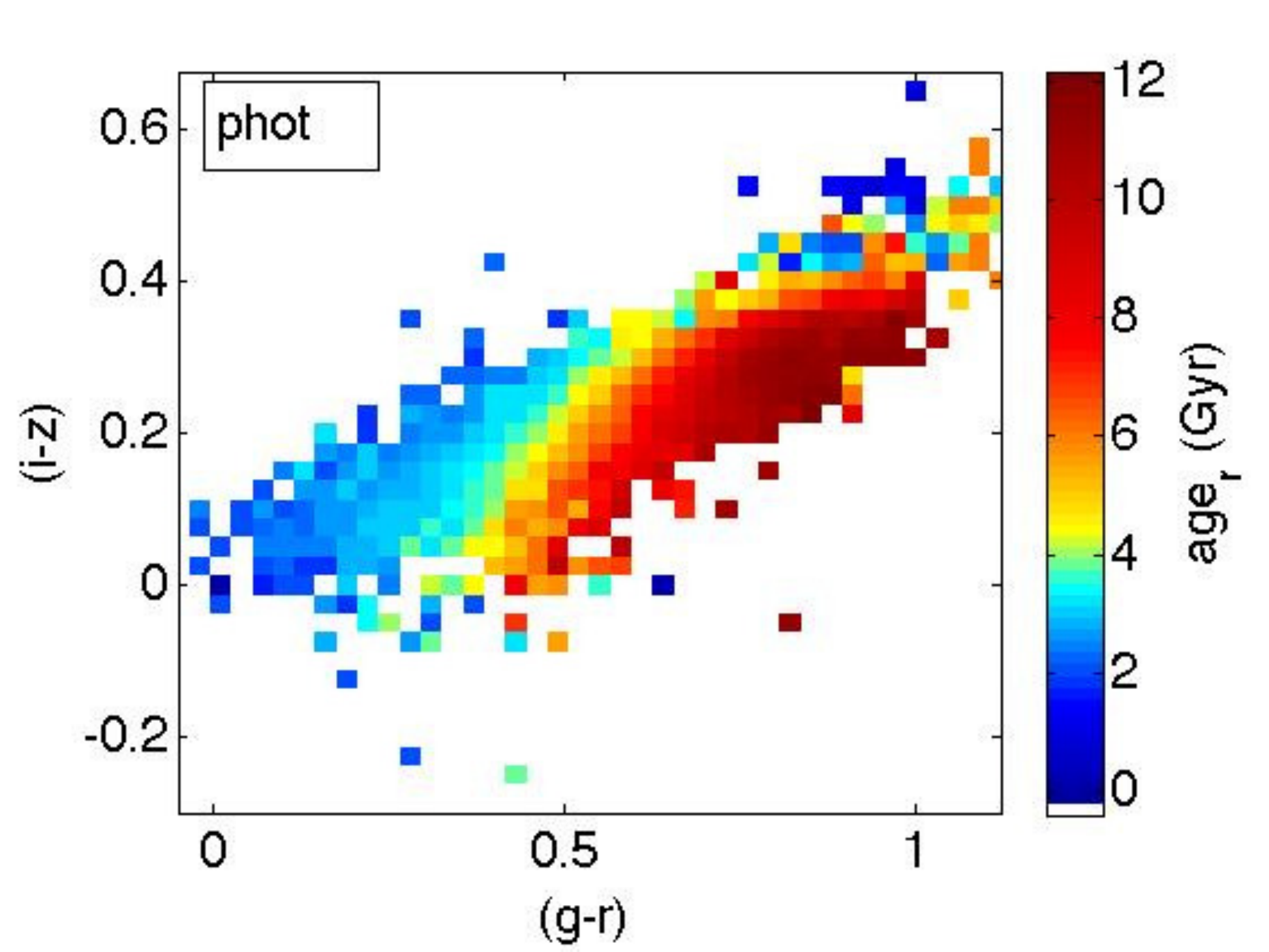}
\includegraphics[width=5cm]{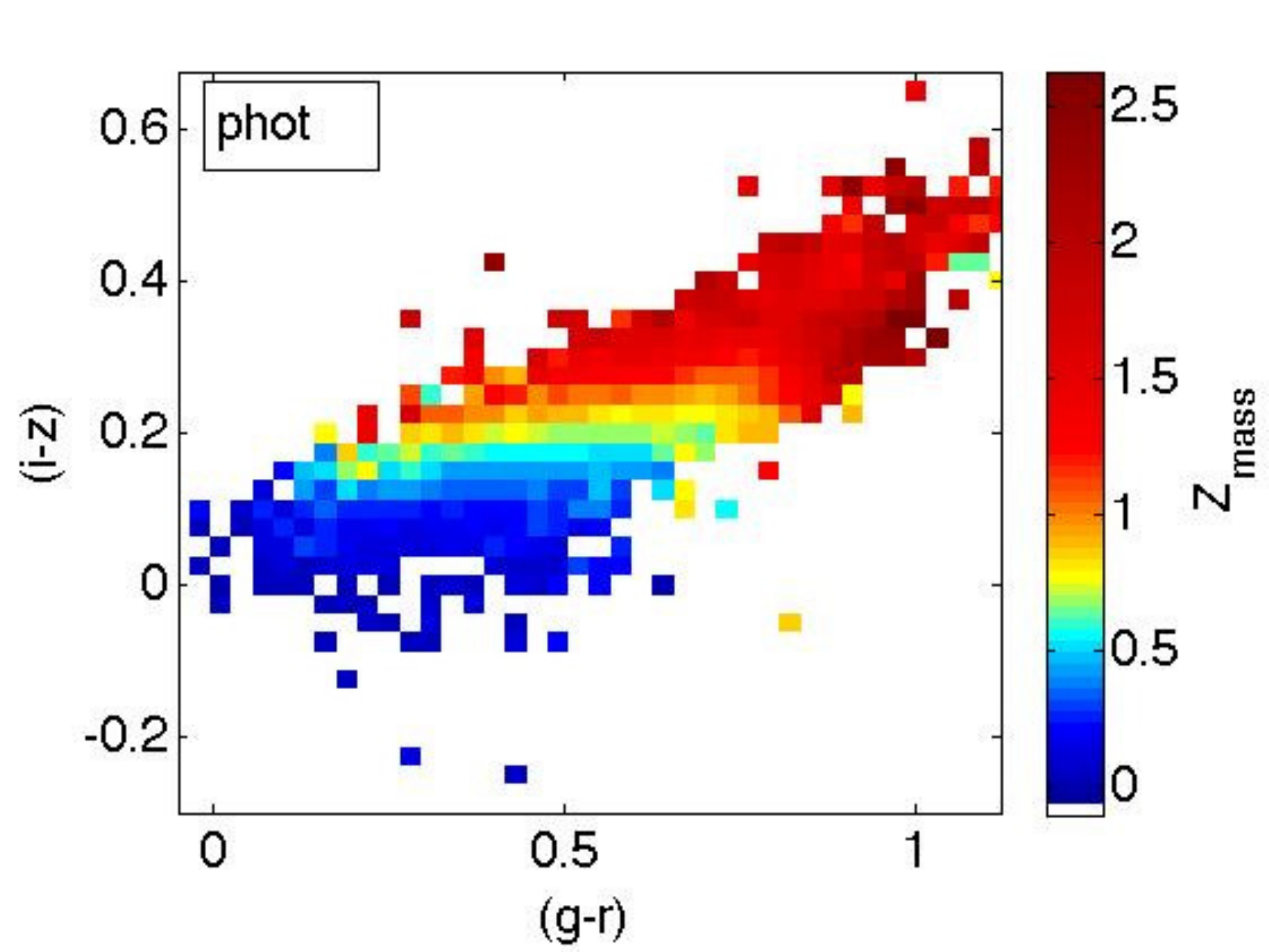}\\
\includegraphics[width=5cm]{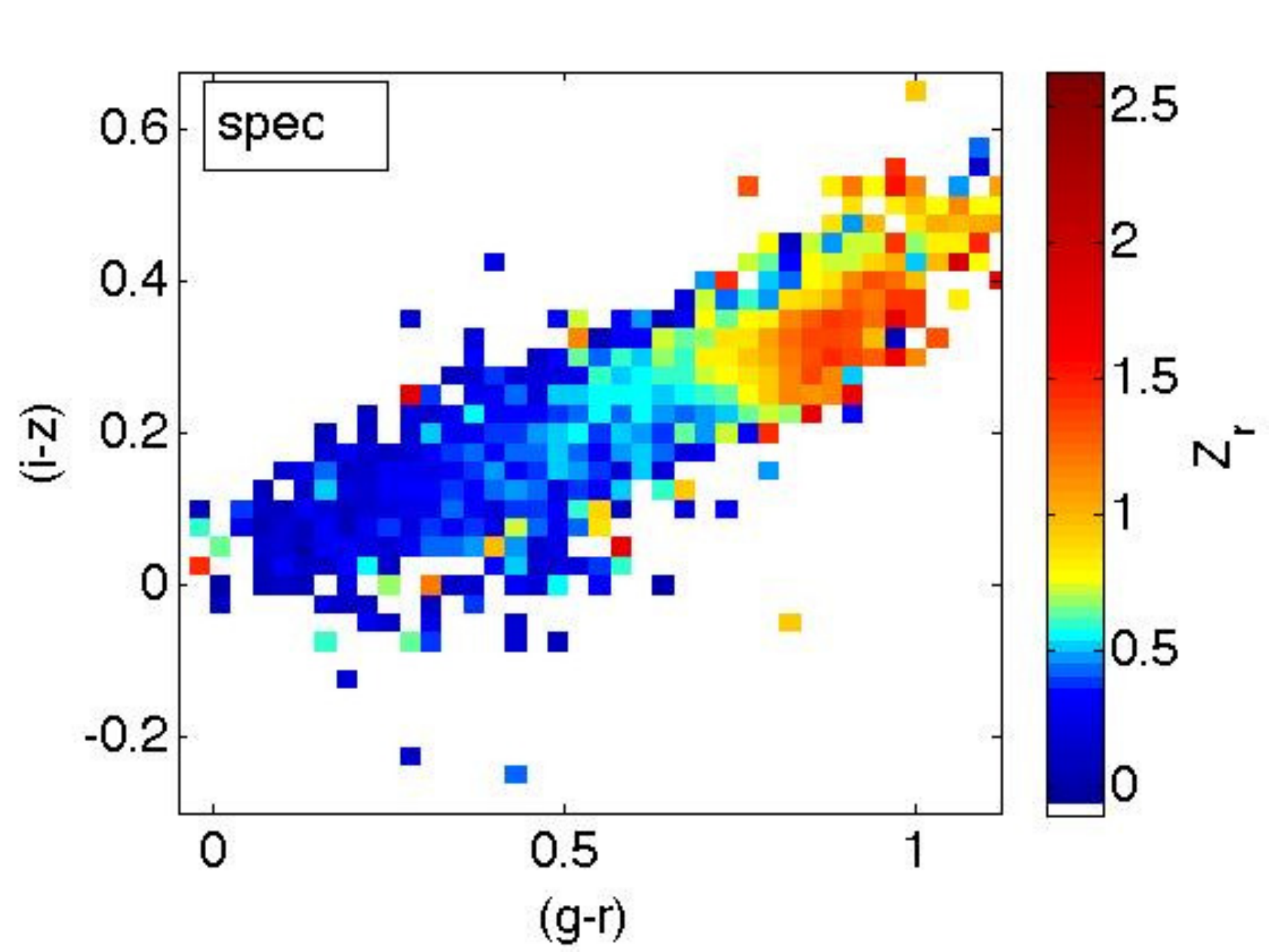}
\includegraphics[width=5cm]{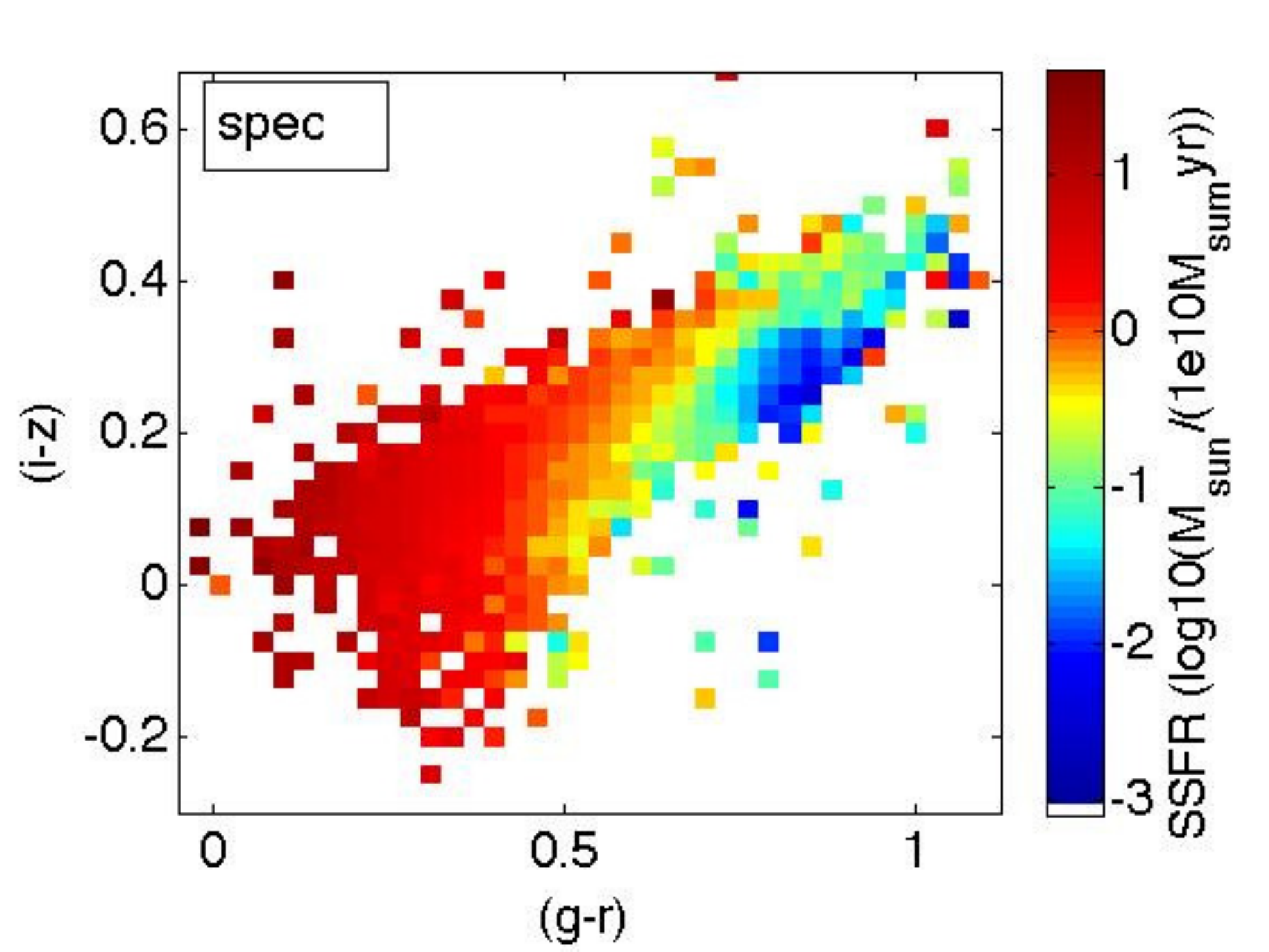}
\includegraphics[width=5cm]{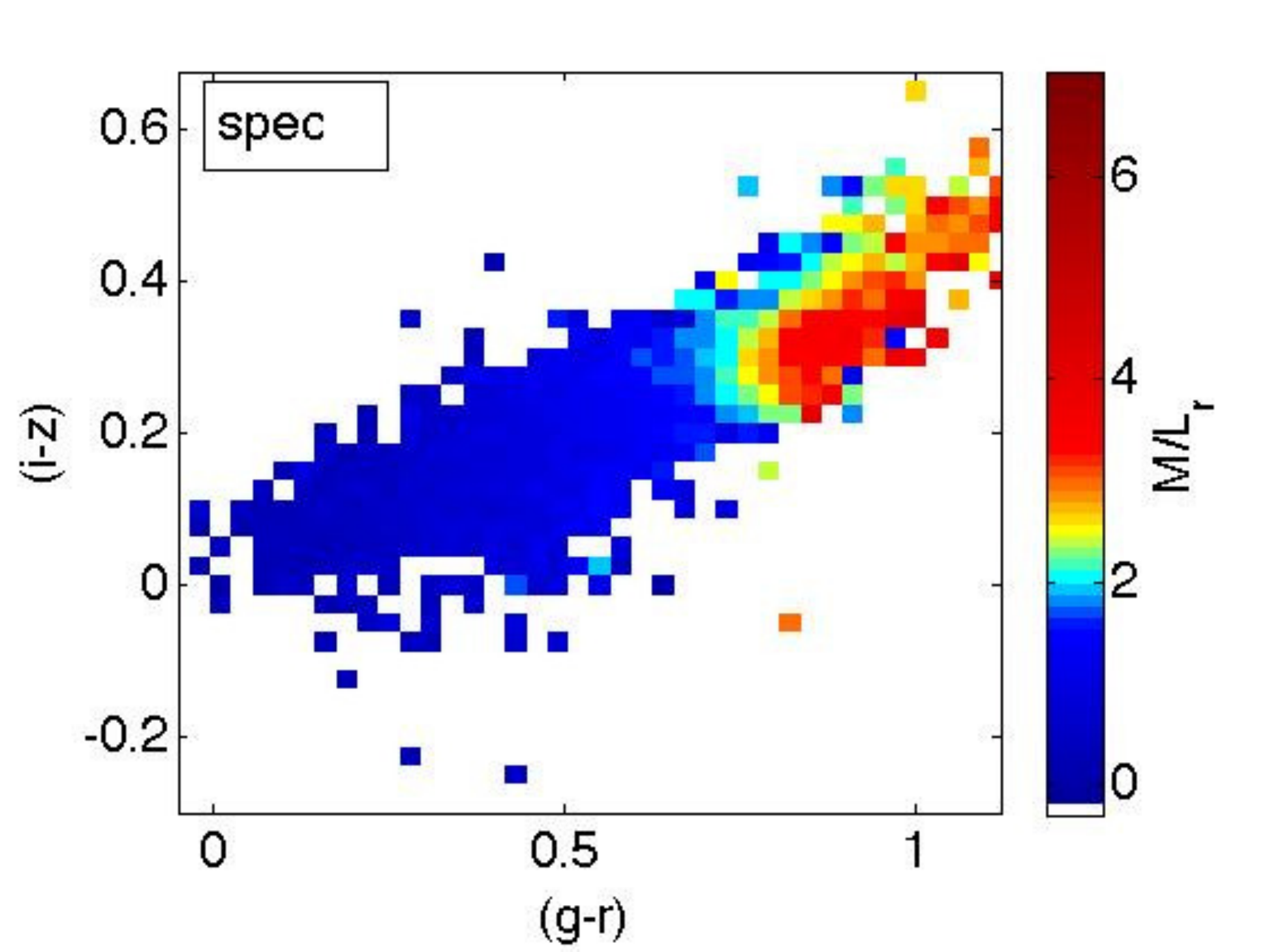}\\
\includegraphics[width=5cm]{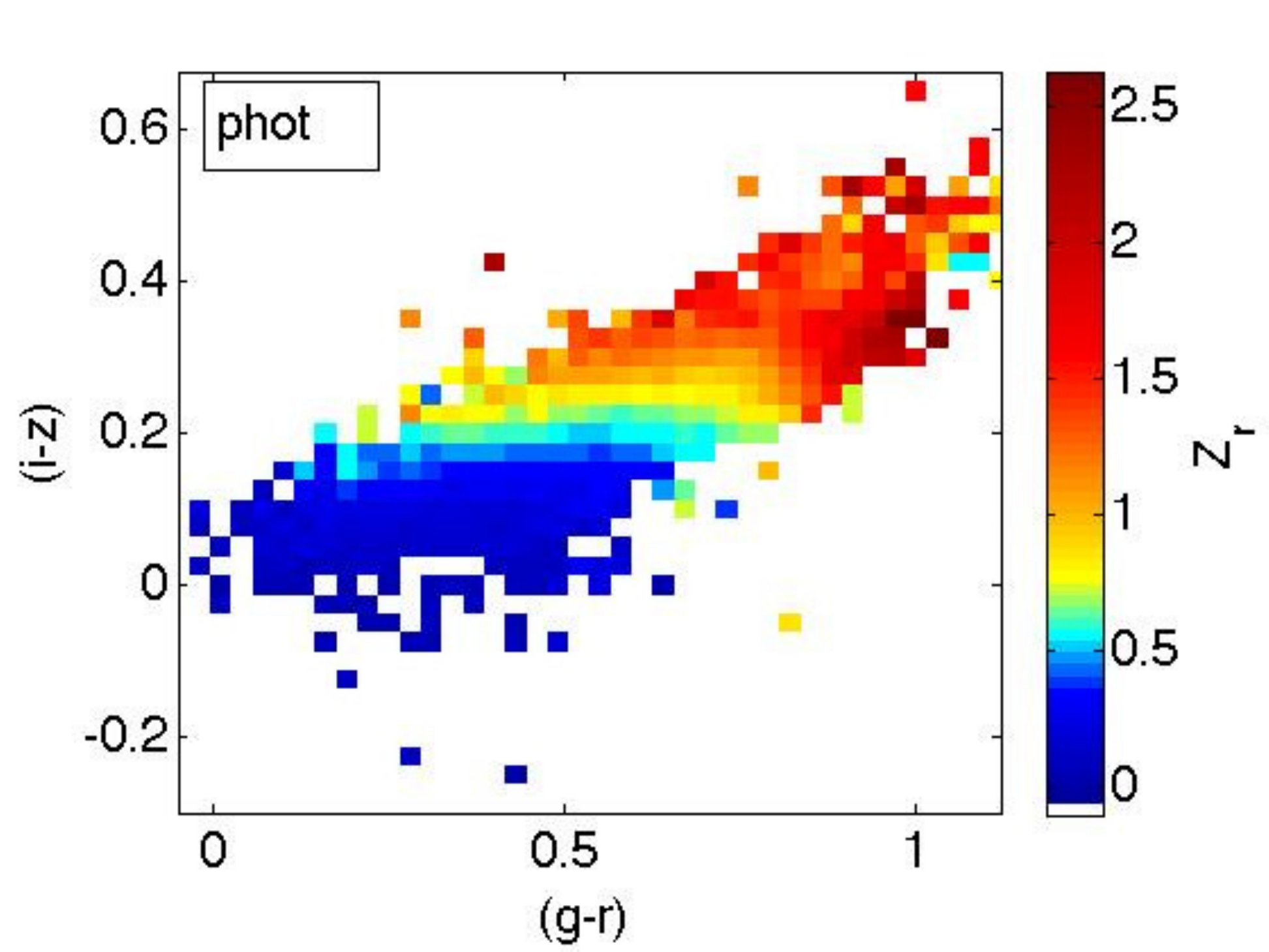}
\includegraphics[width=5cm]{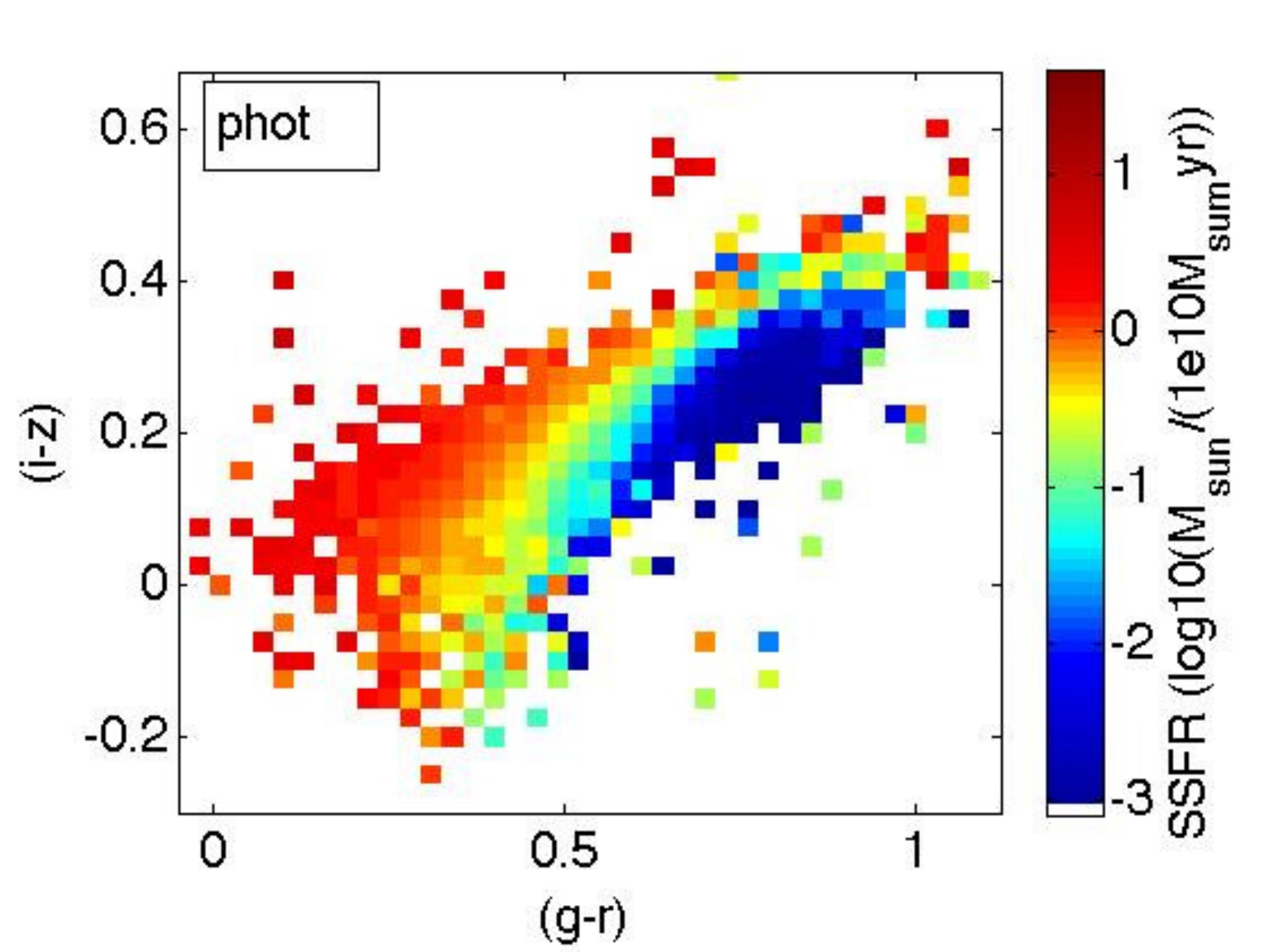}
\includegraphics[width=5cm]{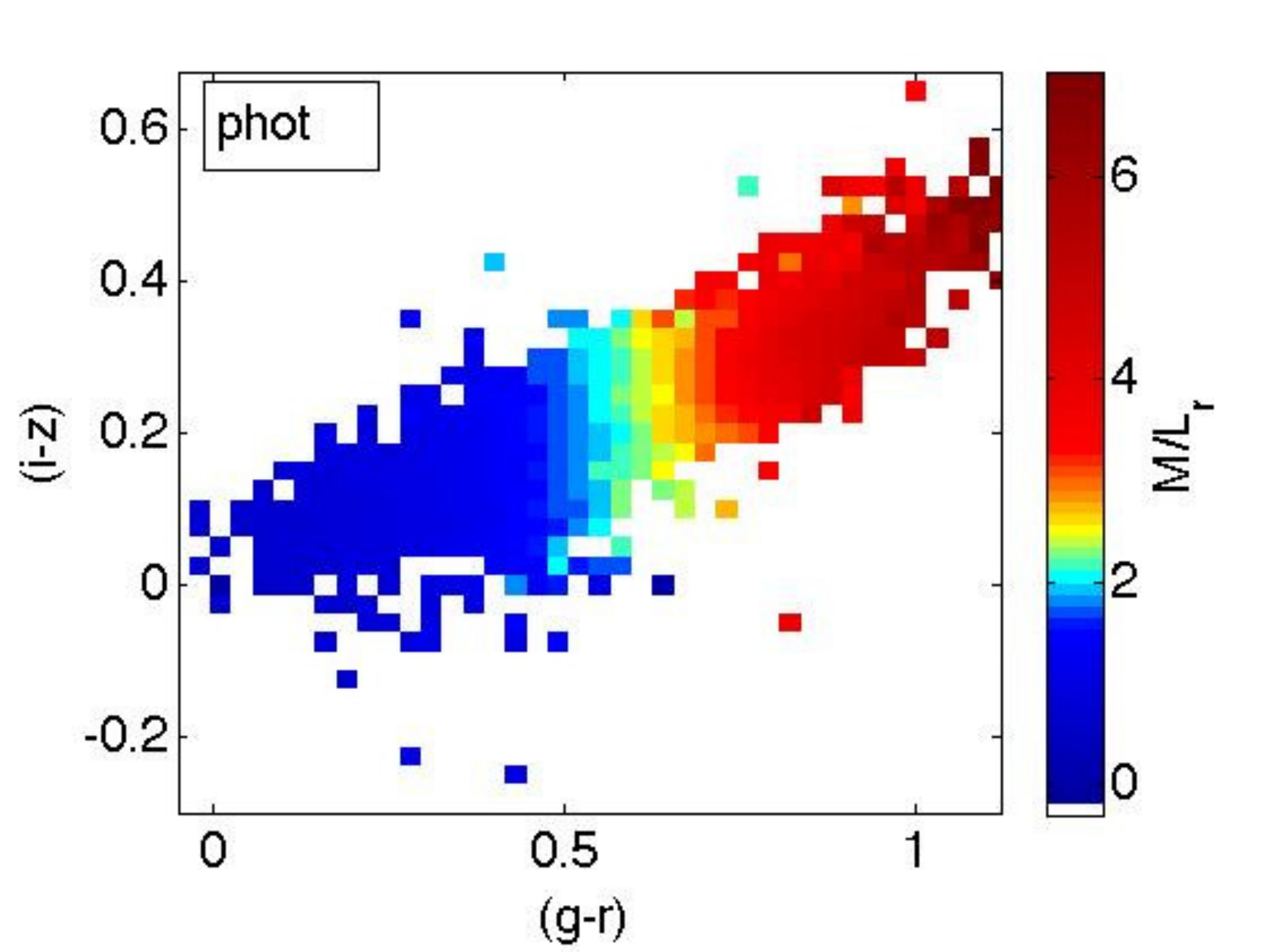}\\
\includegraphics[width=5cm]{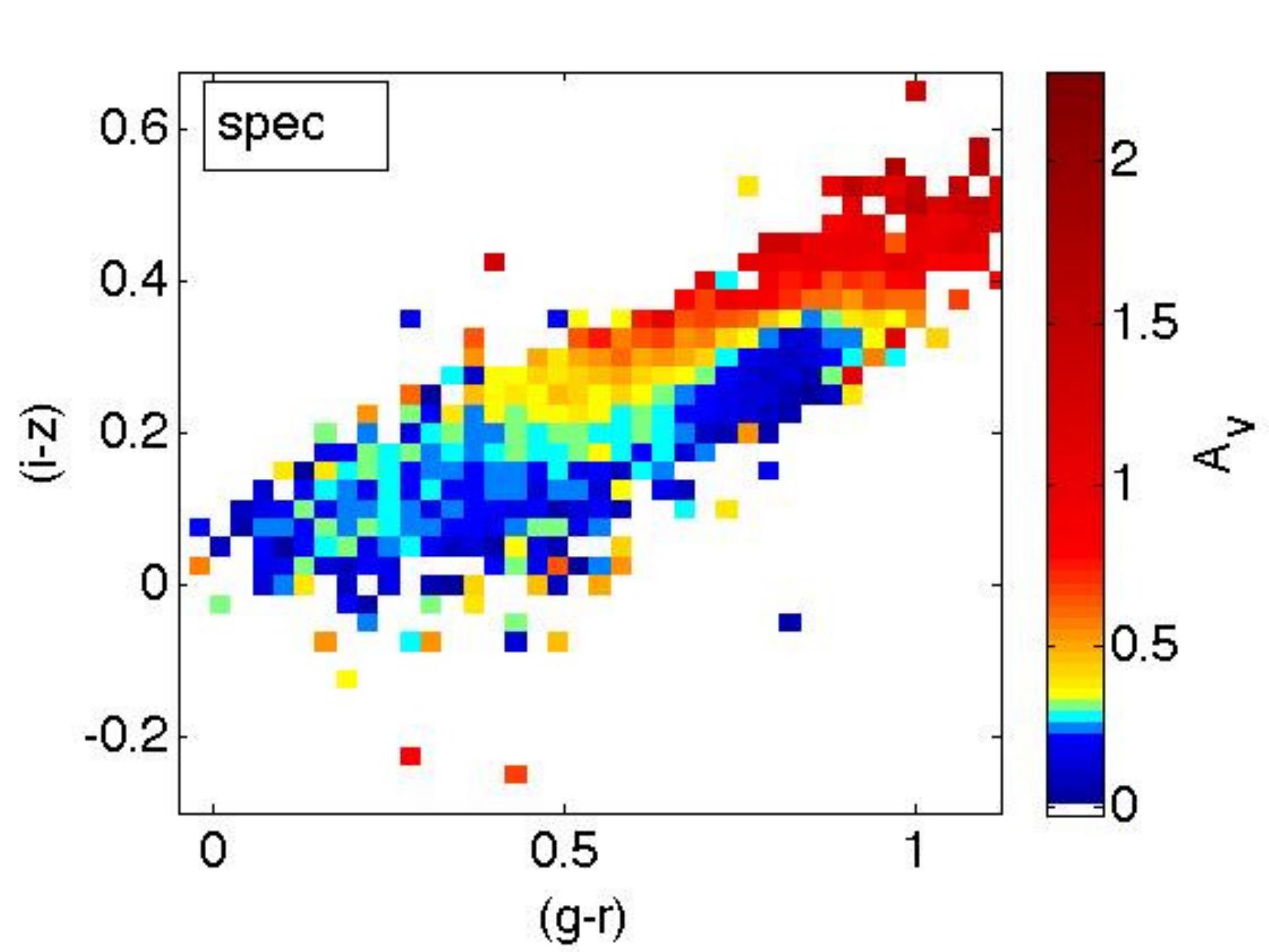}
\includegraphics[width=5cm]{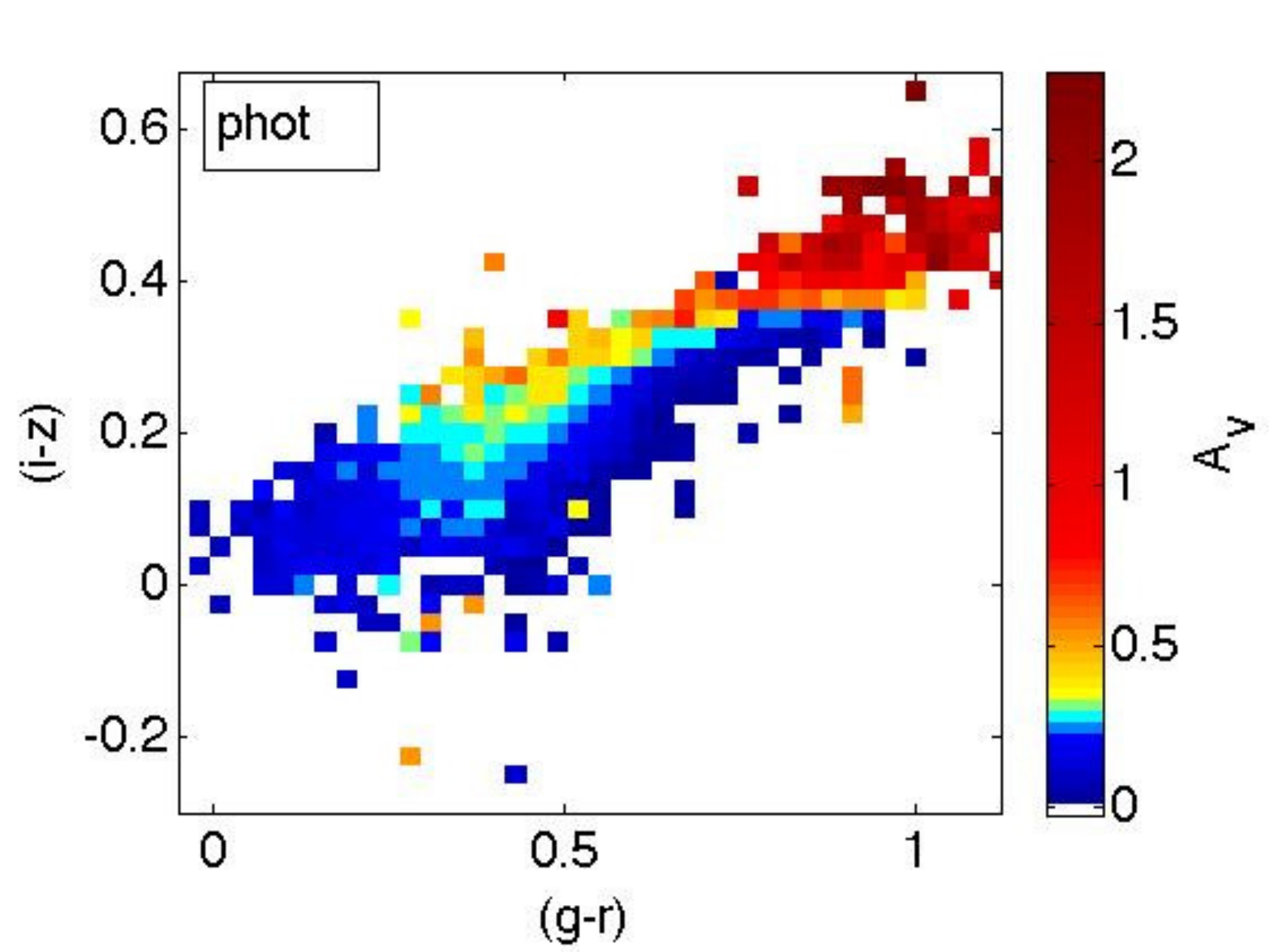}
\caption{Galaxy properties for the spectroscopic comparison sample as we derived from $u$,$g$,$r$,$i$,$z$-band photometry (phot) and \protect\cite{2005MNRAS.358..363C,2005MNRAS.362...41G} derived from spectra (spec). Color coded pixels show the average values of stellar mass weighted ages, $\text{age}_{mass}$, $r$-band luminosity weighted ages, $\text{age}_{r}$, stellar mass weighted metallicities, $\text{z}_{mass}$, $r$-band luminosity weighted metallicities, $\text{z}_{lum}$, specific star formation rates, $SSFR$, $r$-band mass to light ratios, $M/L_{r}$, and total $V$-band extinctions, $A_{V}$, in the ($i$-$z$) vs. ($g$-$r$) diagram. Spectroscopic quantities are derived either using spectral synthesis techniques ($\text{age}_{mass}$, $\text{z}_{mass}$, $\text{age}_{r}$) or an absorption line index analysis ($\text{z}_{lum}$).}
\label{fig:es2}
\end{figure*}

There is no question about the superior information content of optical spectra with respect to broad band photometry. Photometry is nevertheless an important probe of stellar populations of galaxies as it is easy to obtain and because spectroscopic data often suffer from aperture bias. As mentioned in Sect. \ref{sec:obs} we compiled spectroscopically determined galaxy parameters from the literature for testing the performance of our photometric model. The parameters considered are stellar mass weighted ages, $age_{mass}$, luminosity weighted ages, $age_{r}$, stellar mass weighted metallicities, $z_{mass}$, luminosity weighted metallicities, $z_{lum}$, specific star formation rates, $SSFR$\footnote{Specific star formation rates are taken as the ratio of the average star formation rate during the last $10^{8}$yr and the galaxy's stellar mass. An average star formation rate over the last $10^{8}$yr as determined from the stellar continuum has proven to be successful in reproducing star formation rates as estimated from emission lines \citep{2007MNRAS.381..263A}.}, $r$-band mass to light ratios, $M/L_{r}$, and $V$-band extinctions, $A_{V}$. To be able to make a fairer comparison between photometrically and spectroscopically derived quantities we set the maximum of $age_{mass}$ and minimum of $A_{V}$ from \cite{2005MNRAS.358..363C} to 13.5Gyr and 0.0mag, respectively, as opposed to 18Gyr and -0.9mag in the catalogue. Note, however, that $age_{mass}$ derived from spectroscopy may still be systematically older because the star formation histories from which $age_{mass}$ are derived include SSPs with ages up to 20Gyr. Additionally, we set the minimum $SSFR$ to -3 in $\text{log10}(M_{sun}/(10^{10}\text{Gyr}))$ as non star forming galaxies in our photometric model have $SSFR$=-Inf in $\text{log10}(M_{sun}/(10^{10}\text{Gyr}))$. Fig. \ref{fig:es} shows a comparison of derived parameters in the ($i$-$z$) vs. ($g$-$r$) diagram. The overall agreement between the spectroscopically and the photometrically determined quantities appears to be mediocre. The main differences in the behavior in ($i$-$z$) vs. ($g$-$r$) are that: 1) photometric metallicities increase with ($i$-$z$) and are almost independent of ($g$-$r$), while spectroscopic metallicities also depend on the latter; 2) very young objects, as determined from photometry, are present at ($g$-$r$,$i$-$z$)=(1.0,0.5), but these are not seen if spectroscopic ages are used. A direct comparison between the derived parameters, see Fig. \ref{fig:es2}, does show a significant scatter as well as some systematic differences. As the differences are rather ``irregular'' we refrain from an attempt to parameterize them.

Under the assumption that the spectroscopically derived quantities are of higher accuracy than those derived from photometry we can probe the quality of the parameters included in the fitting. Results of such a test are shown in Table \ref{tab:prop} in terms of the standard deviation in the spectroscopically derived quantity at fixed photometrically derived values. The use of $u$,$g$,$r$,$i$,$z$-band photometry alone leads to a low standard deviation in all properties except the SSFR. This problem is solved if a prior as a function of $M_{r}$ is introduced and additionally a slight improvement can be achieved by introducing $a/b$ in the fitting. Note however that $M_{r}$ can be estimated using $u$,$g$,$r$,$i$,$z$-band photometry with an accuracy of about 1mag, a sufficiently good estimate to reach almost the same accuracy as when $M_{r}$ itself is used. Using only $M_{r}$, or $M_{r}$ and $a/b$, we can still constrain several quantities as the model galaxy properties have a strong mass/luminosity dependence as well as some structural dependences \citep{2011MNRAS.413..101G}, but the standard deviation is considerably higher in many of the parameters.

Moreover, with a similar technique we can test whether the exclusion of certain bands in the fitting can improve the quality of the derived parameters. We find that this is the case. Removal of one or two bands can indeed decrease the standard deviation of the comparison with the spectroscopy by 5-10\% for each of the parameters tested. However, an increase in the accuracy of one parameter typically comes at the expense of a diminished accuracy in another. For example, excluding $u$ and $g$ leads to a better estimate in the mass weighted metallicity but makes the accuracy of the $SSFR$ much worse. As no overall improvement is achieved by excluding one or two bands, we prefer to keep all bands in the fitting.

\begin{table*}
\caption{Dependence of the standard deviation in stellar mass weighted age, $age_{mass}$, luminosity weighted age, $age_{r}$, stellar mass weighted metallicity, $z_{mass}$, luminosity weighted metallicity, $z_{lum}$, specific star formation rate, $SSFR$, $r$-band mass to light ratio, $M/L_{r}$, and $V$-band extinction, $A_{V}$, compared with the spectroscopically determined values. The parameters used in the computation of the maximum likelihoods are $u$,$g$,$r$,$i$,$z$-band photometry, ugriz, a prior based on $M_{r}$, prior($M_{r}$), a prior based on a Monte Carlo estimate of $M_{r}$ from $u$,$g$,$r$,$i$,$z$-band photometry, prior(ugriz$\rightarrow M_{r}$), and the isohotal axis ratio $a/b$.}
{\small
\begin{tabular}{l|l|l|l|l|l|l|l}
\hline
& & & & rms\\
fitting parameters & $age_{mass}$&$age_{r}$ & $z_{mass}$& $z_{lum}$& $SSFR$ & $M/L_{r}$ & $A_{V}$\\
 & (Gyr)&(Gyr) & ($z_{\odot}$)& ($z_{\odot}$)& (dex) & ($M_{\odot}/L_{\odot}$) & (mag)\\
\hline
ugriz    & 2.8 & 1.6 & 0.33 & 0.43 & 0.87 & 0.49 & 0.21\\
prior($M_{r}$)   & 2.9 & 2.3 & 0.31 & 0.44 & 0.50 & 0.89 & 0.27\\
a/b + prior($M_{r}$)  & 3.2 & 2.3 & 0.38 & 0.48 & 0.73 & 1.01 & 0.24\\
ugriz + prior(ugriz$\rightarrow M_{r}$)   & 2.8 & 1.5 & 0.34 & 0.44 & 0.75 & 0.50 & 0.21\\
ugriz + a/b + prior(ugriz$\rightarrow M_{r}$)  & 2.8 & 1.5 & 0.34 & 0.44 & 0.49 & 0.50 & 0.21\\
ugriz + prior($M_{r}$)  & 2.9 & 1.6 & 0.32 & 0.44 & 0.49 & 0.50 & 0.21\\
ugriz + a/b + prior($M_{r}$) & 2.8 & 1.5 & 0.32 & 0.44 & 0.49 & 0.50 & 0.21\\
\hline
\end{tabular}
}
\label{tab:prop}
\end{table*}

\subsection{Insights from spectral fitting}
\label{sec:ifs2}

While the main focus of this paper is testing model predictions for $u$,$g$,$r$,$i$,$z$-band photometry an efficient test can be achieved by comparing observed and modelled spectra instead of colors due to the much larger information content. A drawback is that we already expect there will be some issues with the modelling because the chemical content of a galaxy is described by a single parameter in our model, i.e., the metallicity, an approximation that is not expected to reproduce all possible line ratios in galaxies.

We randomly select 1000 galaxies in our sample for which we obtain fully reduced flux calibrated SDSS spectra. These have a median signal to noise ratio per pixel of 19 in the $r$-band and are fitted to our model library in the wavelength range $3700\mbox{\AA}$ to $7000\mbox{\AA}$ (rest frame), as described in the Appendix \ref{sec:d}. Sky lines and emission lines are masked out in the fit and the line of sight velocity distribution is taken into account.

The median reduced $\chi^{2}$ of the fit to the 1000 galaxies is 1.4 for the BC03hr models with dust. A visual inspection of observations and best-fitting models shows that not only issues with lines, but also with the continuum fit exist. We note that the median reduced $\chi^{2}$ can be substantially lowered if the fitted region is restricted to a shorter wavelength interval like $5000\mbox{\AA}$ to $5500\mbox{\AA}$ instead of $3700\mbox{\AA}$ to $7000\mbox{\AA}$ \footnote{Other spectral fitting methods that do not rely on a set of predefined models (in our case 50000) have been able to produce better fits to SDSS spectra \citep[c.f.][]{2005MNRAS.358..363C}.}.

\section{DISCUSSION}
\label{sec:dis}

Before starting the discussion about the model performance and prediction it is worth taking a look at how the models are constructed.

\subsection{Model ingredients}

As opposed to \cite{2009ApJ...699..486C,2010ApJ...708...58C} and \cite{2010ApJ...712..833C} our aim has not been to judge each model ingredient and we thus find it sufficient to say that some, but certainly not all, SSP models can be successfully used to predict $u$,$g$,$r$,$i$,$z$-band photometry of the local galaxy population. Caution should be taken in the choice of model. Of the models we tested (see Table \ref{tab:t1}), \emph{we recommend the \cite{2003MNRAS.344.1000B} high resolution models for the purpose of predicting optical broad band photometry of galaxies.}

Does our library of star formation histories make sense? Given the distribution of average values for our models it seems we have covered essentially all possibilities with ages from 0-13Gyr, metallicities from 0.00-0.05 in Z and with extinctions from 0-3mag. Furthermore we note that Eq. \ref{eq:gav} produces star formation histories that resemble results from modelling of spectra of nearby galaxies by \cite{2004Natur.428..625H}. As mentioned in the introduction the inclusion of some stochasticity is expected to improve the realism of the models. In the context of the amplitude of the fluctuations in star formation over time we note the following. By studying the ensemble of UV to $H\alpha$ flux ratios in star forming dwarf galaxies \cite{2009ApJ...692.1305L} investigated the starburst mode of dwarf galaxies. They conclude that starbursts with amplitudes of $\sim4$ above the quiescent mode, which last for about 100Myr and occur every 1-2Gyr, could explain the whole population. The majority of our models do not have starburst episodes that exceed 4 times the local time average (see Fig. \ref{fig:sfh}).

Our dust model is rather simple. However, given the success in reproducing the optical colors of the observed galaxy population (after some fine-tuning of constants, see Fig. \ref{fig:new}), along with the observations that partially support our dust model (see Section \ref{sec:dust}) we believe it captures some of the essentials of obscuration by dust. It also turns out that the distribution of $z$-band extinctions for our models looks rather similar to the values for the galaxy sample of \cite{2003MNRAS.341...33K}.

In terms of line emission Fig. \ref{fig:em1} suggests that modelling $H\alpha$ and $H\beta$ is not sufficient for the modelling of optical colors. If GALEV treats line emission adequately, this model further tells us that line emission only has a second order effect on $integrated$ $u$,$g$,$r$,$i$,$z$-band photometry of nearby galaxies ($M_{r}\le17$), but caution should be taken at higher redshifts since migration of strong emission lines between filter bands does cause color shifts. Future studies improving the accuracy of the estimation of emission line strengths are on the way \citep{2011AJ....141..133G} and will further improve the quality of stellar population models.

Given the limited sensitivity of $u$,$g$,$r$,$i$,$z$-band photometry to the IMF as shown in Fig. \ref{fig:imf} we conclude that the IMF slope, and consequently the stellar mass, cannot be $measured$ with integrated optical broad band colors of galaxies at least in the range explored here (single slope IMFs from 2.1-2.9, Salpeter=2.35). This result is expected considering that most light from a stellar population with a ``normal'' IMF comes from stars of a rather limited mass range that reside close to the main sequence turnoff, on the giant branches or that are supergiants.

Is the number of models sufficiently large? When determining galaxy properties along with errors from colors using, e.g., a Bayesian maximum likelihood, the density of models in color space is of importance. Judging from Fig. \ref{fig:age}-\ref{fig:d} our library is sufficiently large for this purpose except at the very edges of the grid if using ($u$-$g$) vs. ($r$-$i$) or ($g$-$r$) vs. ($i$-$z$). However, if more than two colors are used the required density of the model library increases and caution has to be taken. The errors in Fig. \ref{fig:sfh} do not appear to be underestimated, except for one galaxy falling outside the model grid, and the method we use thus seems to work properly even for 10 colors.

The galaxy spectra are rather poorly reproduced with our model library, as mentioned in Section \ref{sec:ifs2} both due to issues with lines and the continuum.
%, though a significant improvement with respect to SSP models is achieved.
The fact that the \emph{reduced} $\chi^{2}$ can be lowered by considering a shorter wavelength range probably indicates that our model struggles to reproduce the overall continuum shape, but can preform better over shorter wavelength intervals. Spectral lines may be poorly reproduced due to the simple one parameter treatment of element abundances. We should on the other hand be able to adequately model the continuum, especially considering its close connection to colors. However, in the modelling of galaxy spectra it is often assumed that the continuum is unreliable due to, e.g., flux calibration issues, and continuum variations are therefore removed with polynomial fits \citep[cf.][]{2007IAUS..241..175C,2009A&A...501.1269K}. It is thus not completely clear whether the poorly fitted spectra mainly reflect problems with the models or the observations.

\subsection{Model predictions}

It is well known that optical colors can be used to separate star forming galaxies from quiescent ones. However, metallicity and dust extinction also have major impact. Our models suggest that, by using optical colors only, it is possible to reduce the age-metallicity degeneracy and constrain ages and metallicities fairly well over a surprisingly large part of the parameter space (see Fig. \ref{fig:age}-\ref{fig:z}). Furthermore, an interesting prediction of the model is that, to some extent, the amount of internal extinction can be measured using optical colors (Fig. \ref{fig:d}). Is this really the case? We expect extinction to have the strongest effect in edge on disk galaxies. We therefore selected two subsamples of galaxies, with high and low estimated extinction, respectively (Fig. \ref{fig:off}). It does indeed look like the high extinction bin contains a lot more edge on disk galaxies than the low extinction bin and, moreover, dust is seen in many more cases in the high extinction bin. It thus appears as if our color based estimate is reasonable.

Fig. \ref{fig:es} shows a rather good agreement between spectroscopically and photometrically determined galaxy properties in terms of average ages, metallicities, effective extinctions, specific star formation rates, and mass to light ratios. The strong correlation between ($i$-$z$) and metallicity in the photometric model (Fig. \ref{fig:z} and \ref{fig:es}) is not as pronounced in ($i$-$z$) vs. spectroscopic metallicity (Fig. \ref{fig:es}), suggesting that this particular model prediction may not be robust. We note, however, that the correlation between ($i$-$z$) and metallicity is present in all models we have considered (see Table \ref{tab:t1}). The wavelength dependence of the luminosity weighting, i.e. the wavelength dependence of the relative contribution of light from young and old stellar populations, could influence the metallicity estimate. If the spectroscopic metallicities are based on strong absorption lines line like e.g. the Mg line at $\sim5100\mbox{\AA}$, they are likely more influenced by young stellar populations and blue horizontal branch stars than the $i$ and $z$-band which have effective wavelengths of 7472 and 8917$\mbox{\AA}$. The source of the discrepancy in metallicity is thus not resolved.

As the amount of dust present is coupled to a galaxy's specific star formation rate in our model, galaxies with young stellar populations are often heavily obscured. The introduction of the prior (see Sect. \ref{sec:bay}) reduces the degeneracy between old and dust-free and young and dusty galaxies. However, a small number of old galaxies (ages determined by spectroscopy) still have young stellar populations according to our photometric model (Fig. \ref{fig:es}). Figure \ref{fig:es2} reveals some systematic differences between the parameters derived from photometry and spectroscopy along with some scatter. However, deviations are to be expected given the large difference in the amount of information contained in the two kinds of data (5 data points along the spectral energy distribution for $u$,$g$,$r$,$i$,$z$ with respect to $\sim$2000 for SDSS spectra). To probe galaxy properties with optical photometry thus stands as a viable option, especially useful when limited spectral information is available such as in the case of severe aperture bias. We would also like to point out that spectroscopically derived quantities are not necessarily ``correct'' as model dependencies may be present.

The exclusion of some bands can slightly diminish the differences between photometrically and spectroscopically derived quantities (see Sec. \ref{sec:ifs2}). This may be caused by the wavelength dependent luminosity weighting, but could also reflect that certain colors do not trace all parameters we study.

An interesting result is that the newer CB07 models perform worse than the BC03 models in reproducing the colors of the nearby galaxy population. We would like to point out that this is not the same as saying that the ingredients of the BC03 model are somehow better than the newer version. We speculate that the offsets in the CB07 models may be a backlash of trying to optimize the model for predictions over a larger wavelength range. In BC03 as well as CB07, models based on the empirical library of stellar spectra by \cite{2003AA...402..433L} do improve the modelling with respect to models based on stellar atmosphere models of \cite{1997AAS..125..229L}. This suggest that more uncertainties are introduced through a complete modelling than through an empirical approach \cite[see also][]{2009MNRAS.394L.107M,2011MNRAS.418.2785M,2011ApJ...737....5P}.

A subclass of galaxies at ($g$-$r$,$i$-$z$)=(0.4,0.0) can not be modelled by any of the models considered (see Fig. \ref{fig:off}). Most of these galaxies are rather faint in the $z$-band ($<m_{z}>\sim17.0$ as compared to $<m_{z}>\sim15.5$ for the entire sample), which has led us to suspect that a part of this population could simply be an artifact caused by photometric errors. The fraction of galaxies with small $z$-band photometric errors $\sigma_{m_{r}}<0.05$ in the region ($0.3<(g-r)<0.4$,$-0.05<(i-z)<0.05$) is a factor of two smaller than for the overall sample. However, this is not sufficient to explain the entire population. Another explanation is that these galaxies are even more metal poor than the most metal poor models (see Fig. \ref{fig:z}) or, alternatively, that the SSP models fail at low metallicity.

The star formation histories, along with average ages, metallicities and $r$-band optical depth shown in Fig. \ref{fig:sfh} look reasonable at first glance. A quiescently looking galaxy, which in fact does not have any emission lines in its SDSS spectra, should according to the models have some amount of recent star formation. However, non star forming models lie within the errors. We thus conclude that the model output is reasonably reliable -- within the limits of broad band photometry -- in providing information about a galaxy's star formation history and dust content.

\section{SUMMARY}
\label{sec:sum}

If an adequate range of star formation histories, chemical enrichments and dust obscuration are being considered, some of the currently available stellar population models perform well in reproducing the integrated optical $u$,$g$,$r$,$i$,$z$-band photometry of nearby galaxies (see Fig. \ref{fig:new}). However, strong differences between the various models are seen and it is therefore necessary to carefully consider the choice of model before proceeding with the analysis. \emph{Our preferred model library is available on request.} Optical broad-band colors are insensitive to the slope of the IMF, but can on the other hand put constraints on star formation histories, chemical enrichments, and dust extinctions as verified through a comparison with spectroscopic data. The accuracy of the constraint depends on the galaxy's position in color space. This is mainly due to overlap of reddened and dust-free models in certain parts of the color space.

\section{ACKNOWLEDGEMENT}

We would like to thank Jay Gallagher for useful comments during the course of this work.

K.S.A.H. and T.L. are supported within the framework of the Excellence Initiative by the German Research Foundation (DFG) through the Heidelberg Graduate School of Fundamental Physics (grant number GSC 129/1).

K.S.A.H. acknowledges his funding from the University of Heidelberg through a Landesgraduiertenf\"orderung (LGFG) fellowship for doctoral training.

Funding for the SDSS and SDSS-II has been provided by the Alfred P. Sloan Foundation, the Participating Institutions, the National Science Foundation, the U.S. Department of Energy, the National Aeronautics and Space Administration, the Japanese Monbukagakusho, the Max Planck Society, and the Higher Education Funding Council for England. The SDSS Web Site is http://www.sdss.org/.

    The SDSS is managed by the Astrophysical Research Consortium for the Participating Institutions. The Participating Institutions are the American Museum of Natural History, Astrophysical Institute Potsdam, University of Basel, University of Cambridge, Case Western Reserve University, University of Chicago, Drexel University, Fermilab, the Institute for Advanced Study, the Japan Participation Group, Johns Hopkins University, the Joint Institute for Nuclear Astrophysics, the Kavli Institute for Particle Astrophysics and Cosmology, the Korean Scientist Group, the Chinese Academy of Sciences (LAMOST), Los Alamos National Laboratory, the Max-Planck-Institute for Astronomy (MPIA), the Max-Planck-Institute for Astrophysics (MPA), New Mexico State University, Ohio State University, University of Pittsburgh, University of Portsmouth, Princeton University, the United States Naval Observatory, and the University of Washington.

\appendix
\section{GALAXY COLORS}
\label{sec:a}

\begin{figure}
\includegraphics[width=8.5cm]{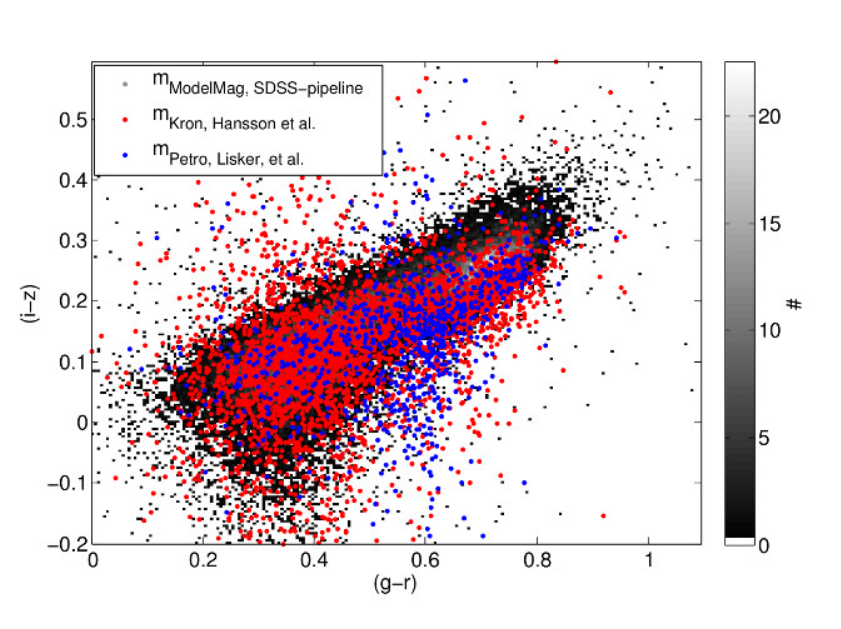}
\includegraphics[width=8.5cm]{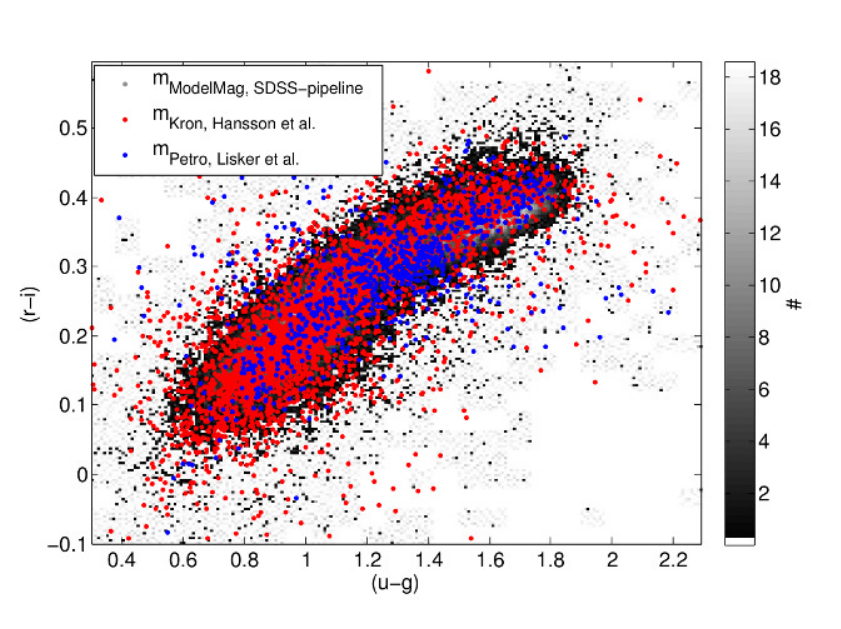}
\caption{SDSS pipeline modelMags for the galaxy sample used in this paper (grey 2D histogram), Kron magnitudes of SDSS galaxies from Hansson et al. (in prep.) and Petrosian magnitudes from \protect\cite[][Meyer et al. (in prep.)]{2008AJ....135..380L,2009ApJ...696L.102J} in the ($r$-$i$) versus ($u$-$g$) and ($i$-$z$) versus ($g$-$r$) planes. All magnitudes were corrected for Galactic foreground extinction.}
\label{fig:phot}
\end{figure}

We test the colors of the galaxy sample we use in this paper against two independent samples derived from SDSS data. The first sample from \citet[][]{2008AJ....135..380L,2009ApJ...696L.102J} and Meyer et al. (in prep.), hereafter S1, consists of Virgo cluster galaxies for which colors are measured within 2 effective radii or 1 Petrosian radius. The second sample is from Hansson et al. (in prep)., hereafter S2, and consists of an essentially volume limited sample of galaxies brighter than $M_{r}\sim-16$ within 50Mpc. Colors for these galaxies are measured within 2.5Kron radii, as determined from the $r$-band image using SourceExtractor \citep{1996AAS..117..393B}. For these two samples as well as for the sample we use in this paper, hereafter S3, ($r$-$i$) vs. ($u$-$g$) and ($i$-$z$) vs. ($g$-$r$) colors are given in Fig. \ref{fig:phot}. The general agreement is good, but differences are seen. S3 have on average higher values of ($i$-$z$) at ($g$-$r$)=0.7. We test whether this can be explained by the difference in how the magnitudes are computed by looking at the change in colors when adopting Petrosian magnitudes (PetroMag) instead of modelMags. The median change in ($i$-$z$) is -0.06mag and for the other colors it is between -0.02mag and 0.00mag. Because the definition of Kron magnitudes is similar to Petrosian magnitudes we conclude that this offset can be explained by the difference in magnitude definitions. S1 also have many more galaxies in the region around ($g$-$r$,$i$-$z$)=(0.6,0.0). We find that all these galaxies are fainter than about $M_{r}=-16$ and as such are therefore absent in S3.

\section{SPECTRAL FITTING}
\label{sec:d}

The spectra are deredshifted using the redshift provided by the SDSS according to
\begin{equation}
\lambda_{rest}=\lambda_{obs}\sqrt{\frac{1-v/c}{1+v/c}}
\end{equation}
where $\lambda_{rest}$ and $\lambda_{obs}$ are the rest wavelength and observed wavelength, respectively, $v$ is the velocity corresponding to the Doppler shift and $c$ is the speed of light.

The best models in our library of BC03hr spectra reproducing the data are found by means of calculating the reduced $\chi^{2}$ between each model and the observations. In the fitting process we make use of the spectral region from $3700\mbox{\AA}$ to $7000\mbox{\AA}$. Regions affected by the four brightest skylines (OI5577, OI6300, OI6354 and NaD5890) are masked out. The model library is at a slightly lower resolution ($\sigma_{mod}=3\mbox{\AA}/2.35$) than the data ($\sigma_{obs}\sim2.5\mbox{\AA}/2.35$ at $\lambda=5000\mbox{\AA}$) and we therefore degrade the resolution of the observed spectra to FWHM=$3\mbox{\AA}$ by convolving with a Gaussian with $\sigma=\sqrt{\sigma_{mod}^{2}-\sigma_{obs}^{2}}$. As the model library is fairly large a direct fit to all models would be rather computationally expensive. We therefore carry out a two step fitting making use of binned models produced in the following way. The rms between one model and all other models is computed. The average of the 225 models\footnote{The number of models in each bin was chosen for speed optimization ($\sqrt{50000}\sim224$).} with the lowest rms is then computed, constituting the first binned model. The procedure is repeated but all models already included in the binned spectra are excluded. After 223 repetitions we have 223 binned models, each of which, except the last one, points to 225 other models. In the fitting procedure we first find the best binned model and secondly the best model in the bin. Such a fitting need not be equivalent to a $\chi^{2}$ fit of all models since the best-fitting model not necessarily needs to be contained in the best-fitting bin. Nevertheless, we find this approach preferable since it accelerates the fitting procedure by a factor of about $10^{2}$.

Two things need to be considered before comparing the models with the observations, namely the line of sight velocity distributions of the stars and the emission lines. We assume that the former can be modelled by a Gaussian distribution. For each spectrum the minimum $\chi^{2}$ is found after convolving the spectrum with Gaussians having dispersions from 100 to 285 km/s\footnote{A lower limit of 100km/s is adopted to avoid ending up with observations at higher resolution than the models at short wavelengths (SDSS spectra have $\delta \lambda/\lambda \sim2000$). With an upper limit of 285km/s we can model all except some of the most massive galaxies in the Universe \citep{2007AJ....133.1741B}.} in 12 logarithmically spaced steps. The procedure is repeated 25 times after having perturbed each pixel of the observed spectra with its errors. The best-fitting dispersion is taken to be

\begin{equation}
\sigma = \frac{\sum{\sigma_{i}\exp(-\frac{1}{2}\chi_{i}^{2})}}{\sum{\exp(-\frac{1}{2}\chi_{i}^{2})}}
\end{equation}

where $\sigma_{i}$ is the best-fitting dispersion of the $i$th Monte Carlo realization, and where $\chi_{i}^{2}$ is the lowest reduced $\chi^{2}$ of the $i$th Monte Carlo realization.

The emission lines we consider in the fit are those 20 emission lines considered in \cite{2002AJ....123..485S}. These lines are expected to be the strongest in star forming galaxies according to the models of \cite{2003AA...401.1063A}, and furthermore include the strongest emission lines expected due to AGN activity. For simplicity all 20 lines are masked out in the fit.

\section{BAYESIAN MAXIMUM LIKELIHOOD MODELLING}
\label{sec:e}

\begin{figure}
\includegraphics[width=8.5cm]{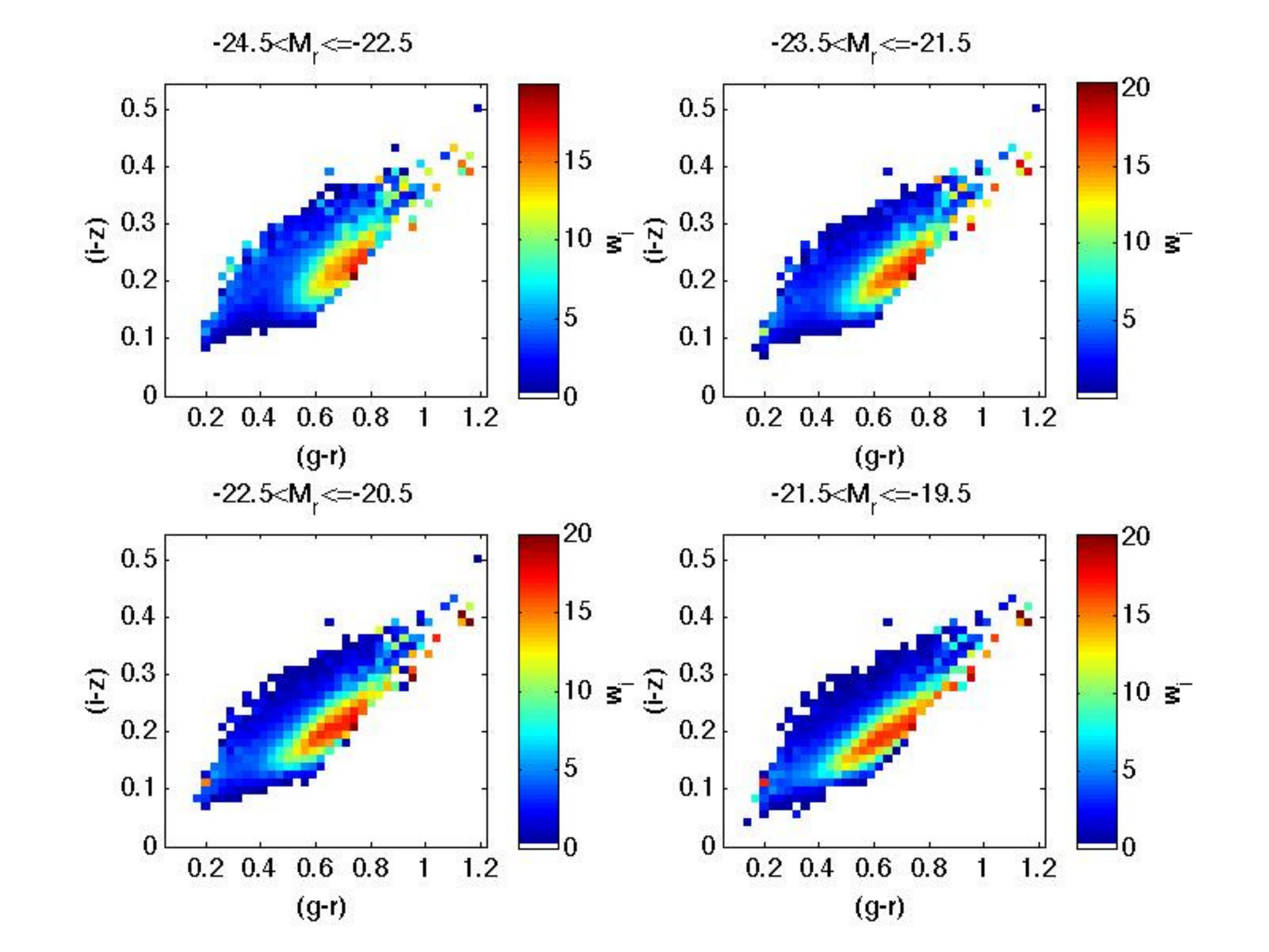}
\includegraphics[width=8.5cm]{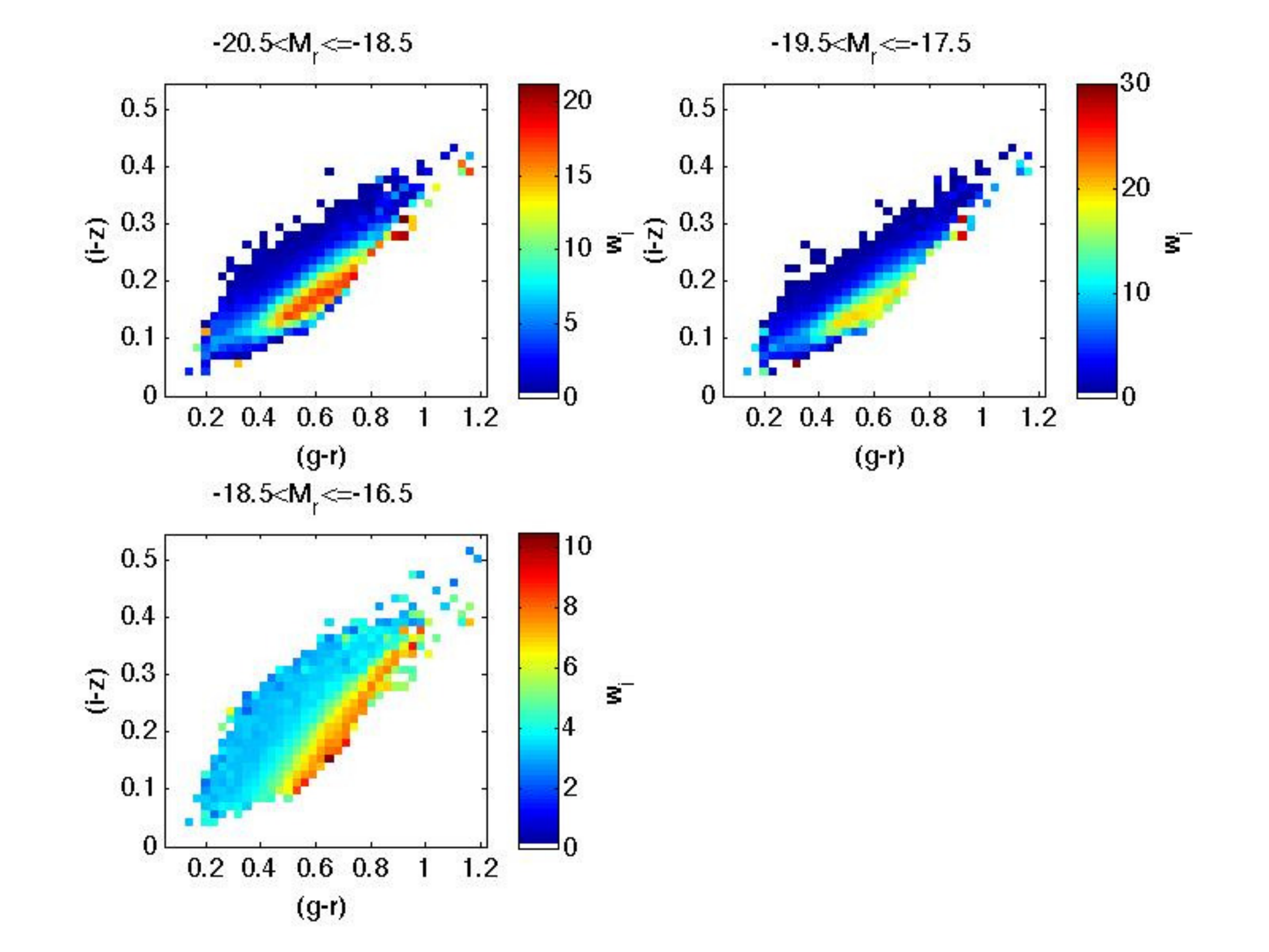}
\caption{Distribution of Bayesian weights, $w_{i}$, in the ($i$-$z$) vs. ($g$-$r$) diagram (color coded pixels) for the seven bins in $M_{r}$ adopted.}
\label{fig:bayt}
\end{figure}

Assuming the errors are Gaussian the star formation history, stellar mass weighted stellar metallicity and effective extinction can be computed according to
\begin{equation}
a = \frac{\sum \limits_{i=1}^{i=n}\sum \limits_{j=1}^{j=m} w_{i}a_{i}e^{-\frac{1}{2}(\frac{c_{i,j}-c_{o,j}}{\sigma_{c_{o,j}}})^{2}}}{\sum \limits_{i=1}^{i=n}\sum \limits_{j=1}^{j=m} w_{i}a_{i}}
\end{equation}
where $a$ is any of the properties mentioned above, $i$ denotes a model, $o$ an observation, $c_{j}$ a color, $\sigma_{c_{o,j}}$ the error in an observed color and the summations are done over all $n$ models and all $m$ colors. The prior, i.e. the weights, $w_{i}$, are taken as the ratio of the number of models in the library and the number of galaxies in the \cite{2011MNRAS.413..101G} catalog\footnote{We only make use of the part of the simulation box known as milli-millenium. The sample contains more than 50000 galaxies which ought to be sufficient.} for bins\footnote{-24.5--22.5,-23.5--21.5,...,-18.5--16.5mag. Note that even in the absence of spectroscopic distances $M_{r}$ can be estimated based on the observed galaxy structure.} in $M_{r}$ where the ratios are computed for models and galaxies within 0.1dex in stellar mass weighted age and 0.3dex in stellar mass weighted z of $i$. Figure \ref{fig:bayt} illustrates how the Bayesian approach influence the model grid in terms of age and metallicity.

\bibliographystyle{mn2e}
\bibliography{ugriz}
\end{document}